\documentclass{article}
\usepackage{microtype}
\usepackage{graphicx}
\usepackage{enumitem}
\usepackage{booktabs}
\usepackage{makecell}
\usepackage{caption}
\usepackage{subcaption}
\usepackage{multirow}
\usepackage{titlesec}
\usepackage{siunitx}
\usepackage{xurl}
\usepackage{tikz}

\usepackage[normalem]{ulem} 

\usepackage{amsmath}
\usepackage{amssymb}

\usepackage[section]{placeins}   
\usepackage{dblfloatfix}         

\newcommand{\ZZ}{\mathbb{Z}}                
\newcommand{\Cay}{\operatorname{Cay}}       
\newcommand{\Qn}{\ensuremath{Q_n}}          
\usepackage{hyperref}

\usepackage[accepted]{mlsys2025}

\mlsystitlerunning{Workload-Static Optical Circuit Switching for Scalable LLM Inference Decode}
\defcitealias{ualinkspec}{UALink Spec.}

\begin{document}

\twocolumn[
\mlsystitle{Workload-Static Optical Circuit Switching for Scalable LLM Inference Decode}

\begin{mlsysauthorlist}
\mlsysauthor{Maximilian Thürmer}{salience}
\mlsysauthor{Nicolas Scotto di Perto}{salience}
\mlsysauthor{Eloise Stein}{salience}
\mlsysauthor{Natanijel Vasic}{salience}
\mlsysauthor{Mikaeil Bagherli}{salience}
\mlsysauthor{Zhenguo Wu}{nvidia}
\mlsysauthor{Benjamin Klenk}{nvidia}
\mlsysauthor{Holger Fröning}{hawaii}
\mlsysauthor{Johannes Feldmann}{salience}
\end{mlsysauthorlist}

\mlsysaffiliation{salience}{Salience Labs Ltd, 100 Park Drive, Oxfordshire, OX14 4RY, United Kingdom}
\mlsysaffiliation{hawaii}{Hardware and Artificial Intelligence Lab, Heidelberg University, Germany}
\mlsysaffiliation{nvidia}{NVIDIA Corporation, Santa Clara, CA, USA}

\mlsyscorrespondingauthor{Maximilian Thürmer}{max@saliencelabs.ai}

\mlsyskeywords{LLM, Optical Circuit Switching, Inference, Decode, Scalability}

\vskip 0.3in

\begin{abstract}
LLM inference decode executes fine-grained collectives whose latency directly limits per-user interactivity, yet scale-up fabrics are still built from electronic packet switches designed for training. We observe that, for a given model deployment and parallelization, the communication structure of decode is fixed, and propose workload-static optical circuit switching (OCS): circuits are configured once per deployment, co-designed with model parallelism and with topology-aware collective schedules that turn spare accelerator ports into fewer communication phases or higher bandwidth. We evaluate this design with \textsc{gxnet}, a graph-based workload emulator coupled to analytical and flit-level network models grounded in hardware measurements. In a 576-accelerator scale-up domain serving 5T-parameter MoE models with up to 5M tokens of context, all-OCS fabrics improve interactivity by up to 65\,\% and energy efficiency by up to 119\,\% over 4:1 oversubscribed packet-switched fabrics.
\end{abstract}
]

\printAffiliationsAndNotice{}

\section{Introduction}

Large language model inference increasingly stresses not only accelerator throughput, but also latency, energy efficiency, and system cost. While training and prefill are dominated by large matrix operations and comparatively large communication transfers, autoregressive decode generates one token or verifies only a few at a time and repeatedly executes fine-grained communication whose latency directly affects user-facing responsiveness~\cite{yu2022orca,kwon2023vllm,zhong2024distserve,agrawal2024sarathi}.

At the same time, growing model size, mixture-of-experts architectures, and long-context inference increase parameter, activation, and KV-cache footprints~\cite{kwon2023vllm,aminabadi2022deepspeedinference,deepseekv4}. As these requirements outgrow per-accelerator memory capacity, individual model instances increasingly span many accelerators~\cite{aminabadi2022deepspeedinference,bhatia2025helix,yang2025contextparallelism}, creating pressure to extend tightly coupled scale-up fabrics to larger domains and beyond conventional rack boundaries~\cite{shou2025infinitehbd,zuo2025cloudmatrix,nvidia2026verarubinpod}.

The consequence is a tension between scale and latency: decode benefits from low token-to-token latency and therefore low round-trip time within the scale-up fabric~\cite{singhania2026}. 
However, extending electronic-packet-switched scale-up networks beyond a single rack typically requires additional switching levels, longer physical paths, or both, increasing end-to-end communication latency. 

\begin{figure}[t]
  \centering
  \includegraphics[width=1.0\linewidth]{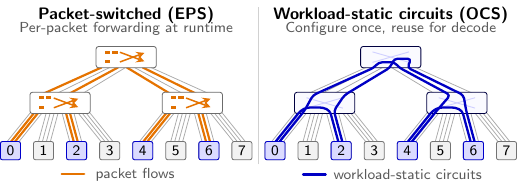}
  \caption{Schematic two-level scale-up fabrics connecting a highlighted collective group (ranks 0, 2, 4, and 6). EPS forwards packets independently at runtime; OCS circuits are planned once for the deployment's collective demands and reused throughout decode. Gray connections denote other physical links. Paths are illustrative rather than a bandwidth-matched comparison.}
  \label{fig:intro}
\end{figure}

Optical links are attractive for spanning larger accelerator domains, while optical circuit switching (OCS) can avoid repeated electronic packet switching (EPS) at intermediate levels~\cite{jouppi2023tpuv4}. 
Figure~\ref{fig:intro} contrasts conventional packet-switched fabrics with our workload-static OCS approach: circuits are configured for the model deployment and reused throughout decode rather than reconfigured within the token-generation loop. 
Efficient use of such circuits requires physical topology, routing, collective schedules, device placement, and model parallelization to be co-designed~\cite{wang2023topoopt,wang2024railonly,jouppi2023tpuv4}.

Prior work has addressed distributed training~\cite{narayanan2021megatron,jiang2024megascale}, workload and network simulation~\cite{won2023astrasim2,agrawal2024vidur}, inference serving and disaggregation~\cite{yu2022orca,kwon2023vllm,agrawal2024sarathi,zhong2024distserve,patel2024splitwise,qin2025mooncake}, and expert- or context-parallel inference~\cite{deepseekv3,zhu2025megascaleinfer,zuo2025cloudmatrix,yang2025contextparallelism,bhatia2025helix}. 
These approaches improve accelerator utilization and throughput, but do not remove the latency sensitivity of low-batch or high-interactivity\footnote{We use \emph{interactivity} to denote the per-user token generation rate, i.e., tokens per second per active user, the inverse of token-to-token latency during autoregressive decode.} decode. 

OCS has been explored for datacenter networks~\cite{poutievski2022jupiter,liu2023lightwave}, scale-out fabrics~\cite{wang2023topoopt}, and distributed training~\cite{jouppi2023tpuv4,khani2021sipml,liao2025mixnet,shou2025infinitehbd}, typically for provisioning, topology optimization, or workload-phase-dependent reconfiguration. 
Decode poses a different challenge: communication occurs at fine temporal granularity, while OCS reconfiguration is generally too slow to follow the token-generation loop. This makes per-token optical reconfiguration unattractive for latency-critical inference.

The co-design of decode parallelization, collective communication, and optical scale-up topology remains underexplored. For inference decode, the central question is not whether OCS can follow token-level communication changes dynamically, but whether the communication pattern can be regularized and scheduled such that an optical topology can be configured once and reused throughout serving.

This work studies workload-static OCS as a scale-up network design point for large-scale LLM inference decode. The OCS fabric is configured at deployment time and kept fixed during token generation, acting as a workload-specific static optical topology that realizes the logical communication structure of distributed decode. For a fixed model, context regime, accelerator allocation, service-level objective (interactivity target) and parallelization strategy, communication patterns are known at deployment time. Even in mixture-of-experts inference, token routing only modulates volume,
not connectivity: production all-to-all kernels communicate with every peer of a fixed
expert-parallel group during each call, with statically allocated per-peer buffers,
and with batches padded to pre-captured accelerator execution graph sizes~\cite{trtllm, vllm-a2a-onesided}.

Evaluating distributed AI workloads on compute systems requires exploring model architectures, parallelization strategies, accelerator properties, collective algorithms, and network topologies that are difficult to benchmark directly at full scale. Physical access to hundreds of future-generation accelerators is limited, and production traces are tied to specific hardware and software stacks. A flexible modeling framework is therefore needed to study the design space while grounding assumptions in kernel and communication behavior measured on real hardware.

This work makes the following contributions.
\begin{enumerate}
\item We introduce \emph{workload-static OCS} for LLM inference decode: optical circuits are configured once per deployment and co-designed with model parallelization and topology-aware collective schedules to realize recurring communication demand with low-latency logical overlays.
\item We introduce \textsc{gxnet}, a graph-based distributed LLM workload emulation framework that transforms model-level computation graphs into distributed execution traces with explicit compute, memory, and communication operators.
\item We combine analytical and event-based network modeling for all-EPS, hybrid EPS/OCS, and all-OCS scale-up fabrics. The analytical model supports broad design-space exploration, while a flit-level simulator estimates collective execution time, routing behavior, credit effects, and link utilization.
\item We evaluate 1T- and 5T-parameter MoE decode across 128--512 serving accelerators in a 576-accelerator scale-up domain. All-OCS improves interactivity by up to 65\,\% and system-level energy efficiency by up to 119\,\% over all-EPS, while sensitivity analysis over bandwidth, radix, and accelerator latency identifies the regimes in which workload-static OCS is most beneficial.
\end{enumerate}

More broadly, we argue that inference specialization should extend beyond accelerators to the scale-up network: communication structure known before execution need not be rediscovered dynamically by a general-purpose packet fabric.

\section{Background and Related Work}
\label{sec:background}

There is a large corpus of prior literature which addresses many separate pieces of the work presented here. Some work is devoted to related use cases - e.g. training rather than inference. Other contributions focus on a subset of the design space addressed here - for instance ignoring the co-optimization of collective schedules and underlying network topology:

Large scale distributed LLM training has received wide attention in the past. The parallelization strategies were studied~\cite{narayanan2021megatron} and cast into analytical, roofline-based models~\cite{isaev2023calculon}. Other approaches which seek to model computation and communication separately are partially based on real hardware traces or synthetic ones that are then feeding a network simulator~\cite{won2023astrasim2}. These approaches, however, have been developed for training. Calculon~\cite{isaev2023calculon} models the scale-up and scale-out networks with single-level latency and bandwidth parameters and assumes bandwidth-optimal collective implementations, which captures step time for large, bandwidth-dominated collectives but not the small, latency-bound collectives of decode. Network simulators such as ASTRA-sim~\cite{won2023astrasim2} resolve communication in more detail, but do not treat parallelization strategy, collective schedule, and optical circuit topology as a joint design space. Inference-specific simulators such as Vidur~\cite{agrawal2024vidur} and LLMServingSim~\cite{cho2024llmservingsim} address serving-level questions such as scheduling, batching, and deployment configuration; their communication model is either profiled on existing hardware~\cite{agrawal2024vidur} or inherited from ASTRA-sim~\cite{cho2024llmservingsim}, so the scale-up fabric is not a design variable.

Inference serving and disaggregation methods have been widely described~\cite{yu2022orca,kwon2023vllm,trtllm,zhong2024distserve,patel2024splitwise,qin2025mooncake}. In particular, advances in prefill/decode disaggregation~\cite{zhong2024distserve,patel2024splitwise} have identified the difference in computation, memory access and network access footprints between prefill and decode phases as a root cause of inefficient inference serving. Consequently, recent  work explores dedicated hardware for these two phases of inference and extends this approach by further decomposing the decode phase into attention and feed-forward-network portions, resulting in heterogeneous inference hardware~\cite{nvidia2026lpx,zhu2025megascaleinfer}, while others argue that shifting phase ratios favor a single balanced accelerator~\cite{openai2026jalapeno}. In these cases, the network architectures remain packet-switched~\cite{openai2026jalapeno,microsoft2026maia} or use fixed direct-connect topologies~\cite{nvidia2026verarubinpod,google2026tpu8} whose physical connectivity is not tailored to the workload, with many accelerator links reaching into the scale-up network.

Extensive work has examined the role of OCS in AI compute clusters ever since the inception of Google's TPUv4~\cite{jouppi2023tpuv4}. There has been work focusing on optimizing the network topology through OCS~\cite{wang2023topoopt,liao2025mixnet} and investigations into the required reconfiguration speeds of OCS control planes and optical transceiver relock times to exploit the very predictable and long-running nature of large collective operations resulting from AI training workloads such that an OCS operates more like an optical burst switch (OBS)~\cite{wu2025actina}. Harvest~\cite{rahman2026harvest} and its precursors~\cite{hammer2025shortcircuit,addanki2025lightbends} go further and synthesize optimal circuit schedules within individual collectives of scale-up domains, which requires fast OCS reconfiguration and tight coupling of the OCS and accelerator control planes, advancing the OCS towards something close to an optical packet switch. Other work has tried to leverage the dual network approach, scale-up inside the rack/box, scale-out between them - to reshape scale-out traffic within the scale-up domain such that the scale-out fabric reduces to rails~\cite{wang2024railonly} that can in turn be realized with OCS~\cite{ding2025photonicrails}. Apart from Harvest, all this work addresses training use cases and focuses on collective schedules for bandwidth optimality as a natural consequence. 

In latency-sensitive decode, bandwidth-optimal collective schedules are not necessarily latency-optimal, 
and resulting collective schedules on scale-up networks are already showing an increasing specialization by expected traffic volume as a consequence~\cite{trtllm-ar-strategy}. In decode, tensor, KV-cache, and expert parallelism can induce frequent collectives with small payloads, exposing accelerator-to-accelerator round-trip time rather than only aggregate bandwidth. Optimized communication kernels based on remote-store operations reduce software overhead, but not the underlying sensitivity to fabric latency. In parallel, collective acceleration engines~\cite{google2026tpu8,meta2026mtia} are moving collective execution into dedicated accelerator hardware. Serving stacks already select collective implementations, but through fixed, offline-tuned heuristics~\cite{trtllm-ar-strategy}. Exposing communication explicitly to an automated mapping and scheduling search is a step towards co-optimizing serving and collective implementation~\cite{openai2026jalapeno}, but the repercussions on the ideal network fabric serving the chosen collectives, or even network topology co-optimization for the chosen collective, remain open. The work presented here exploits this specialization more systematically by extending it from the collective schedule to the scale-up topology. It extends the idea of short-circuiting ring collectives~\cite{hammer2025shortcircuit} but, in contrast to per-collective reconfiguration~\cite{rahman2026harvest}, derives one topology for the composite decode schedule, i.e., for all collectives of inference decode at once, which the OCS then holds throughout serving, removing reconfiguration from the decode critical path.
\section{Workload-Static OCS Topology Design}
\label{sec:ocs-design}

Workload-static OCS maps the recurring collective communication demands of distributed decode onto optical circuits configured once at deployment time. 
Different model regions, notably attention and MoE/FFN, induce distinct logical connectivity requirements over the same multi-port accelerators. 
We jointly plan the corresponding logical overlays, collective schedules, and their embedding into the physical fabric, deriving fixed OCS configurations and precomputed accelerator routing tables.
Communication may involve data relay through intermediate accelerators, allowing the fixed circuits to support multiple logical topologies. 
We consider rail-optimized two-level scale-up fabrics~\cite{wang2024railonly}, whose physical construction is detailed in Appendix~\ref{app:network-topology}.

\subsection{Static Communication Demand in Decode}
\label{sec:static-demand}

\begin{table}[t]
\centering
\caption{Deployment-time communication demand per decode step for Megatron 5T (128 layers) on 512 accelerators, using
Helix parallelization $(\mathrm{KVP},\mathrm{TPA},
\mathrm{EP},\mathrm{TPE})=(4,128,64,8)$ and FP4 (AR: all-reduce, AG: all-gather, A2A: all-to-all). Region, axis, collective, groups $G$ of $R$ ranks, and calls per group are fixed for deployment; batch size $B$ only changes the payload per call.}
\label{tab:decode-demand}
\footnotesize
\setlength{\tabcolsep}{3pt}
\begin{tabular}{@{}llcrrrr@{}}
\toprule
 & & & & \multicolumn{3}{c}{Payload per call} \\
\cmidrule(l){5-7}
Axis & Coll. & $G\times R$ & Calls & $B{=}1$ & $B{=}4$ & $B{=}16$ \\
\midrule
\multicolumn{7}{@{}l}{\emph{Attention}} \\
\quad world & AR        & $1\times512$  & 128 & 4\,KiB   & 16\,KiB  & 64\,KiB  \\
\quad KVP   & A2A       & $128\times4$  & 128 & 66\,B    & 264\,B   & 1056\,B  \\
\quad TPA   & none      & $4\times128$  & 0   & --       & --       & --       \\
\multicolumn{7}{@{}l}{\emph{FFN}} \\
\quad EP    & AG        & $8\times64$   & 128 & 256\,KiB & 512\,KiB & 768\,KiB \\
\quad TPE   & AR        & $64\times8$   & 128 & 4\,KiB   & 8\,KiB   & 12\,KiB  \\
\multicolumn{7}{@{}l}{\emph{LM head}} \\
\quad world & AG        & $1\times512$  & 1   & 25\,KiB  & 100\,KiB & 400\,KiB \\
\bottomrule
\end{tabular}
\end{table}

We consider a prefill/decode-disaggregated deployment in which the scale-up fabric serves autoregressive decode. 
At deployment time, an inference framework selects the model, accelerator allocation, parallelization scheme (such as Helix or DP/TP/EP) and degrees (e.g. TP4 EP8), and operating targets such as maximum context length and interactivity. 
For a fixed configuration, the communication demand of each decode step is known structurally:
participating rank groups, mesh axes, collective primitives, and invocation counts are determined by the model and parallelization.
Payloads may depend on batch size and expert routing, but the communication support, the set of possible communicating rank pairs, required by the topology remains fixed.

Inference frameworks already select a collective implementation from a set of viable options: TensorRT-LLM picks between one-shot, two-shot and NCCL all-reduce from offline-tuned tables keyed on tensor-parallel group size, hidden size and token count. 
As scale-up domains grow, additional factors, such as placement-dependent latency, will have to be factored in. Workload-static switching as described here is a more systematic approach to optimize network communication under the same constraints. 

We use Helix-style distributed decode~\cite{bhatia2025helix}, which induces different communication demands across model regions.
Attention uses KV-parallel (KVP) and tensor-parallel (TPA) axes, whereas the MoE/FFN region uses expert-parallel (EP) and tensor-parallel (TPE) axes over the same accelerators. Table~\ref{tab:decode-demand} illustrates the resulting demand: its structural columns
are deployment-static, while batch size changes the payload per collective invocation. 
The KVP all-to-all exchanges partial attention outputs and log-sum-exp statistics across KV-cache shards, as required by Helix-style attention parallelism. Some axes induce no communication in a given region. 
In this configuration, context length affects KV-cache computation and memory traffic, but not the collective structure or payload.

Token routing may change traffic volume across peers, but need not change the communication support. 
Our evaluated expert axis uses fixed-group collectives with buffers padded to the busiest
expert-parallel rank, so FFN payloads need not scale linearly with batch size (Appendix~\ref{app:gxnet-moe}). 
Other implementations can regularize routing-dependent traffic through padding or bucketing, at the cost of additional traffic (Appendix~\ref{app:moe}). 
Thus, the fabric can be planned against recurring collective demands rather than topology-agnostic worst-case connectivity.

\begin{figure*}[htbp]
    \centering
    \begin{subfigure}[b]{\linewidth}
        \centering
        \includegraphics[width=\linewidth]{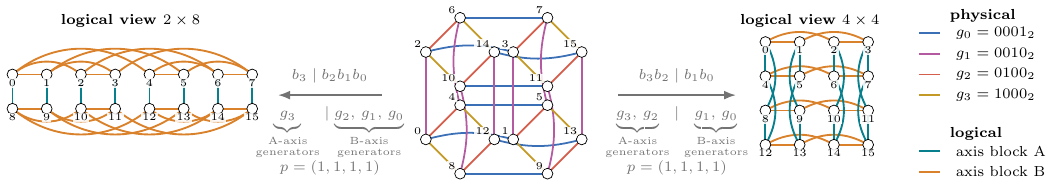}
        \caption{Same $Q_4$ host connectivity under logical dimension partitions $b_3 \mid b_2 b_1 b_0$ and $b_3 b_2 \mid b_1 b_0$, yielding logical shapes $2\times8$ and $4\times4$, respectively.}
        \label{fig:cayley-port-tradeoff_a}
    \end{subfigure}
    
    \vspace{0.5em}
    
    \begin{subfigure}[b]{0.45\linewidth}
        \centering
        \includegraphics[width=\linewidth]{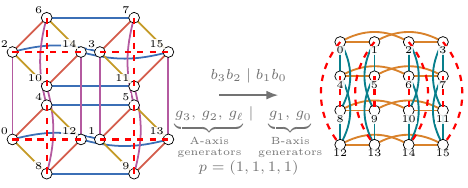}
        \caption{Latency shaping: an additional fold generator $g_\ell=12=1100_2=11\mid00_2$ augments axis-A connectivity and can reduce the collective phase count $h_i$.}
        \label{fig:cayley-port-tradeoff_b}
    \end{subfigure}
    \hfill
    \begin{subfigure}[b]{0.45\linewidth}
        \centering
        \includegraphics[width=\linewidth]{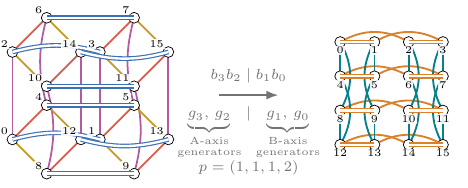}
        \caption{Bandwidth shaping: an additional circuit duplicates the $g_0=1=0001_2=00\mid01_2$ edge family, increasing its multiplicity $p_i$ without changing logical connectivity.}
        \label{fig:cayley-port-tradeoff_c}
    \end{subfigure}
    
    \caption{Workload-static OCS topology design example with base generator set $G=(g_3,\,g_2,\,g_1,\,g_0)$ inducing a $Q_4$ host graph under a shared per-accelerator port budget $L$. The same physical host graph supports multiple logical communication structures (a). 
    One additional port relative to the $Q_4$ base can be allocated either to shortcut connectivity for lower collective depth (b) or to parallel circuits for higher effective bandwidth (c).}
    \label{fig:cayley-port-tradeoff}
\end{figure*}

\subsection{From Communication Demand to Logical Topology}
\label{sec:logical-topology}

We convert the static communication demand of an inference decode forward pass into logical communication graphs. Table~\ref{tab:decode-demand} gives a worked example of such a demand pattern. 
Each graph corresponds to one mesh axis in one decode region; its vertices are the participating accelerator ranks, and its edges are the direct communication links available to the collective schedule. 
The topology planner embeds these graphs into the physical accelerator links.

Since different decode regions induce different logical meshes over the same accelerators, these graphs must be embedded jointly into a shared physical host graph. 
Its circuits remain fixed during serving, while different regions may reuse them through different logical views.

For power-of-two accelerator counts, we describe logical overlays through Cayley-style graphs. 
Ranks are labeled by bit strings, and generators connect ranks whose labels differ by a fixed mask. 
Unit generators yield cube-like structures; additional generators introduce folds or chords that reduce diameter. 
Folds extend the idea of short-circuiting ring collectives~\cite{hammer2025shortcircuit}: instead of reconfiguring the OCS per collective step, the shortcut circuits are provisioned once at deployment, so their cost is accelerator port degree rather than reconfiguration delay on the critical path.
Different mesh axes can thus be viewed as different partitions or colorings of one shared host graph, allowing several phase-specific logical topologies to be served by one workload-static set of optical circuits.
Figure~\ref{fig:cayley-port-tradeoff}(a) illustrates this for a \(Q_4\)-style host graph: the same physical connectivity induces different logical mesh views by changing the partition of rank-label bits. Appendix~\ref{app:shared-host} gives the formal construction.

\subsection{Port Allocation as Latency/Bandwidth Trade-off}
\label{sec:port-allocation}

The previous subsections turn a decode workload into a set of axis requirements over a shared accelerator set. 
We now expose the main degree of freedom of OCS-based scale-up fabrics with multi-port accelerators: each accelerator has a finite port budget \(L\), and every port can be spent either to reduce the number of serialized communication phases or to increase bandwidth along an already selected logical edge.

For one logical axis $i$, assume a regular overlay of degree $d_i$ and uniform edge multiplicity $p_i$, i.e., each logical edge on that axis is realized by $p_i$ parallel circuits. 
We estimate a collective schedule with a Hockney-style model
\begin{equation}
T_i \approx h_i \lambda_i + \frac{v_i}{p_i R},
\label{eq:hockney-axis}
\end{equation}
where $h_i$ is the number of serialized schedule phases, $\lambda_i$ is the per-phase accelerator-to-accelerator latency along the selected physical path, $v_i$ is the bottleneck volume, and $R$ is the bandwidth of one physical link. 
For an EPS path, $\lambda_i$ includes the latency of the electronic switch stages traversed; for an OCS path, it includes endpoint, link, and optical pass-through latency. 
Folds reduce $h_i$ by shortening the logical graph; parallel circuits increase $p_i$ and reduce serialization time, as shown in Figures~\ref{fig:cayley-port-tradeoff}(b) and~\ref{fig:cayley-port-tradeoff}(c).

More generally, let $H=(V,E)$ denote the physical host graph and $p_e$ the multiplicity of edge $e$.
Since each circuit consumes one port at its endpoint, the per-accelerator port budget is
\begin{equation}
\sum_{e \in \delta(u)} p_e \leq L \qquad \forall u \in V ,    
\label{eq:port-budget}
\end{equation}
where \(\delta(u)\) is the set of edges incident to accelerator \(u\). 

For disjoint axis-specific edge families of degree $d_i$ and uniform multiplicity $p_i$, this simplifies to $\sum_i d_i p_i \le L$.
In the shared-host construction, physical edges reused across different logical views are counted only once.
Thus, the OCS fabric exposes an explicit exchange rate between latency and bandwidth: ports spent on additional generators make the graph shallower ($h_i \downarrow$), while ports spent on multiplicity make selected edges wider ($p_i \uparrow$). 

A fold that reduces the schedule by $\Delta h$ phases saves approximately $\Delta h\lambda_i$, provided the schedule exploits it. Increasing multiplicity from $p_i$ to $p_i+1$ instead saves $v_i/[Rp_i(p_i+1)]$. 

With a locally linear volume model $v_i \approx c_i m$, where $m$ is the number of tokens per forward pass and $c_i$ the bottleneck volume per token, the crossover occurs at
\begin{equation}
m_i^\star \approx \frac{\Delta h \lambda_i R p_i(p_i+1)}{c_i}.
\label{eq:fold-width-crossover}
\end{equation}

Below this threshold, reducing schedule depth is preferable; above it, additional bandwidth becomes more valuable. 
High-interactivity decode therefore favors latency-shaped topologies for many, but not necessarily all, axes.

\subsection{Collectives as Topology-Specific Schedules}
\label{sec:topology-aware-collectives}

A logical topology improves collective performance only if the communication schedule exploits its connectivity. 
We describe each collective implementation by a logical overlay, a validity predicate specifying which overlays support the collective, and an ordered schedule of communication phases. 
Each phase specifies communicating rank pairs, the logical edges used, payload slices, and dependencies on preceding phases.
The schedule determines the serialized phase count, bottleneck volume, and link utilization for the analytical model, as well as the communication events initiated by a network simulator (such as \emph{Simlette}, see Section \ref{sec:network-model}).

We have implemented a \emph{planner}, conceptually part of the inference deployment framework, which jointly searches over axis partitions, folding generators, edge multiplicities, and compatible collective schedules under the shared accelerator port budget. 
It selects the topology--schedule combination minimizing predicted communication time for the deployment's target message-size regime. 
Folds reduce latency only when schedules exploit their additional connectivity, while parallel circuits improve bandwidth only when schedules distribute traffic across them.
Appendix~\ref{app:cost} details the topology search, and Appendix~\ref{app:schedules} describes the topology-aware collective algorithms.
\section{\textsc{gxnet}: Workload Modeling for Distributed LLM Decode}
\label{sec:workload}

\textsc{gxnet} converts a model-level computation graph and a deployment, i.e., accelerator specification, system size, precision, and parallelization strategy, into a shape-exact distributed decode execution with explicit compute regions, memory traffic, collectives, device groups, and timing dependencies. It provides the deployment-time communication demand for topology planning and the timed execution for network simulation. Appendix~\ref{app:gxnet} provides additional details.

\subsection{Model Graph and Sharding Transformations}

One decode step is a computation graph whose tensors carry shape, element type, and logical device placement.
A decode step is specialized to generating one token (or verifying multiple tokens) per active request against an existing KV cache.
Communication is derived structurally from this placement: wherever adjacent operators require different tensor placements, \textsc{gxnet} inserts a collective, whose group follows from the mesh axes, payload from tensor shape and element width, and occurrence count from the layer structure.
For a fixed model and parallelization, this yields the deployment-time demand of Table~\ref{tab:decode-demand} before execution.

The same sharded graph yields the per-rank weight and KV-cache footprint, which bounds the batch size feasible in accelerator memory. 
For MoE regions, we model dropless routing that is uniform in expectation: routing affects expert compute and FFN collective payloads, but not collective group membership; routing skew is not modeled.

\subsection{Per-rank Execution Trace and Cost Model}

\textsc{gxnet} lowers the communication-explicit graph into a per-rank execution trace, akin to traces recorded on real hardware with tools such as Chakra~\cite{sridharan2026chakra}: the compute regions, memory accesses, collectives, and dependencies a single accelerator carries out or is exposed to during a decode step.
Where a collective schedule (Section~\ref{sec:topology-aware-collectives}) describes how one collective uses a logical overlay, the trace determines when collectives occur, which ranks participate, what payload they carry, and whether their latency is exposed or hidden behind computation.

The dependency representation can express communication--computation overlap.
We use Helix's attention/FFN sharding without HOP-B, i.e., without splitting attention into microbatches whose all-to-all overlaps with the attention of the next, so communication latency is fully exposed in the decode-step timeline.
A sensitivity analysis (Appendix~\ref{app:hopb}) shows that HOP-B does not improve decode-step time in the evaluated configurations: for GPT-OSS-5.3-like, microbatching adds more all-to-all communication than HOP-B can hide, and for Megatron 5T, it yields no net benefit.

Each compute region's execution time is estimated with a roofline model taking the maximum of compute time and HBM-transfer time, using tensor-derived FLOPs and bytes together with size-dependent compute and memory efficiencies. 
Operators between communication points are grouped into fused execution windows and charged one launch overhead per window. 
Exact efficiency functions and accelerator parameters are given in Appendix~\ref{app:gxnet-cost}, and Appendix~\ref{app:accelerators} relates the model to kernel measurements on current hardware.

\textsc{gxnet} is analytical rather than cycle-accurate; projected accelerators use specification-level throughput and bandwidth parameters, while measured kernel durations can replace analytical estimates for existing hardware.

\subsection{Coupling \textsc{gxnet} to the Network Model}

\textsc{gxnet} interfaces with the network model at two points.
Before execution, it emits the complete communication demand of a decode step, against which the OCS fabric of Section~\ref{sec:ocs-design} is mapped and routed once at setup. In practice, an inference framework such as vLLM~\cite{kwon2023vllm} or SGLang~\cite{zheng2024sglang}, which already holds this information, would perform this step.
During execution, \textsc{gxnet} \emph{registers} each collective with its group, primitive, payload, axis, and earliest start time. It does so until all further events in its scheduler are blocked by dependencies to pending/registered collective operations. It then calls an \emph{advance} function to delegate control to the network simulator. The network simulator returns the collective's actual start and completion time on the specified network topology and for the given collective schedule. Once the completion time is known, dependent operators in \textsc{gxnet}'s internal scheduler are released. Whether a collective is exposed on the decode critical path is thus determined by the coupled execution, not by summing independently timed messages.

All collective instances whose dependencies are satisfied may be in flight simultaneously; Simlette advances them jointly and therefore captures contention and credit stalls between concurrent groups, whereas the analytical backend estimates collectives independently.
A run produces a timed decode-step timeline, from which we derive token-to-token latency, per-user interactivity, and system throughput, and, with the power model of Appendix~\ref{app:power-model}, energy-normalized throughput.

\section{Communication Model and Grounding}
\label{sec:network-model}

Evaluating workload-static OCS fabrics requires estimating the completion time of collective communication kernels on hardware and topologies that are not yet available at scale. We use two models for this: a fast analytical model to evaluate candidate topologies and schedules, and a flit-level simulator to capture effects that the analytical model misses, such as credit stalls and link contention. Both use latency parameters based on measurements from current hardware.

\subsection{\emph{Simlette}: a Flit-Level Event-Based Simulator}
\label{sec:Simlette}

We evaluate collective schedules and routing on EPS and OCS fabrics using \emph{Simlette}, a discrete-event simulator built on SimPy~\cite{simpy}.
Inspired by the UALink specification~\cite{ualinkspec}, it models memory-semantic transactions at flit granularity, capturing serialization, link contention, and credit stalls during execution.
\emph{Simlette} supports single-level and two-level spine-leaf fabrics, dynamic EPS packet switching and spraying, and workload-static OCS routing without runtime reconfiguration.

The simulator separates workload mapping into four independent layers: physical topology, worker allocation, collective schedule, and routing (Appendix~\ref{app:simlette-layers}).
This modular design allows us to vary each component independently while keeping the others fixed, enabling controlled comparisons of fabrics, placements, schedules, and routing strategies.

\subsection{UALink Transactions and Credit-Based Flow Control}
\label{section:credit_based_flow_control}

\emph{Simlette} implements the transaction and credit-based flow control system of the UALink specification, the same mechanisms used in practice for direct accelerator-to-accelerator communication. Each transaction pairs a request (e.g., a read or a write) issued by an Originator with the response returned by a Completer. One or more 64-byte Transaction Layer (TL) flits are grouped into a 640-byte Data Link (DL) flit before being put on the link, and switching and flow control operate at DL-flit granularity. Each port maintains independent credit channels for request and response traffic, so that traffic in one direction can never block the other. A port must hold a credit on the corresponding channel before sending a DL flit; credits are returned once the receiver has processed the data, so a buffer is never sent more data than it can store.

\subsection{Latency Decomposition}
\label{section:latency_decomposition_time_consumption}

The fixed per-hop latency is divided into four components: endpoint latency inside the accelerator (processor to cache to scale-up port), protocol latency of the scale-up port including credit transmission, link propagation delay, and switch forwarding delay. 

For the latency comparison, we apply common endpoint, protocol, and propagation parameters, while distinguishing EPS forwarding (approx. 300\,ns~\citepalias[Sec.~10.9.2]{ualinkspec} unloaded pin-to-pin latency) from OCS optical pass-through (approx. 10\,ns).

\begin{figure}[t]
\centering
\includegraphics[width=\columnwidth]{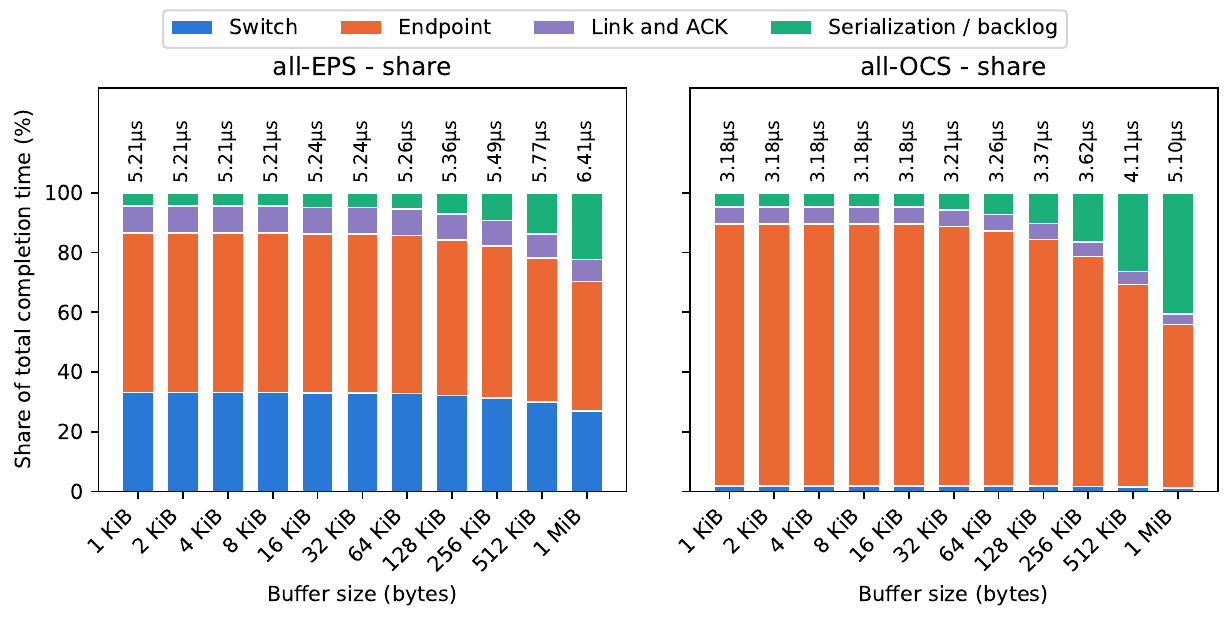}
\caption{Latency decomposition of a 512-accelerator all-reduce (parallel RS+AG schedule) on a single-level fabric (36 ports per accelerator, 400\,Gb/s links).
Total completion time is shown above each bar. Simulations were run using optimized mode (each message is sent as a single flit; see Appendix~\ref{app:sim-modes}).}
\label{fig:latency-decomposition}
\end{figure}

Figure~\ref{fig:latency-decomposition} shows the latency of a 512-accelerator all-reduce on an example single-level fabric.
The buffer size is the amount of data, in bytes, that each accelerator holds at the end of the collective. For the low-batch payloads of Table~\ref{tab:decode-demand} (4 to 64\,KiB), switch latency accounts for about a third of each EPS bar, while OCS removes almost all of it (2\,\% of the bar).

For buffers of 1 to 64 KiB, which cover the low-batch payloads of Table~\ref{tab:decode-demand}, the completion time is 3.18--3.26 $\mu$s for OCS, compared with 5.21--5.26 $\mu$s for EPS. As the payload grows beyond 64 KiB, serialization takes over on OCS, reaching 41\,\% of the time at 1 MiB (2.07 $\mu$s), where OCS remains faster with 5.10 $\mu$s compared with 6.41 $\mu$s for EPS.

\subsection{Analytical Model and Consistency Checks}
\label{sec:analytical-model}

The analytical model times each candidate port allocation of Section~\ref{sec:port-allocation} by Eq.~\eqref{eq:hockney-axis}, with the bottleneck volume $v$ obtained as the sum over all phases of the data carried by the most loaded edge (Appendix~\ref{app:analytical-derivation}). \emph{Simlette} then builds the circuits of the allocation with the lowest predicted time and simulates the collectives on them flit by flit, including credit waits and the load on each link.

The analytical model assumes that phases do not overlap, that no transfer shares its link with another, and that a sender never waits for buffer space at the receiver. Without contention and with unlimited buffers, \emph{Simlette} reproduces the analytical completion time for all message sizes and radices on a single-level OCS fabric (Appendix~\ref{app:analytic_vs_simlette}). When the buffer per port is too small to keep a link busy for one hop, senders wait for credits and \emph{Simlette} results are up to $1.9\times$ slower; when several flows reach the same switch port, as for the two-shot all-reduce on EPS, the analytical model underestimates completion time by up to $2\times$. 

The analytical model is accurate for contention-free schedules with sufficient buffering. 
\emph{Simlette} captures additional delays from credit stalls and concurrent traffic, which the analytical model omits. 
We therefore use the analytical model for design-space exploration and \emph{Simlette} to evaluate the selected configurations with a more detailed representation of contention and credit effects.

\subsection{Grounding}
\label{sec:grounding}

We parameterize \emph{Simlette} using hardware measurements and published interconnect specifications (Appendix~\ref{app:latency-profile}). 
Custom accelerator-to-accelerator ping-pong kernels estimate end-to-end communication latency, while pointer-chasing microbenchmarks characterize the local memory hierarchy~\cite{luo2025hopper}. 
On NVIDIA Blackwell B300 for example, memory-hierarchy measurements yield access latencies of approximately 14\,ns (shared memory), 17\,ns (L1), and 140\,ns (L2), compared with a measured accelerator-to-accelerator remote-access round trip of 1.61\,$\mu$s. These measurements constrain the accelerator-side latency budget, while the attribution of the remaining latency to protocol, interconnect, and switching components requires additional modeling assumptions (Appendix~\ref{app:latency-profile}).
We characterize accelerator-to-accelerator latency using several ping-pong kernel implementations, and we select the minimum measured duration as a minimum upper bound.  Appendix~\ref{app:latency-profile} details further measurements and implementation-dependent variation.

End-to-end measurements do not uniquely determine the latency contribution of every internal component. 
We therefore state the assumed component latencies explicitly and examine their influence through the sensitivity analysis of Appendix~\ref{app:sensitivity}. 
\emph{Simlette} produces deterministic completion times for the specified hardware and workload parameters; run-to-run jitter and tail-latency distributions are outside the scope of this evaluation.

\section{Evaluation}
\label{sec:eval}

\begin{figure}
    \centering
    \includegraphics{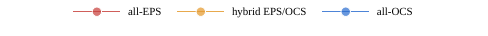}\\[-2pt]
    \subcaptionbox{Megatron 5T MoE\label{fig:per-batch-both-megatron}}[0.48\linewidth]{%
        \includegraphics{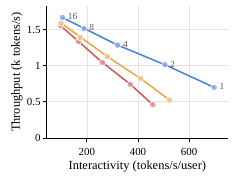}}\hfill
    \subcaptionbox{GPT-OSS-5.3-like\label{fig:per-batch-both-gpt-oss}}[0.48\linewidth]{%
        \includegraphics{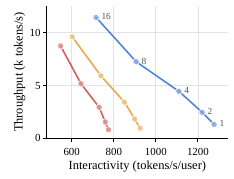}}
    \caption{Per-user interactivity vs.\ total system token throughput Pareto curves for two large LLMs provisioned for a context of 5 million tokens. The models are spread across 512 Rubin-like (RX00) accelerators, keeping 64 accelerators (eight per rack) in cold spare. Marker color distinguishes the physical topologies (Appendix~\ref{app:physical-fabrics}), and labels on the all-OCS curve give the batch size. Each point describes a separately provisioned and planned point in the deployment design space.}
    \label{fig:per-batch-both}
\end{figure}

We first present results for two model architectures, Megatron 5T MoE and GPT-OSS-5.3-like, for very large context lengths of 5 million tokens. The choice of large models with large context is motivated by current hardware limitations and the resulting need to scale to fabrics larger than what a single datacenter rack can provide, even with accelerators such as the upcoming Vera Rubin platforms~\cite{nv-rubin}. Model architectures and accelerator properties are given in Appendices~\ref{app:model-details} and~\ref{app:accelerators}.

\subsection{Optimized Interactivity-Throughput Envelope}
Figure~\ref{fig:per-batch-both} plots achieved interactivity, the per-user throughput metric, against total system throughput for various batch sizes. A batch of 16 concurrent users is the largest that the 512 serving accelerators can hold for Megatron before KV cache pages have to be swapped out of an accelerator's HBM. The three network scenarios are two-level networks using only EPS (all-EPS), OCS on the second level (hybrid EPS/OCS), or only OCS (all-OCS); the first two are 4:1 oversubscribed towards the second level, while all-OCS is non-blocking (Appendix~\ref{app:physical-fabrics}). 
The EPS-based fabrics use 4:1 leaf-to-spine oversubscription to maintain a practical system-power envelope. A sensitivity experiment with GPT-OSS-5.3-like at batch size 1 shows that removing oversubscription requires four times as many spine switches (288 vs. 72), yet reduces collective completion time by only 3.2\,\% (13.61 to 13.18\,$\mu$s). Consistently, the bandwidth sensitivity in Figure~\ref{fig:sensitivity} from Appendix~\ref{app:sensitivity} shows that even quadrupling per-port bandwidth barely changes the all-OCS interactivity advantage (64.8\,\% to 62.8\,\%), supporting the interpretation that the observed gains are primarily latency-driven rather than an artifact of EPS oversubscription.

The Pareto front for each scenario is found by searching the Helix parallelization space (Appendix~\ref{app:gxnet-search}), so the optimal parallelization for a given batch size may differ between fabrics: reduced collective completion time can make a higher degree of parallelism along a mesh dimension viable. For instance, batch size 1 on Megatron calls for a Helix scheme (KVP4, TPA128, EP16, TPE32) on all-EPS, reaching 458 tokens/s/user, while all-OCS reaches 698 tokens/s/user with (KVP4, TPA128, EP8, TPE64). The larger FFN tensor-parallel group is enabled by our collective schedule optimization for large groups, whereas a group of 64 on this physical topology forces EPS traffic through both levels, and thus three switching devices.

\subsection{Bottlenecks Change with Batch Size}
In high-interactivity/low-batch regimes of Figure~\ref{fig:per-batch-both}, collective communication volume is small and network latency contributes substantially to decode-step time, so the difference between all-EPS and all-OCS fabrics is significant: for GPT-OSS-5.3-like, 775 versus 1277 tokens/s/user, 1.65$\times$ the interactivity. In practice, achievable values may be lower due to unmodeled system properties such as accelerator-to-accelerator jitter or GEMM inefficiencies for certain operand aspect ratios; on the other hand, this analysis does not factor in EPS-specific disadvantages such as tail latency due to buffer congestion and arbitration delays. The lower power consumption of OCS fabrics further amplifies their advantage, so the energy per generated token decreases through both lower system power and faster token generation.

\begin{figure}[t]
  \centering
  \includegraphics{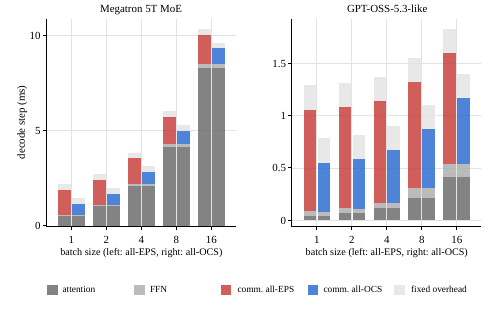}
  \caption{Decode-step time over batch size for the fabric-specific Pareto-optimal configurations of Figure~\ref{fig:per-batch-both}, all-EPS (left bar) and all-OCS (right bar) of each pair. For Megatron, attention time grows with batch and dominates from batch 4 on, so the constant OCS saving in communication shrinks in relative terms; for GPT-OSS-5.3-like, communication remains the dominant term. Note the different vertical scales.}
  \label{fig:step-breakdown}
\end{figure}

As batch size increases, compute kernel time and network transaction sizes grow, which reduces the share of exposed latency in the token generation time (time per output token, TPOT). Figure~\ref{fig:step-breakdown} shows that this effect is strong for Megatron but weak for GPT-OSS-5.3-like. Attention time scales with the KV bytes read per decode step, i.e., with the product of batch size and context length. Megatron keeps 1\,MiB of KV cache per token of context, GPT-OSS-5.3-like only 48\,KiB, since its interleaved sliding-window attention reads the full context in only half of its layers, with 16 instead of 128 KV heads (Table~\ref{tab:model-props}, Appendix~\ref{app:model-details}); the hidden size, 8192 in both models, does not explain the difference. At 5M context, Megatron's attention time therefore grows from 0.53\,ms at batch 1 to 8.3\,ms at batch 16, while communication stays nearly constant, so the OCS saving of 0.71 to 0.74\,ms per step falls from 34\,\% to 7\,\% of the all-EPS step. For GPT-OSS-5.3-like, attention reaches only 0.41\,ms, and communication remains the largest term throughout. The remainder of each step is the fixed invocation cost of its fused execution windows (Section~\ref{sec:workload}), identical across fabrics and batch sizes.

\subsection{Scaling and Energy Efficiency}
\begin{figure}[t]
  \centering
  \includegraphics{legend_fabrics.pdf}\\[-2pt]
  \includegraphics{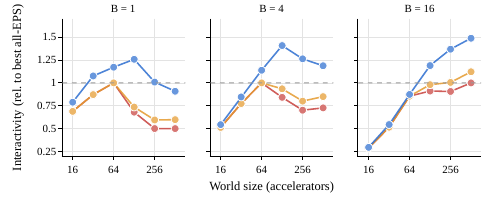}
  \caption{System size (world size) scaling of Megatron 5T MoE at a fixed context of 64K tokens for batch sizes 1, 4, and 16. Interactivity is shown relative to the best all-EPS result at each batch size (dotted line).}
  \label{fig:system-scaling}
\end{figure}

Figure~\ref{fig:system-scaling} shows the scaling with system size for each of the three physical topologies; each point is again the result of a parallelization search. The maximum context is fixed at 64K tokens to allow scaling across a range of world sizes; after all, growing scale-up domains is not always motivated by higher interactivity but may also be required to provide more low-latency, high-bandwidth memory. At batch sizes 1 and 4, interactivity first rises with world size and then falls once collectives span the second level: all-EPS peaks at 64 accelerators, all-OCS only at 128, 26\,\% and 41\,\% higher than the all-EPS peak. At batch 16, all-OCS fabrics show the most favorable interactivity, and thus throughput, scaling, reaching 1.5$\times$ the interactivity of all-EPS at 512 accelerators.

\begin{figure}
  \centering
  \includegraphics{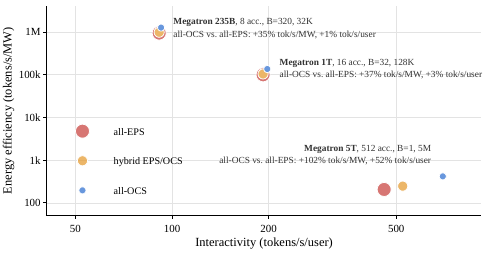}
  \caption{Energy efficiency vs.\ interactivity across models of different size and context length, illustrating the benefits of workload-static, all-OCS switching in large-scale inference decode.}
  \label{fig:model-scaling}
\end{figure}

Finally, Figure~\ref{fig:model-scaling} takes a broader perspective across three scenarios of model size and maximum context length. Small models with small context are usually the free-tier variants that AI providers offer, while large models with long context are a paid service because they aim to provide high-value work such as agent supervision and long-term reasoning. Instead of total system throughput, the y-axis shows tokens per second per megawatt, the perspective of datacenter operators whose token volume, and thus profitability, is usually limited by available power. The proposed scheme is of general utility across the three very different regimes evaluated here: owing to the lower per-port power of OCS (Appendix~\ref{app:power-model}), all-OCS fabrics outperform their EPS counterparts in throughput per unit energy by 35 to 37\,\% even for the 8- and 16-accelerator deployments of Megatron 235B and 1T, where interactivity improves by only 1 to 3\,\%, and by up to 102\,\% in the high-value, large-model regime where the interactivity advantage rises to  52\,\%; across the GPT-OSS family, the energy efficiency gain ranges from 32 to 119\,\% (Appendix~\ref{app:gpt-family}). GPT-OSS-5.3-like at batch 1 is also the baseline of the sensitivity analysis over per-port bandwidth, accelerator radix, and switch and accelerator latency in Appendix~\ref{app:sensitivity}.
\section{Discussion and Conclusion}
\label{sec:discussion}

Distributed LLM decode exhibits recurring collective communication patterns that can be served by optical circuits configured once per deployment. 
We show that jointly optimizing circuit connectivity, accelerator port allocation, and topology-aware collective schedules allows OCS-based fabrics to support these patterns without runtime reconfiguration. 
Allocating ports to reduce communication phases or increase effective bandwidth improves modeled interactivity and energy efficiency over EPS-based fabrics, particularly in latency-sensitive decode regimes.
For Megatron 5T MoE, all-OCS fabrics improve modeled interactivity by up to 52\,\% and energy efficiency by
up to 102\,\% over the oversubscribed all-EPS baseline; for GPT-OSS-5.3-like, the corresponding improvements
reach 65\,\% and 119\,\%.

Given a physical fabric and accelerator port budget, our planner constructs candidate logical overlays from Cayley-graph generators. 
It jointly explores mesh-axis partitions, shortcut connections that reduce collective schedule depth, and parallel circuits that increase bandwidth. 
Compatible collective schedules are evaluated using an analytical latency--bandwidth model, and the configuration minimizing workload-weighted communication time is mapped onto physical ports as a deployment-static circuit topology.

These benefits of our approach are conditional on the workload and hardware assumptions. 
Their magnitude depends on the communication exposed on the decode critical path, accelerator endpoint latency, model architecture, and the power and oversubscription assumptions of the compared fabrics. With co-packaged optics (CPO), accelerators may see a rise in both per-port data rates, e.g., through WDM-capable optical PHYs such as OCI~\cite{ocimsa2026}, and port count (accelerator radix). Both can be shown to improve the benefits of workload-static switching in terms of system performance and energy efficiency (Appendix~\ref{app:sensitivity}). 

Our main evaluation excludes compute--communication overlap, although HOP-B microbatching is examined separately (Appendix~\ref{app:hopb}). 
We also assume balanced expert routing and do not model routing skew or stochastic execution-time variations.

More broadly, our results motivate treating scale-up topology, model parallelization, and collective schedules as a joint optimization problem, including for optically switched fabrics. 
Future work should validate the predicted performance and energy benefits on physical hardware, extend the optimization to overlapping computation and communication, and assess robustness to dynamic traffic and routing imbalance. 
Workload-static OCS thus offers a promising alternative to general-purpose packet switching when communication requirements are predictable at deployment time.

\bibliography{main}
\bibliographystyle{mlsys2025}

\section*{AI Usage Disclosure}

Assistance of AI was extensively used in the context of this work. 
It was used to assist in writing code for \textsc{gxnet}.
It was used to assist in writing code for the \emph{Simlette} network simulator.
It was used to help in formalizing the mathematical concepts underlying the algorithmic ideas of mesh fabric creation and collective schedule implementation.
It was used to check public availability of performance and power data used in this work.
It was used to draft text snippet suggestions for certain paragraphs that were then manually edited and revised.
It was used for consistency checks between data and text. 
It was used for a consistent formatting style of figures.

\appendix

\renewcommand{\topfraction}{0.9}
\renewcommand{\dbltopfraction}{0.9}
\renewcommand{\floatpagefraction}{0.8}
\renewcommand{\dblfloatpagefraction}{0.8}
\setcounter{topnumber}{4}
\setcounter{dbltopnumber}{3}
\setcounter{totalnumber}{6}
\clearpage
\section{Workload-Static Topology and Collective Design}
\label{app:formalism}

This appendix details the workload-static topology design and collective schedules. It covers graph construction, joint topology–schedule optimization, and variable MoE traffic, followed by worked examples, fabric comparisons, alternative port allocations, and limitations.

Throughout this appendix we use the notation of Section~\ref{sec:ocs-design} and drop the axis index $i$ where only one axis is considered:
$h$ is the number of serialized schedule phases, $\lambda$ the per-phase latency, $v$ the bottleneck volume, $p$ the edge multiplicity, $R$ the bandwidth of one physical link, and $m$ the number of tokens per forward pass.
We denote the payload of one collective call by $\sigma$, with $\sigma = c\,m$ for a per-token volume $c$; we also call it the \emph{buffer size} of the collective, i.e., the larger of its input and output buffers (Section~\ref{sec:static-demand}). It differs from the \emph{messages} a schedule sends: each phase transfers per-port messages whose sizes are determined by the schedule and, e.g., for coset-factored schedules, are not uniform across phases.

\subsection{Shared-Host Graph Construction}
\label{app:shared-host}

\begin{figure*}[t]
\centering
\includegraphics[width=\textwidth]{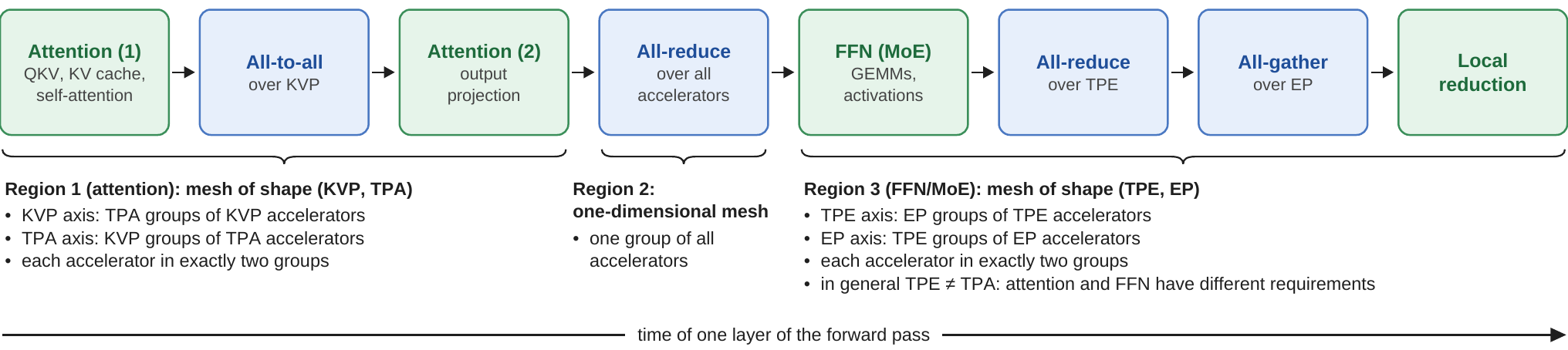}
\caption{One layer of a decode step under Helix parallelism (simplified); the full decode step repeats it for every layer, followed by one all-gather of the logits (not shown). Each region arranges the same accelerators into its own device mesh; every collective runs over one axis of its region's mesh.}
\label{fig:helix-regions}
\end{figure*}

Under Helix parallelism~\cite{bhatia2025helix}, every layer of a decode step passes through a sequence of \emph{regions}, each of which arranges the same $N$ accelerators into its own logical \emph{device mesh} (Figure~\ref{fig:helix-regions}).
The attention region uses a mesh of shape (KVP, TPA) and ends in an all-to-all over its KV-parallel axis; the output projection is followed by an all-reduce over a one-dimensional mesh spanning all accelerators; and the FFN/MoE region uses a mesh of shape (TPE, EP), with a tensor-parallel all-reduce and an expert-parallel all-gather.
Each dimension of a mesh is a \emph{mesh axis}, and its extents form the \emph{mesh shape}.
The accelerators that differ only in their coordinate along one axis form a \emph{collective group} of that axis, so in a two-axis mesh every accelerator belongs to exactly two collective groups, one per axis.

Helix is one of several parallelization schemes for decode, and its region structure is more demanding than that of the common TP/EP scheme.
There, attention and FFN typically share one mesh and one tensor-parallel axis; under Helix, the attention and FFN meshes split the same accelerators along different axes, since TPE generally differs from TPA.
The shared-host allocation must therefore serve several conflicting mesh shapes with one host graph, in which every mesh axis uses a subgraph and the subgraphs of different meshes share edges.
The subgraphs have to be chosen in conjunction with the \emph{collective schedules}, since a densely connected logical topology only pays off where a schedule can exploit its additional edges.
We build every subgraph from a hypercube spanning the collective groups of its axis, the regular topology in which each of $2^a$ accelerators has one link per dimension to $a$ neighbors.
Hypercubes suit this purpose because the hypercubes of all mesh axes are sub-cubes of one hypercube spanning all accelerators, so a single wiring serves every mesh shape.
To them we add either folds, i.e., shortcuts across the cube that lower its diameter, or parallel circuits that raise its bandwidth, both at the cost of additional ports.
Choosing them for all meshes at once is a joint optimization problem, in which the alignment of folds across meshes (Appendix~\ref{app:folds}) and the joint choice of partitions and folds (Appendix~\ref{app:planner}) become necessary.

\paragraph{Bit labels, cosets, and the hypercube invariant}

We label each of the $N = 2^n$ accelerators with its index in binary.
A \emph{generator} is an $n$-bit mask $g$ that induces the edge $x \leftrightarrow x \oplus g$ at every node.
A set $S$ of generators thus defines the Cayley graph $\Cay((\ZZ_2)^n, S)$ of the group of $n$-bit labels under bitwise XOR: its vertices are the labels, and every label $x$ is connected to $x \oplus s$ for each $s \in S$~\cite{godsil2001agt}.
Since every label is its own inverse under XOR, the edges are undirected.
Three standard properties of such graphs are used throughout: for distinct non-zero generators, the graph is $|S|$-regular, i.e., every node has one neighbor per generator; it is vertex-transitive, i.e., it looks identical from every node, so per-node port counts and schedules carry over to all nodes; and it is connected exactly when $S$ spans $(\ZZ_2)^n$.
We write labels and masks as bit strings, most significant bit first, so that a label reads $x = b_{n-1}\dots b_0$; the generator $g = 111$, for instance, sets bits 0, 1 and 2 and connects every label $x$ to $x \oplus 111$, i.e., to the label with these three bits flipped.
The unit masks $e_j = 2^j$ generate the hypercube \Qn{}, whose degree and diameter are both $n$ and whose graph distance is the Hamming distance (Table~\ref{tab:notation}).

\begin{table*}[t]
\centering
\caption{Notation for topologies and folds used in this appendix. Example: the block fold $Q_5[3,2] = K_8 \,\square\, K_4$ has degree $7 + 3 = 10$ and diameter 2.}
\label{tab:notation}
\small
\begin{tabular}{@{}lp{0.72\textwidth}@{}}
\toprule
notation & meaning \\
\midrule
Hamming weight $|g|$ & number of 1-bits of a mask; the Hamming distance of two labels $x, y$ is $|x \oplus y|$, the number of bit positions in which they differ \\
$Q_n$ & hypercube on $n$ bits; its generators are the $n$ unit masks, i.e., the masks of Hamming weight 1 \\
$Q_n{+}g$ & $Q_n$ with the additional generator $g$, a fold \\
$FQ_n$ & folded cube $Q_n{+}1\dots1$, in which every node is joined to its antipode \\
$K_s$ & complete graph on $s$ nodes \\
$G \,\square\, H$, $K_s^\ell$ & Cartesian product: two nodes are adjacent if they agree in one factor and are adjacent in the other; $K_s^\ell = K_s \,\square\, \cdots \,\square\, K_s$ with $\ell$ factors \\
block, sub-block & the bit positions an axis owns; a subset of them \\
block subgroup & all masks supported on the bits of a block, i.e., all ways of flipping only these bits \\
coset (of a block) & the labels $x \oplus s$ for a fixed $x$ and all $s$ in the block subgroup, i.e., all labels that agree with $x$ outside the block; each coset is one collective group of the axis that owns the block \\
complete layer on $k$ bits & all $2^k - 1$ non-zero masks supported on a sub-block of $k$ bits; it turns every coset of that sub-block into $K_{2^k}$ \\
$Q_a[k_1,\dots,k_\ell]$ & block fold: the $a$ bits of an axis split into sub-blocks of $k_j$ bits, each carrying a complete layer; equals $K_{2^{k_1}} \,\square\, \cdots \,\square\, K_{2^{k_\ell}}$, with degree $\sum_j (2^{k_j} - 1)$ and diameter $\ell$ \\
uniform block fold & all sub-blocks of equal size, $k_j = a/\ell$, e.g., $Q_6[3,3] = K_8 \,\square\, K_8$ \\
mixed-radix block fold & sub-blocks of different sizes, e.g., $Q_7[4,3] = K_{16} \,\square\, K_8$; reading a label as a number whose digit $j$ is the value of sub-block $j$, every digit has its own base $2^{k_j}$ \\
\bottomrule
\end{tabular}
\end{table*}

A 2-D mesh of shape $(2^{a_0}, 2^{a_1})$ partitions the bit positions into two \emph{blocks}.
The coordinate of a node along an axis is the value of that axis's bits, and a collective group is the set of nodes that agree on all bits outside the block.
Each group is therefore a coset of the block subgroup: there are $N/s_i$ of them, each of size $s_i = 2^{a_i}$.
This is why we restrict axis extents to powers of two: grid coordinate and label slice become the same object.
Other sizes could be handled by planning the enclosing power of two and marking the surplus labels as absent, which we do not implement.

Unit masks over disjoint blocks together form the unit-mask set of the larger cube.
For $N=16$, the meshes $(2,8) = Q_1 \,\square\, Q_3$ and $(4,4) = Q_2 \,\square\, Q_2$ are therefore the same 32-edge cube $Q_4$, colored into axes in two different ways (Figure~\ref{fig:cayley-port-tradeoff}(a)).
One fixed wiring thus serves every power-of-two mesh shape at diameter $n$. Every other configuration the shared-host planner considers keeps this cube as its base and only adds to it, either additional generators that fold it (Appendix~\ref{app:folds}) or parallel circuits that widen its edges.

The planner treats the bit partition of each mesh as its placement decision and chooses it jointly with the folds (Appendix~\ref{app:planner}).

\begin{table}[t]
\centering
\caption{Per-axis degree ladder for an axis of size $2^a$ at degree $d$.}
\label{tab:ladder}
\small
\setlength{\tabcolsep}{3pt}
\begin{tabular}{@{}llll@{}}
\toprule
degree $d$ & topology & group & diameter \\
\midrule
$d = a$      & hypercube $Q_a$        & $(\ZZ_2)^a$ & $a$ \\
$d > a$      & folded cube & $(\ZZ_2)^a$ & $< a$ \\
$2 < d < a$  & circulant               & $\ZZ_{2^a}$ & intermediate \\
$d = 2$      & ring                    & $\ZZ_{2^a}$ & $2^{a-1}$ \\
\bottomrule
\end{tabular}
\end{table}

Table~\ref{tab:ladder} summarizes the topologies a single axis could take.
Reducing the degree below $d = a$ differs qualitatively from increasing it: the axis can no longer remain in $(\ZZ_2)^a$ and has to change group.
The shared-host planner therefore stays in the cube family, $d \ge a$;
rings and circulants appear only in the stacked-link allocation of Appendix~\ref{app:stacked-link} and for a ring schedule on a single mesh (Appendix~\ref{app:worked}).

\paragraph{Folding generators, the support rule, and alignment}
\label{app:folds}

\begin{figure}[t]
\centering
\includegraphics[width=\columnwidth]{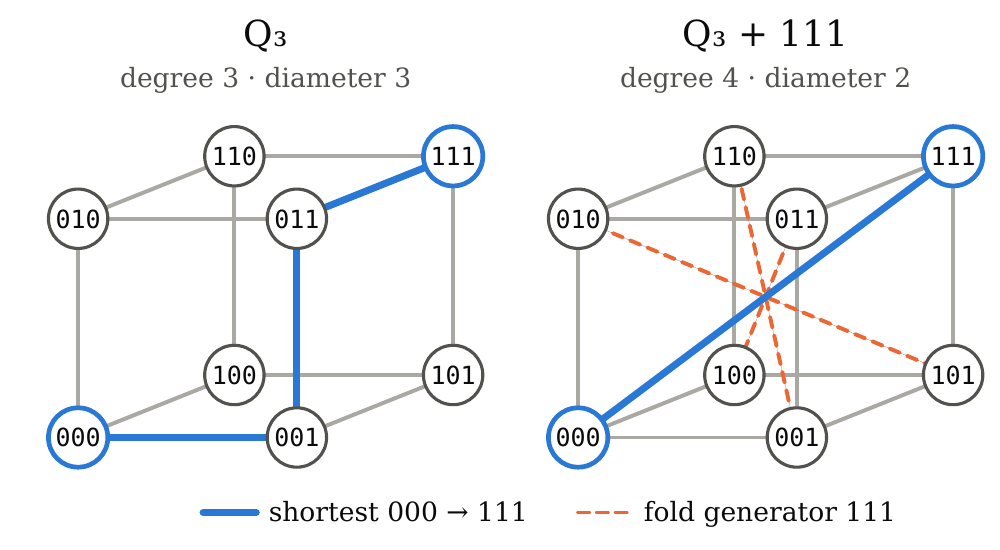}
\caption{Folding a hypercube. Left: $Q_3$ with generators $\{001,010,100\}$ has degree 3 and diameter 3; $000 \to 111$ takes three hops. Right: the fold generator $g = 111$ adds one port per node and joins every node to its antipode, so $000 \to 111$ takes one hop and the diameter drops to 2.}
\label{fig:foldq3}
\end{figure}

The hypercube uses $n$ ports per accelerator, one per unit mask.
Any further port, i.e., a host degree $D > n$, can be spent on an additional generator of Hamming weight at least two.
Whereas a unit mask connects labels that differ in a single bit, a generator of weight $w$ connects labels that differ in $w$ bits at once, a pair the cube alone separates by $w$ hops.
Such generators act as chords that \emph{fold} the cube.
Adding $g = 111$ to an axis of size 8 makes antipodal nodes adjacent, raising the degree from 3 to 4 and lowering the diameter from 3 to 2 (Figure~\ref{fig:foldq3}).
Whether a fold helps a given axis is governed by its \emph{support}: a folding generator only helps a mesh axis if all of its set bits lie inside that axis's bit block.
If its support spans two blocks, the circuits are still built and occupy ports, but the collectives of that mesh leave them idle.

\begin{figure}[t]
\centering
\resizebox{\columnwidth}{!}{
\begin{tikzpicture}[
  font=\scriptsize,
  v/.style={circle, draw=#1, fill=white, line width=0.6pt, inner sep=0.5pt, minimum size=11pt, font=\tiny\ttfamily},
  e/.style={draw=#1, line width=0.7pt},
  f/.style={draw=#1, line width=0.7pt, dash pattern=on 2pt off 1.5pt},
  lab/.style={align=center, font=\scriptsize},
  hd/.style={align=center, font=\scriptsize\bfseries}]
  \definecolor{axA}{HTML}{2A78D6}\definecolor{axB}{HTML}{E0662C}
  \node[hd] at (3.3,3.35) {$2\times8$:\ \ $b_3 \mid b_2 b_1 b_0$\quad diameter $1+2=3$};
  \node[v=axA] (a0) at (0.6,0.9) {0}; \node[v=axA] (a1) at (0.6,2.0) {1};
  \draw[e=axA] (a0) -- (a1);
  \node[lab, text=axA] at (0.6,0.1) {axis A: $b_3$\\no fold};
  \begin{scope}[xshift=1.45cm, yshift=0.7cm]
    \foreach \z in {0,1}{\foreach \y in {0,1}{\foreach \x in {0,1}{
      \node[v=axB] (b\z\y\x) at ({1.5*\x+1.15*\z},{1.4*\y+0.8*\z}) {\z\y\x};}}}
    \foreach \z in {0,1}{\foreach \y in {0,1}{\draw[e=axB] (b\z\y0) -- (b\z\y1);}}
    \foreach \z in {0,1}{\foreach \x in {0,1}{\draw[e=axB] (b\z0\x) -- (b\z1\x);}}
    \foreach \y in {0,1}{\foreach \x in {0,1}{\draw[e=axB] (b0\y\x) -- (b1\y\x);}}
    \foreach \a/\b in {000/011,001/010,100/111,101/110} \draw[f=axB] (b\a) -- (b\b);
  \end{scope}
  \node[lab, text=axB] at (2.8,0.1) {axis B: $b_2 b_1 b_0$\\fold $0011$, diameter 2};
  \node[lab, text=black!55] at (5.75,1.45) {$1100$ idle:\\spans axes\\A and B};
  \begin{scope}[yshift=-3.0cm]
    \node[hd] at (3.3,2.45) {$4\times4$:\ \ $b_3 b_2 \mid b_1 b_0$\quad diameter $1+1=2$};
    \foreach \ax/\xo/\c/\t in {A/0.6/axA/{axis A: $b_3 b_2$\\fold $1100$, diameter 1}, B/3.9/axB/{axis B: $b_1 b_0$\\fold $0011$, diameter 1}}{
      \begin{scope}[xshift=\xo cm]
        \node[v=\c] (\ax00) at (0,0.8) {00}; \node[v=\c] (\ax01) at (1.6,0.8) {01};
        \node[v=\c] (\ax10) at (0,1.85) {10}; \node[v=\c] (\ax11) at (1.6,1.85) {11};
        \draw[e=\c] (\ax00) -- (\ax01); \draw[e=\c] (\ax10) -- (\ax11);
        \draw[e=\c] (\ax00) -- (\ax10); \draw[e=\c] (\ax01) -- (\ax11);
        \draw[f=\c] (\ax00) -- (\ax11); \draw[f=\c] (\ax01) -- (\ax10);
        \node[lab, text=\c] at (0.8,0.2) {\t};
      \end{scope}}
  \end{scope}
  \begin{scope}[yshift=-3.45cm]
    \draw[line width=0.7pt, black!70] (1.2,0) -- (1.7,0); \node[anchor=west] at (1.75,0) {cube edge};
    \draw[line width=0.7pt, black!70, dash pattern=on 2pt off 1.5pt] (3.5,0) -- (4.0,0); \node[anchor=west] at (4.05,0) {fold inside the axis};
  \end{scope}
\end{tikzpicture}}
\caption{Folds shared by two meshes on one host: $N=16$, $Q_4$ with the folds $0011$ and $1100$, under the partitions of Figure~\ref{fig:cayley-port-tradeoff}(a). Each panel shows one collective group per axis; every other group of the axis is a translate of it. Solid lines are cube edges, dashed lines folds whose support lies inside the axis. The mask $0011$ folds axis~B of both meshes; the mask $1100$ folds axis~A of the $4\times4$ mesh and is idle for the $2\times8$ mesh, whose partition splits bits 3 and 2.}
\label{fig:shared-folds}
\end{figure}
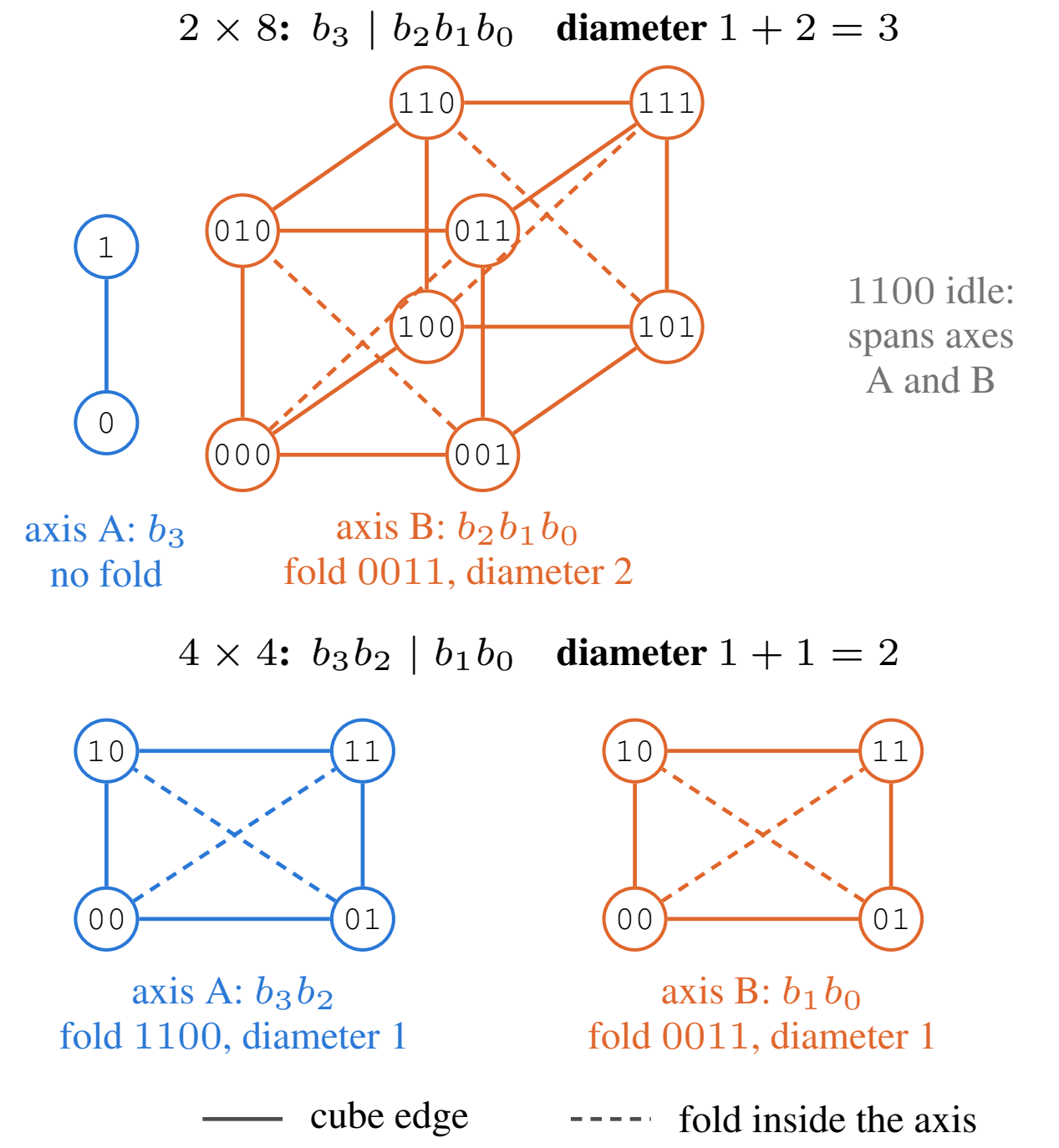

Since a generator adds one port on every node, folds are a global resource shared by all meshes, and they are the main source of coupling between them.
For $N=16$ with the concurrent mesh shapes $(2,8)$ and $(4,4)$ and their partitions $b_3 \mid b_2 b_1 b_0$ and $b_3 b_2 \mid b_1 b_0$ of Figure~\ref{fig:cayley-port-tradeoff}(a), the mask $0011$ lies inside axis~B of both meshes and reduces the diameter of both from 4 to 3.
The mask $1100$ helps only $(4,4)$, whose axis~A it turns into $K_4$; it leaves $(2,8)$ at diameter 4, since that partition places bits 3 and 2 on different axes.
Together, the two masks bring $(4,4)$ to diameter 2 and $(2,8)$ to 3 (Figure~\ref{fig:shared-folds}).
Choosing partitions such that one chord serves the latency-bound axes of several meshes is therefore a genuine optimization problem, and the reason why partitions are part of the search.

Two cases simplify the problem.
With a single mesh there is nothing to align to, so an axis could leave the cube family, e.g., become a degree-2 ring that frees ports for edge multiplicity;
our planner offers this only to the ring schedule, which it can evaluate, since an analytical cost model exists, but which the simulator does not execute (Table~\ref{tab:single}). This is because in the latency-bound inference decode regime, the ring algorithm typically is the worst of all schedule choices.
A \emph{logical-only} axis, i.e., a mesh dimension that carries no collective, such as a replica or pipeline-stage axis, receives neither circuits nor collective groups.
Its bits still label nodes so that the other axes can index their cosets by them, but the ports it would otherwise occupy remain available.

\paragraph{Realizing the host graph: edge coloring}
\label{app:colouring}

Mapping the host graph onto the OCS planes is an edge coloring problem in which colors correspond to ports.
For a Cayley graph of $(\ZZ_2)^n$, this coloring is immediate: the edges $\{x, x \oplus g\}$ of one generator $g$ form a perfect matching, so coloring every edge by its generator is proper and uses exactly $D$ colors, and each of the $p_e$ circuits of a multiplied edge receives its own copy of that color.
Every aligned host graph therefore fits the port budget of Eq.~\eqref{eq:port-budget} without a coloring search, whatever the weights of its generators.
Only when the combined demand graph is not a Cayley graph of $(\ZZ_2)^n$, for instance for meshes that are not aligned to one host, can it require one color more than its maximum degree (Vizing's theorem); a port then has to be reserved.

\subsection{Joint Topology–Schedule Optimization}
\label{app:cost}

\paragraph{The impact of folds and multiplicity}
\label{app:pricelists}

The planner spends the ports of Eq.~\eqref{eq:port-budget} on two resources: fold generators, which cost one port per node and reduce the phase count $h$ of the axes whose schedules can use them, and parallel circuits, which raise the multiplicity $p$ of an axis.
Each axis is evaluated using Eq.~\eqref{eq:hockney-axis}, where $h$ counts the phases the selected schedule actually traverses, not the graph diameter, and an axis completes at the pace of its least-multiplied generator, its \emph{effective multiplicity} $p_i$.
An edge shared by several axes takes the largest multiplicity any of them requires; together with the global nature of folds, this is the only coupling between meshes.
The planner minimizes $\sum_i w_i T_i$ subject to Eq.~\eqref{eq:port-budget}, where $w_i$ counts how often axis $i$'s collective occurs per decode step.

\begin{figure}[t]
\centering
\includegraphics[width=\columnwidth]{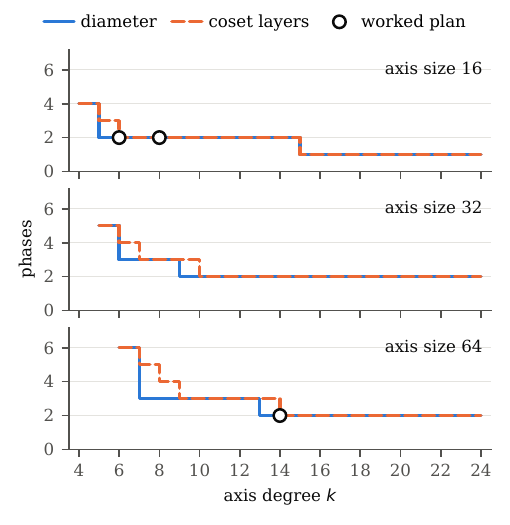}
\caption{Diameter frontier and schedulable frontier for axis sizes 16, 32 and 64: the fewest phases a folded cube of a given degree allows, for any schedule (diameter) and for the coset-factored schedules (coset layers). Both are staircases with plateaus that lengthen with axis size. On the axis of size 64, diameter 2 is reached at degree 13, but two coset layers only at degree 14. Circles mark the folded axes of the worked plan (Table~\ref{tab:plan}).}
\label{fig:frontier}
\end{figure}

\paragraph{The diameter frontier and the schedulable frontier}
\label{app:frontier}

The number of phases a given degree can buy is a staircase (Figure~\ref{fig:frontier}).
The diameter frontier bounds every schedule; the schedulable frontier of the coset-factored schedules (Appendix~\ref{app:allreduce}) lags behind it, because these schedules need generator sets that factor into non-overlapping layers:
on an axis of 64 accelerators, diameter 2 is reached at degree 13, but two coset layers only at degree 14.
Both frontiers are precomputed per axis size.

\paragraph{Search structure and the planner}
\label{app:planner}

\begin{figure*}[t]
\centering
\includegraphics[width=\textwidth]{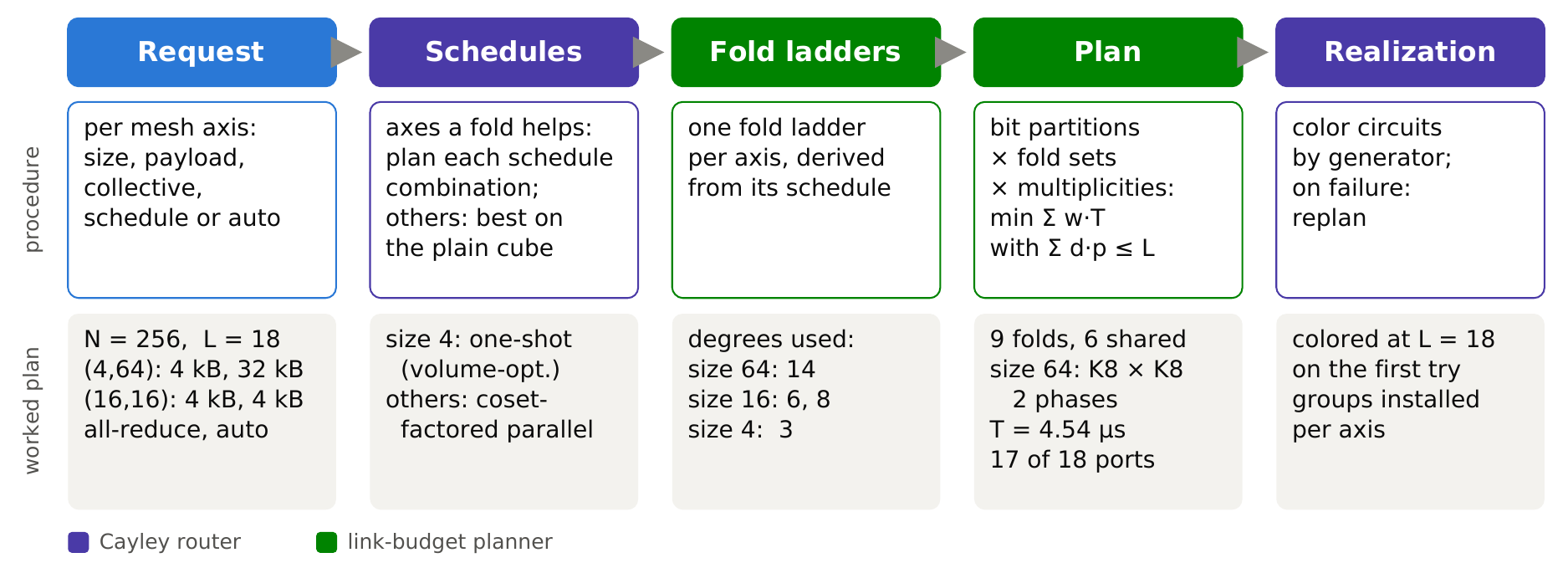}
\caption{Fabric planning. Top: the five stages, colored by the component that runs them; the router selects schedules and realizes the circuits, the planner builds each axis's fold ladder and searches partitions, folds and multiplicities. Bottom: the worked plan of Table~\ref{tab:plan} ($N=256$, $L=18$), which is colored onto ports at the full budget on the first attempt.}
\label{fig:planner-flow}
\end{figure*}

The planner proceeds in the stages of Figure~\ref{fig:planner-flow}: it selects candidate schedules per axis, derives each axis's ladder of candidate folds from its schedule, searches bit partitions, fold sets and multiplicities jointly, and colors the resulting circuits onto ports.
For large instances, the search is heuristic.
The planner is deterministic, so a plan is a pure function of its request and can be cached and reused.

\paragraph{Binding the cost model to execution}
\label{app:conformance}

The cost model is only trustworthy if $h$ and $v$ are properties of the actual schedule rather than of an idealization of it.
We therefore specify every collective implementation as a triple of an executable schedule, a validity predicate, and a closed-form cost (Section~\ref{sec:topology-aware-collectives}), and check that the three agree.
The closed-form cost takes the topology and the payload and returns $(h, v, u)$: serialized phases, bottleneck-link bytes, and ports driven per node.
It does not receive $\lambda$, $R$ or $p$; the planner alone owns these deployment parameters and forms $T$ by Eq.~\eqref{eq:hockney-axis}.
We run the resulting check for every collective schedule the planner can select: under the assumptions of the closed form, i.e., a fixed per-phase latency $\lambda$, link bandwidth $R$, multiplicity $p = 1$ and no competing traffic, the simulated completion time of the schedule has to match $h\lambda + v/(pR)$.

The check prevents two failure modes that would otherwise be undetectable.
First, crediting recursive doubling with a folded diameter it cannot use would make the selector prefer it on every folded cube.
Second, charging RS+AG a full message per phase instead of its geometric two-shot total would make the selector never choose it.
The validity predicate keeps invalid combinations of schedule and topology out of the search entirely, rather than having the cost model extrapolate a value for a schedule that would not run.

The scope of this check is internal consistency between the closed form and a simulated execution of the same schedule under the same idealizing assumptions; it is not a hardware validation.
It currently covers $p = 1$ only: with parallel circuits, the closed form is known to over-predict the simulated time, so plans that use $p > 1$ are, if anything, estimated pessimistically.
Grounding of the constants themselves ($\lambda$, $R$, endpoint and protocol terms) is treated separately in Section~\ref{sec:grounding}.

\subsection{Topology-Aware Collective Schedules}
\label{app:schedules}

\paragraph{All-reduce operating points}
\label{app:allreduce}
Two lower bounds hold for any all-reduce schedule: the diameter of the topology, since every node's result depends on all inputs, and $\lceil \log_{d+1} N \rceil$ phases at degree $d$.
The achievable minimum is the larger of the two.

\begin{table*}[t]
\centering
\caption{Serialized phases $h$ and bottleneck volume $v$ (cell: $h$ / $v$, with $v$ in units of the buffer size $\sigma$) of each all-reduce schedule, as returned by its closed-form cost. min.\ $h$: the lower bound on the phase count, i.e., the larger of the diameter and $\lceil \log_{d+1} N \rceil$. Bold marks the schedules that attain it.}
\label{tab:allreduce}
\setlength{\tabcolsep}{4pt}
\begin{tabular}{@{}lccccc@{}}
\toprule
topology (deg, diam) & min.\ $h$ & rec.\ doubling & one-shot & coset-fact. & coset-fact.\ par. \\
\midrule
$Q_3$ (3, 3)          & 3 & \textbf{3} / $3\sigma$ & \textbf{3} / $12\sigma$ & \textbf{3} / $3\sigma$ & \textbf{3} / $\sigma$ \\
$FQ_3$ (4, 2)         & 2 & 3 / $3\sigma$          & \textbf{2} / $6\sigma$  & \textbf{2} / $2\sigma$ & \textbf{2} / $\sigma$ \\
$Q_4$ (4, 4)          & 4 & \textbf{4} / $4\sigma$ & \textbf{4} / $32\sigma$ & \textbf{4} / $4\sigma$ & \textbf{4} / $\sigma$ \\
$Q_4{+}1100$ (5, 3)   & 3 & 4 / $4\sigma$          & \textbf{3} / $20\sigma$ & \textbf{3} / $3\sigma$ & \textbf{3} / $\sigma$ \\
$Q_4{+}1111$ (5, 2)   & 2 & 4 / $4\sigma$          & \textbf{2} / $7\sigma$  & 3 / $3\sigma$          & 3 / $\sigma$ \\
\bottomrule
\end{tabular}
\end{table*}

Recursive doubling cannot use folding edges, since merging partial results with overlapping source sets would count inputs twice.
Our coset-factored schedules respect this constraint: they group the generators of a fold into layers $S_1,\dots,S_\ell$ that tile the group without overlap, $\prod_j (1 + |S_j|) = N$, and exchange over one layer per phase, so a fold of $\ell$ layers needs $\ell$ phases.
On a block fold, these layers are exactly its complete layers (Table~\ref{tab:notation}).
Where no such tiling is as short as the diameter, as on $Q_4{+}1111$, the diameter can still be reached by reusing links across phases at a higher volume, as the one-shot schedule does (Table~\ref{tab:allreduce}).
The parallel variant of the coset-factored schedule keeps the $\ell$ phases and reduces the bottleneck volume from $\ell\sigma$ to $\sigma$.
RS+AG on the same layers trades phases for volume: a reduce-scatter and an all-gather of $\ell$ phases each, with a volume that shrinks geometrically in the \emph{layer radices} $r_j = 1 + |S_j|$, the number of nodes a layer connects,
\begin{equation}
v_{\textsc{rs+ag}} = 2\sigma \sum_{j=1}^{\ell} \prod_{k \le j} r_k^{-1},
\label{eq:rsag-volume}
\end{equation}
which on the plain cube is the familiar $2\frac{N-1}{N}\sigma$; its parallel variant divides this volume by $\ell$ on a uniform fold.

\begin{table}[t]
\centering
\caption{The all-reduce operating points on a chain of $\ell$ coset layers. The volume of parallel RS+AG holds for uniform folds.}
\label{tab:opoints}
\small
\setlength{\tabcolsep}{2.5pt}
\begin{tabular}{@{}llll@{}}
\toprule
schedule & phases & bottleneck vol. & regime \\
\midrule
recursive doubling          & $\log_2 N$ & $\log_2 N \cdot \sigma$ & dominated \\
coset-factored              & $\ell$     & $\ell \sigma$          & latency \\
coset-factored parallel     & $\ell$     & $\sigma$               & latency \\
RS+AG                     & $2\ell$    & Eq.~\eqref{eq:rsag-volume} & bandwidth \\
RS+AG parallel            & $2\ell$    & Eq.~\eqref{eq:rsag-volume}$/\ell$ & bandwidth \\
\bottomrule
\end{tabular}
\end{table}

The planner selects by $\arg\min T$ among the schedules whose validity predicate holds and that are not dominated (Table~\ref{tab:opoints}).
In the latency-bound regime the coset-factored parallel schedule usually wins on the deepest fold the budget affords;
in the bandwidth-bound regime parallel RS+AG does, whose volume is lower than $\sigma$ but which needs twice as many phases.
The worked plan of Appendix~\ref{app:worked} illustrates the former. Figure~\ref{fig:collective-sweep} compares the resulting all-reduce completion times with those of the all-EPS fabric over a range of buffer sizes (Appendix~\ref{app:eps}).

\paragraph{All-to-all}
\label{app:a2a}

The all-to-all generalizes the radix-$r$ Bruck algorithm~\cite{bruck1997alltoall} to multi-port accelerators on the same coset layers and takes $\ell$ phases on a fold of $\ell$ layers.
For axes whose number of bits is prime, such as 128 accelerators, uniform folds are scarce, so we additionally admit balanced mixed-radix block folds such as $Q_7[4,3] = K_{16} \,\square\, K_8$ (Table~\ref{tab:notation}); their parallel schedules are priced by the narrowest layer.
Figure~\ref{fig:collective-sweep} shows the corresponding all-to-all completion times (Appendix~\ref{app:eps}).

\subsection{Variable MoE Traffic and Regularization}

\paragraph{MoE token routing and deterministic all-to-all}
\label{app:moe}

The workload-static argument requires the communication of decode to be fixed at deployment time.
Token routing is the only mechanism in the model that varies from token to token, so we state here what has to be regularized, what does not, and what regularization costs.

With $N_{\mathrm{exp}}$ experts distributed over $N_{\mathrm{ep}}$ expert-parallel groups and top-$k$ routing, each decode step produces a dispatch traffic matrix $W \in \mathbb{N}^{N_{\mathrm{ep}}\times N_{\mathrm{ep}}}$, whose entry $W_{uv}$ counts the rows sent from group $u$ to group $v$;
a combine phase with the transposed pattern follows.
The router determines the entries of $W$, but not its support: no routing decision is excluded a priori, and over any nontrivial number of steps every ordered pair occurs.
The logical graph the circuits must serve is therefore the complete graph on the expert-parallel axis, i.e., an all-to-all, independent of the router.
The topology can thus be static without any regularization.
Regularization is only needed for the schedule, since a fixed phase structure and a fixed planner input both require the per-pair payloads to be fixed.

\subparagraph{Capacity padding.} We assign each expert a fixed capacity
\begin{equation}
C \;=\; \left\lceil \gamma \cdot \frac{k\,m}{N_{\mathrm{exp}}} \right\rceil
\label{eq:capacity}
\end{equation}
with capacity factor $\gamma \ge 1$, and pad every buffer between a source group and a destination expert to $C$ rows.
Every ordered pair then exchanges exactly $C \cdot (N_{\mathrm{exp}}/N_{\mathrm{ep}}) \cdot d_{\mathrm{model}}$ elements per step, independent of the routing outcome, where $d_{\mathrm{model}}$ is the hidden size.
The dispatch becomes a uniform all-to-all over the expert-parallel axis that maps directly onto the Bruck/coset schedule of Appendix~\ref{app:a2a}, with $h$ and $v$ fixed and known.
Tokens exceeding the capacity are handled on an overflow path outside the static schedule, either dropped, as is common in training, or deferred to the next step.
Production inference kernels take the opposite end of this trade-off: they size per-peer receive buffers for the worst case, so that no token is dropped~\cite{trtllm-a2a-ws,vllm-a2a-onesided}.
In our terms, this corresponds to the largest capacity factor, for which padding is only incurred in buffer memory and not necessarily on the wire.

\subparagraph{Bucketing.} Padding is the special case of a more general scheme in which per-pair counts are quantized to a small set of levels $C_1 < \dots < C_q$ and the schedule is planned for the largest.
Bucketing recovers volume when the routing histogram is reliably skewed, at the cost of $q$ cached plans instead of one.
Since plans are pure functions of their request and are cached (Appendix~\ref{app:planner}), the additional planning cost is paid once per bucket at deployment.
Our simulator implements neither routing, capacity nor overflow: it runs the dispatch as a uniform all-to-all of the declared per-pair payload, which is exactly the padded dispatch with $q = 1$.

\subparagraph{Cost of padding in decode.} Padding increases the bottleneck volume from $v$ to $\gamma v$, so by Eq.~\eqref{eq:hockney-axis} the time of the axis becomes $T = h\lambda + \gamma v/(pR)$.
With $v = c\,m$, the relative penalty is
\begin{equation}
\frac{T_{\gamma}}{T_{1}}
\;=\;
1 + \frac{\gamma - 1}{1 + \dfrac{h\lambda p R}{v}}
\;=\;
1 + \frac{(\gamma-1)\,m}{m + \hat m},
\qquad
\hat m = \frac{h \lambda p R}{c},
\label{eq:padcost}
\end{equation}
where $\hat m$ is the token count at which the latency and bandwidth terms of Eq.~\eqref{eq:hockney-axis} are equal.
For $m \ll \hat m$ the penalty vanishes: an axis operating an order of magnitude below $\hat m$ pays less than 10\,\% of $\gamma-1$.
Regularization is therefore affordable in exactly the regime that requires it.
At small $m$, the routing histogram is least balanced and $\gamma$ has to be largest, but this is also where $m$ is furthest below $\hat m$.
Conversely, at large batch sizes, where $\gamma$ approaches 1 because the histogram concentrates, the axis has moved toward the bandwidth-bound regime in which padding would be expensive.
Both effects act in the same direction.

\subparagraph{Regularized token routing.} Where an implementation may constrain the router, for instance through balanced assignment or expert-choice routing, the histogram becomes deterministic by construction and $\gamma = 1$ is attainable without dropping tokens.
Since this changes model quality, it is out of scope for this paper; we note it because it removes the only remaining non-determinism in the communication pattern.

\subparagraph{What remains variable.} Regularization fixes the dispatch and combine patterns, but not the compute time of the expert MLPs, which still depends on the realized token counts unless the implementation also computes on padding.
Similar variability arises from speculative decoding, which varies the number of tokens verified per step and thus compute times and payloads, but not which accelerators communicate.
How such variability, and the regularization strategies that bound it, affect the decode step is a question for the workload model (Section~\ref{sec:workload}) rather than for the network model: the circuits serve the collective schedule, whose structure stays fixed, so this variability changes the execution trace but not the network fabric.

\subsection{Worked Examples and Fabric Comparisons}
\label{app:validation}

\subsubsection{Worked Multi-Mesh Plans}
\label{app:worked}

\subparagraph{A latency-bound multi-mesh plan.} We consider $N=256$ ($n=8$) accelerators with $L=18$ ports, $R = 5\times10^{10}$\,B/s (400\,Gb/s) and $\lambda = 0.52\,\mu$s on a single-level OCS fabric.
The fabric serves the mesh $(4,64)$ with all-reduce payloads of 4\,kB and 32\,kB and the mesh $(16,16)$ with 4\,kB on both axes, and the planner selects every schedule automatically.
Every axis is latency-bound, so the budget is spent on degree.
The resulting plan has host degree $D = 17$, consisting of the 8 cube generators and 9 folding generators, uses 17 of the 18 ports, and achieves a total of $T = 4.54\,\mu$s (Table~\ref{tab:plan}; Figure~\ref{fig:planner-flow} traces it through the planner's stages).

\begin{table}[t]
\centering
\caption{Resulting plan for the latency-bound multi-mesh example; for the notation, see Table~\ref{tab:notation}.}
\label{tab:plan}
\small
\setlength{\tabcolsep}{3pt}
\begin{tabular}{@{}llrlrrrr@{}}
\toprule
mesh & axis & size & topology & deg & $h$ & $p$ & axis $T$ \\
\midrule
(4, 64)   & 0 & 4  & $K_4$                         & 3  & 1 & 1 & $0.60\,\mu$s \\
(4, 64)   & 1 & 64 & $Q_6[3,3] = K_8 \,\square\, K_8$   & 14 & 2 & 1 & $1.70\,\mu$s \\
(16, 16)  & 0 & 16 & $Q_4[2,2] = K_4 \,\square\, K_4$   & 6  & 2 & 1 & $1.12\,\mu$s \\
(16, 16)  & 1 & 16 & $Q_4[3,1] = K_8 \,\square\, K_2$   & 8  & 2 & 1 & $1.12\,\mu$s \\
\bottomrule
\end{tabular}
\end{table}

The size-64 axis carries the coset-factored parallel schedule on two complete layers of seven generators each.
In terms of Figure~\ref{fig:frontier}, it operates at degree 14, the first point of the schedulable frontier with two phases, rather than at degree 13, where the diameter frontier already reaches 2 but no two-layer factorization exists.
The size-4 axis is folded into a complete graph and carries the volume-optimized one-shot schedule, which ties with coset-factored parallel there.
Both axes of $(16,16)$ reach two phases without additional ports: the folds they use are a subset of those already present for $(4,64)$, which is why the second axis even receives a degree-8 fold where degree 6 would suffice.
The eighteenth port remains unused, since no further fold reduces a phase count and a multiplicity increase on any axis costs at least three ports.

\begin{table*}[t]
\centering
\caption{The same bandwidth-bound axis, multi-mesh versus single-mesh, under the ring schedule and under parallel RS+AG: $N=64$, $L=36$, $R = 5\times10^{10}$\,B/s, $\lambda = 0.52\,\mu$s, an axis of size 32 carrying 655\,MB (640\,000\,KiB).}
\label{tab:single}
\setlength{\tabcolsep}{4pt}
\begin{tabular}{@{}lllrrr@{}}
\toprule
schedule & case & topology & deg & $p$ & axis $T$ \\
\midrule
ring (planning only)   & multi-mesh $[(2,32),(8,8)]$ & $Q_5$ sub-cube & 5 & 7  & $3660\,\mu$s \\
ring (planning only)   & single mesh $(2,32)$        & ring           & 2 & 17 & $1526\,\mu$s \\
parallel RS+AG       & multi-mesh $[(2,32),(8,8)]$ & $Q_5$ sub-cube & 5 & 7  & $731\,\mu$s \\
parallel RS+AG       & single mesh $(2,32)$        & $Q_5$          & 5 & 7  & $731\,\mu$s \\
\bottomrule
\end{tabular}
\end{table*}

\subparagraph{The cost of alignment.} Table~\ref{tab:single} contrasts a bandwidth-bound axis with and without a second mesh to align to.
With two or more meshes, the fabric is one aligned $Q_6$, so every axis must be a sub-cube and a degree-2 ring is not available.
A ring schedule would have to run over a Gray-code Hamiltonian cycle of the $Q_5$, driving only 2 of the 5 links per node;
each multiplicity unit then costs 5 ports instead of 2, and the multi-mesh plan reaches $p = 7$ where a single-mesh ring reaches $p = 17$.
Alignment costs nothing, however, for the schedule the planner actually selects here.
Parallel RS+AG drives all $d$ links of the cube in every phase, so it gains nothing from a ring and both regimes arrive at the same plan, at half the time of the best ring.
The ring schedule serves only as a reference: the planner can evaluate it, but the simulator does not execute it.

\subsubsection{Collective Performance Across Fabrics}
\label{app:eps}

\begin{figure*}[t]
\centering
\includegraphics{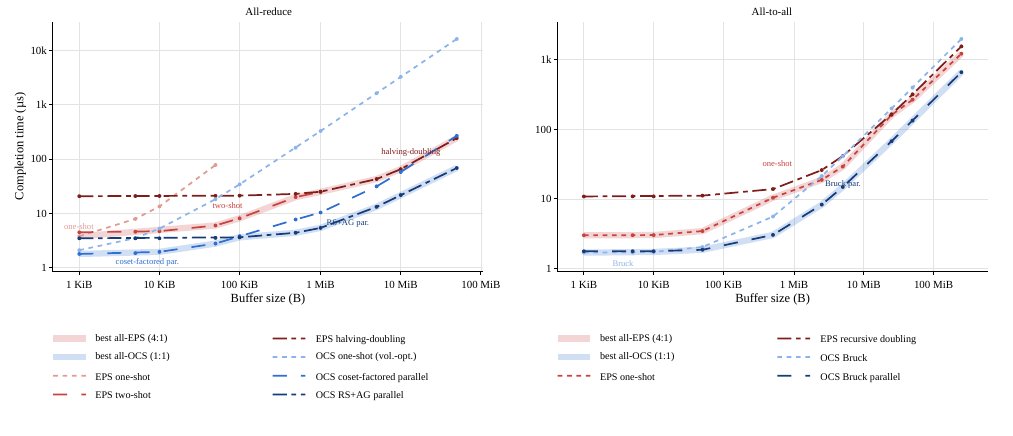}
\caption{Completion time of a single all-reduce (left) and all-to-all (right) spanning 512 accelerators in 16 racks, 36 ports per accelerator, against the buffer size of the collective. All-EPS: two levels, 4:1 oversubscribed towards the second level, dynamic routing. All-OCS: non-blocking, with the topology planned for every buffer size. Lines show the individual schedules, shaded bands the best schedule per fabric and buffer size, labeled with the schedule that attains it.}
\label{fig:collective-sweep}
\end{figure*}

Figure~\ref{fig:collective-sweep} compares single collectives spanning all 512 accelerators on the two fabrics of Section~\ref{sec:eval}: the all-EPS fabric, two levels and 4:1 oversubscribed towards the second level, with dynamic routing and the schedules of current inference frameworks, and the all-OCS fabric, non-blocking, with the topology and schedule planned for every buffer size.

Every all-reduce on both fabrics is charged at least $2\frac{N-1}{N}\sigma$, the volume of a valid sum all-reduce.

For small buffers, both collectives are latency-bound, and all-OCS completes them about twice as fast: a 1\,KiB all-reduce takes 1.78\,$\mu$s instead of 3.77\,$\mu$s, a 1\,KiB all-to-all 1.70\,$\mu$s instead of 3.02\,$\mu$s.
This advantage does not stem from bandwidth and would therefore persist against a non-blocking EPS fabric.
A packet-switched fabric lets any node reach any other, which determines which edges are usable and therefore affects the bandwidth term; it cannot remove serialized phases from a collective, and the switch stages it traverses add to the latency term.
Folding instead converts ports into fewer phases.
For the all-reduce, the coset-factored parallel schedule is best up to 50\,KiB and parallel RS+AG beyond, while the all-EPS fabric moves from one-shot to two-shot and, from 5\,MiB on, halving-doubling.

For large buffers, the all-reduce advantage grows to $3.6\times$ at 50\,MiB (68 instead of 242\,$\mu$s), which mostly reflects the 4:1 oversubscription of the all-EPS fabric that all-OCS avoids (Appendix~\ref{app:physical-fabrics}); a non-blocking EPS fabric would close most of this part of the gap.
For the all-to-all, the advantage peaks at $3.4\times$ at 512\,KiB (3.05 instead of 10.4\,$\mu$s) and is $1.84\times$ at 250\,MiB (0.66 instead of 1.21\,ms).

The following Subsection~\ref{app:a2a-oversub} separates the contributions of schedule, subscription, and switch technology to this gap.

\subsubsection{Contributions of Schedule, Subscription, and Switch Technology}
\label{app:a2a-oversub}

The EPS fabric of our comparison is oversubscribed 4:1 between its leaf (level-1) and spine (level-2) switches, since a fully subscribed EPS spine with pluggable conversion exceeds a reasonable power envelope; the OCS spine needs no conversion and is fully subscribed.
The all-to-all is the collective that suffers most from oversubscription: with $C$ racks, a fraction $(C-1)/C$ of every accelerator's data is addressed to other racks and has to cross the spine, whatever the schedule.
To separate the effect of the schedule from that of the fabric, we also run our parallel Bruck schedule on the EPS fabric, with the logical edges of the fold carried as packets over all physical links of each accelerator.

\begin{figure}[t]
\centering
\includegraphics{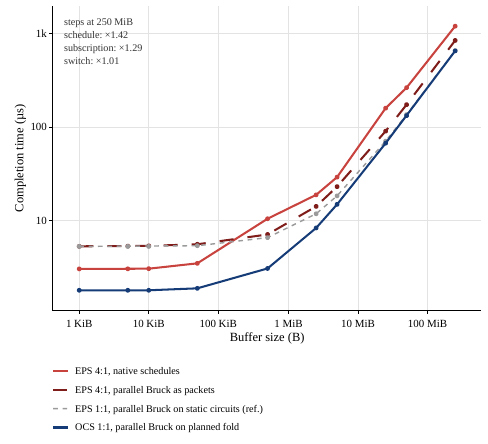}
\caption{All-to-all completion time against buffer size, 512 accelerators in 16 racks of 32 (36 ports per accelerator, 400\,Gb/s per link). Four configurations, each differing from the next in one respect: EPS 4:1 with its native schedules; EPS 4:1 with parallel Bruck carried as packets; EPS 1:1 with parallel Bruck on static circuits, a reference that exceeds the power envelope; OCS 1:1 with parallel Bruck on the planned fold. The native EPS schedule is one-shot at every buffer size; the step ratios at 250\,MiB are given in the upper left.}
\label{fig:a2a-ladder}
\end{figure}

Figure~\ref{fig:a2a-ladder} compares four configurations to examine the contributions of collective scheduling, spine subscription, and switch technology. The transition to the fully subscribed EPS reference also changes transport from packet routing to static circuits, so this step does not isolate subscription alone.
The EPS fabric at 4:1 with its native all-to-all schedules is the deployment that still fits a reasonable power envelope.
Running our parallel Bruck schedule on the same fabric changes only the schedule and accounts for about two thirds of the gap at large buffers: at 250\,MiB, it reduces the completion time from 1.21 to 0.85\,ms, against 0.66\,ms on the OCS fabric.
Raising the EPS spine to full subscription changes only the subscription; this configuration can only be built with static circuits, since circuits routed through an EPS spine are colorable only at full bisection, and we include it as a reference that exceeds a reasonable power envelope.
At 250\,MiB, this step accounts for the remaining third of the gap (0.85 to 0.66\,ms).
The remaining step replaces the EPS switches of that reference by OCS and changes only the switch technology; at large buffers the two perform alike (0.661 and 0.657\,ms at 250\,MiB).
At small buffers they do not: every hop through an EPS switch still pays its switch latency, so the static reference takes 5.3 instead of 1.8\,$\mu$s at 1\,KiB, which is the latency advantage discussed in Appendix~\ref{app:eps}.
At large buffers, the OCS fabric therefore does not gain on bandwidth at equal subscription.
It gains because full subscription is affordable for it, whereas on the EPS fabric the step from 4:1 to 1:1 is exactly the part of the gap that the oversubscribed spine imposes.

\subsection{Shared-Host vs. Stacked-Link Allocation}
\label{app:stacked-link}

\begin{figure}[t]
  \centering
  \includegraphics[width=\columnwidth]{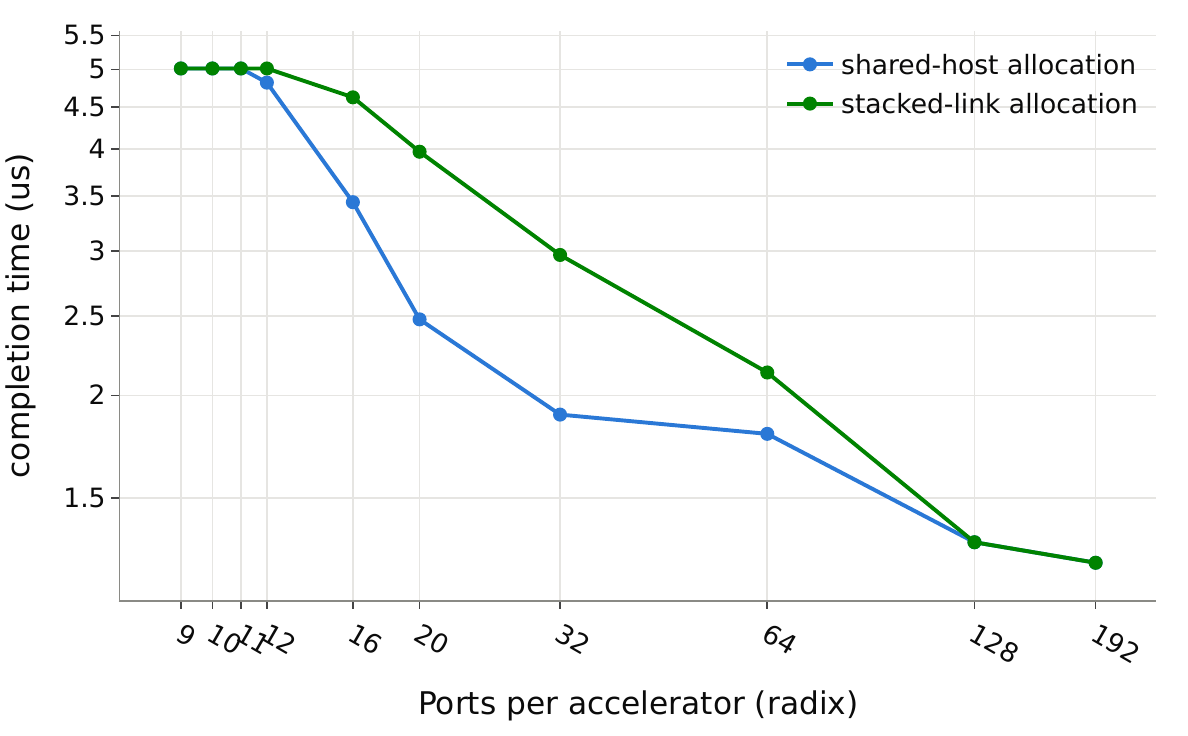}
  \caption{2-D mesh ($8 \times 64$), 8\,kB all-reduce, 512 accelerators: completion time against accelerator radix, stacked-link allocation vs.\ shared-host allocation.}
  \label{fig:sl-vs-sh-2dim}
\end{figure}

An alternative to the shared-host construction of Section~\ref{sec:logical-topology} is to treat each communication axis independently and allocate it a dedicated subset of accelerator ports.
We call this \emph{stacked-link} allocation.
Newer inference hardware architectures devote physical subsets of their scale-up links to a given mesh axis, e.g., TP or EP in TP/EP parallelism~\cite{openai2026jalapeno}.
Stacked-link allocation is the workload-aware variant of this concept, which partitions the ports at deployment time rather than at system design time.

Each axis $i$ receives $L_i$ ports, with $\sum_i L_i\le L$: one port each, and the rest in proportion to the full-connectivity degree $N_i-1$, rounded by largest remainder.
If $\sum_i (N_i - 1) \le L$, every axis is fully connected and the leftover ports become parallel circuits.
For each axis over its $N_i$ ranks, we choose a circulant Cayley graph on $\ZZ_{N_i}$ at degree $L_i$ that minimizes the diameter, using diameter as a latency-oriented proxy for collective completion time.
The search is exhaustive for small axes and sampled for large ones, so for large axes the stacked-link circulant is a heuristic, not necessarily the minimum-diameter one.
Since a circulant is not a cube, only the one-shot schedules run on it.

Figure~\ref{fig:sl-vs-sh-2dim} compares the two allocation approaches on a single-level OCS fabric of 512 accelerators under an 8\,kB all-reduce on a 2-D mesh ($8 \times 64$) with 72 collective groups: 64 groups of 8 accelerators along the first axis, and 8 groups of 64 accelerators along the second.
The two approaches use the $L$ links of each accelerator differently.
Stacked-link fixes the split beforehand: one set of ports is assigned to the 8-rank axis and another to the 64-rank axis.
Each axis can then use its ports only to reduce the diameter of its own graph; until both axes are fully connected, ports assigned to one axis can neither be used by the other nor increase bandwidth.
Shared-host instead uses $\log_2 N = 9$ ports to build the base hypercube, whose generators it splits between the two axes, three for the 8-rank axis and six for the 64-rank one.
The remaining ports form a common pool, and the planner chooses folds and parallel circuits for both axes jointly under that budget: fold generators to remove collective phases when latency is the main bottleneck, parallel circuits to increase bandwidth when it is.
The automatically selected collective schedule then uses whichever the plan provides.
This gives shared-host a clear advantage at intermediate radix, from $L=16$ to $L=64$ in Figure~\ref{fig:sl-vs-sh-2dim}.
At higher radix, the two curves converge again: once $L \ge (8-1)+(64-1) = 70$, stacked-link has enough ports to fully connect both axes independently, and both approaches reach a network diameter of one with the same completion time.

Shared-host allocation is therefore the more efficient in most cases, since one budget serves both axes and surplus ports can go to whichever of latency and bandwidth gains more.
Stacked-link allocation is more flexible in the topology it can realize.
Our current shared-host construction requires $L\ge\log_2N$, since its base host graph is defined over binary rank labels, and therefore assumes power-of-two axis extents;
stacked-link only needs enough ports $L_i$ per axis to connect its own $N_i$ ranks.
Both allocations determine which logical adjacencies and bandwidth are available to a collective, but neither alone determines the resulting execution time.
Their benefits materialize only if the collective schedule can exploit the provided connectivity (Appendix~\ref{app:schedules}).

\subsection{Scope and Limitations}
\label{app:limits}

The multi-mesh search is heuristic and does not establish global optimality.
Only one- and two-dimensional meshes are supported, all meshes of a deployment must span the same power-of-two world size, and non-power-of-two sizes are not supported at all.
The conformance check covers $p = 1$ only (Appendix~\ref{app:conformance}).
The coset-factored schedules require that the partial sums combined in successive phases never overlap, which holds exactly when the generators of a fold tile the group as layers.
On irregular folds, such as the diameter-optimal generator set of degree 13 on an axis of size 64, no such tiling exists, so these schedules need more phases than the diameter.


\section{Details on \textsc{gxnet}}
\label{app:gxnet}

This appendix details how \textsc{gxnet} translates a model graph into a distributed decode-step execution. It describes the graph transformations, MoE load assumptions, compute and memory cost model, and coupling to the network simulator. It then evaluates sensitivity to HOP-B microbatching and explains the search over parallelization configurations used in the evaluation.

\subsection{From Model Graph to Distributed Decode Step}
\label{app:gxnet-passes}

A model graph consists of the token embedding, the decoder layers, and the final norm and LM head. The decoder layers are one repeated block with a repetition count; Megatron 5T repeats one layer, GPT-OSS-5.3-like a pair of a sliding-window and a full-attention layer. Tensor shapes are SymPy expressions in the batch size and in the mesh coordinates of the rank. A sequence of graph-to-graph passes turns the model into a timed distributed decode step:
\begin{enumerate}[leftmargin=*,itemsep=1pt,topsep=2pt]
\item \emph{Decode.} The graph is specialized to generating one token per user against a KV cache that holds the context.
\item \emph{Precision.} Activations, weights, and the KV cache are stored in FP4, and collectives communicate activations in FP4; Appendix~\ref{app:sensitivity} repeats the evaluation with BF16 communication.
\item \emph{Helix sharding.} Attention is placed on a KVP $\times$ TPA mesh, which shards the KV cache by sequence over KVP and by KV head over TPA, and the FFN on an EP $\times$ TPE mesh over the same accelerators. Collectives are inserted as described in Section~\ref{sec:workload} (Figure~\ref{fig:model-graphs}).
\item \emph{Kernels.} Operators are grouped into fused kernels, e.g., the QKV projection, flash-decode attention, and the grouped expert GEMM.
\item \emph{Resources and time.} Each kernel is annotated with its FLOPs and HBM traffic, and then with a duration (Appendix~\ref{app:gxnet-cost}).
\item \emph{Schedule.} Each layer is split into an attention and an FFN region (four regions for GPT-OSS-5.3-like), which execute in order without HOP-B microbatching~\cite{bhatia2025helix} (Appendix~\ref{app:hopb}).
\end{enumerate}
The same sharded graph also yields a per-accelerator memory estimate (weights and KV cache), which bounds the batch size that fits in HBM.

\subsection{MoE Load Model}
\label{app:gxnet-moe}

Since the all-reduce after the attention output projection replicates the expert input on every accelerator, token routing changes no collective (Section~\ref{sec:static-demand}). It only sets how many tokens each expert processes, and with it the expert compute time and the payload of the TPE all-reduce and EP all-gather. We model dropless routing that is uniform in expectation. With $B$ users and top-$k$ routing over $N_{\mathrm{exp}}$ experts, the number of distinct active experts is approximately
\begin{equation}
N_{\mathrm{act}} = N_{\mathrm{exp}}\left(1-\left(\tfrac{N_{\mathrm{exp}}-1}{N_{\mathrm{exp}}}\right)^{Bk}\right),
\end{equation}
rounded to the nearest integer. The active experts are spread over the EP ranks, so each rank holds $\lfloor N_{\mathrm{act}}/\mathrm{EP} \rfloor$ or $\lceil N_{\mathrm{act}}/\mathrm{EP} \rceil$ of them, and its grouped expert GEMM processes $\lceil Bk/N_{\mathrm{act}} \rceil$ tokens per active expert. The collectives after the experts carry the rows of the busiest rank, its share of the routed tokens,
\begin{equation}
R = \left\lceil \left\lceil \tfrac{N_{\mathrm{act}}}{\mathrm{EP}} \right\rceil \tfrac{Bk}{N_{\mathrm{act}}} \right\rceil,
\end{equation}
rounded up once for the rank rather than once per expert. MoE payloads are therefore sublinear in $B$ (Table~\ref{tab:decode-demand}). We do not model routing skew, which would raise expert compute and payload on the most loaded EP rank.

\subsection{Compute and Memory Cost Model}
\label{app:gxnet-cost}

The duration of a kernel is a roofline estimate,
\begin{equation}
t = \max\left(\frac{F}{P_{\mathrm{FP4}}\,\eta_F(F)},\ \frac{D}{W_{\mathrm{HBM}}\,\eta_D(D)}\right),
\end{equation}
where $F$ is the kernel's FLOPs, $D$ the bytes it moves to and from HBM, $P_{\mathrm{FP4}}$ and $W_{\mathrm{HBM}}$ the peak FP4 throughput and HBM bandwidth of Table~\ref{tab:acc-props}, and $\eta_F$ and $\eta_D$ size-dependent efficiencies. For RX00, $\eta_D$ is 0.3 for transfers up to 1\,MB, 0.6 up to 10\,MB, 0.75 up to 100\,MB, and 0.9 above; $\eta_F$ is 0.15 for kernels up to 7\,GFLOP, 0.36 up to 16\,GFLOP, 0.53 up to 127\,GFLOP, and 0.73 above. A grouped expert kernel reads the weights of every active expert on its rank. The efficiency functions are a step approximation of curves measured on H100 (Appendix~\ref{app:accelerators}).

Kernel launches are charged per rank: the first kernel after each collective pays one launch overhead, and the kernels that follow until the next collective run back to back, as in an accelerator execution graph or persistent-kernel implementation. We use 0.8\,$\mu$s, the per-kernel gap of a replayed CUDA graph on B200~\cite{cheng2025mpk}.

\subsection{Execution and Network Coupling}
\label{app:gxnet-execution}

The schedule is compiled into a plan of compute entries with fixed durations and communication entries whose durations are left to the network model. Each accelerator issues its compute entries in order, as from a single stream. Ranks with identical durations and dependencies share an entry, so the plan size depends on the number of distinct rank behaviors, which arise only where the active experts do not divide evenly over the EP ranks.

An entry starts when its last predecessor ends. A communication entry registers its collective with the network model at its start time, and its end stays open until the network model reports it. When no further collective can be registered, the network model advances to the next completion, and the reported end times are substituted into the entries that depend on them. The network model therefore holds every collective that is in flight at a given time, e.g., the concurrent groups of one mesh axis, which share second-level links on the two-level fabrics. \emph{Simlette} simulates these collectives together, so link contention and credit stalls between them lengthen their completion times, and the compute that waits for them starts later. Pricing each collective in isolation would miss both effects; the analytical model does so. Before the walk, the network model receives the traffic demand of Table~\ref{tab:decode-demand}, against which the OCS planner builds circuits (Section~\ref{sec:ocs-design}); a registered collective outside this demand is an error.

Without HOP-B, no communication overlaps with computation, so every collective lies on the critical path and a decode step is the sum of kernel durations, launch overheads, and collective completion times. The OCS saving per step is the reduction in collective completion time. This also matches the public implementations of Helix: TensorRT-LLM provides Helix parallelism, but its documented implementation does not yet include HOP-B's batch-wise pipelining of attention computation and all-to-all communication, which NVIDIA lists as future work~\cite{trtllm-helix-blog}, and the Helix all-to-all backends recently added to vLLM~\cite{vllm-helix-a2a} and SGLang~\cite{sglang-helix-a2a} do not describe such overlap. The following subsection Appendix~\ref{app:hopb} shows that HOP-B would not shorten the evaluated decode steps.

\subsection{Sensitivity to HOP-B}
\label{app:hopb}

\begin{figure*}[t]
  \centering
  \includegraphics[width=\linewidth]{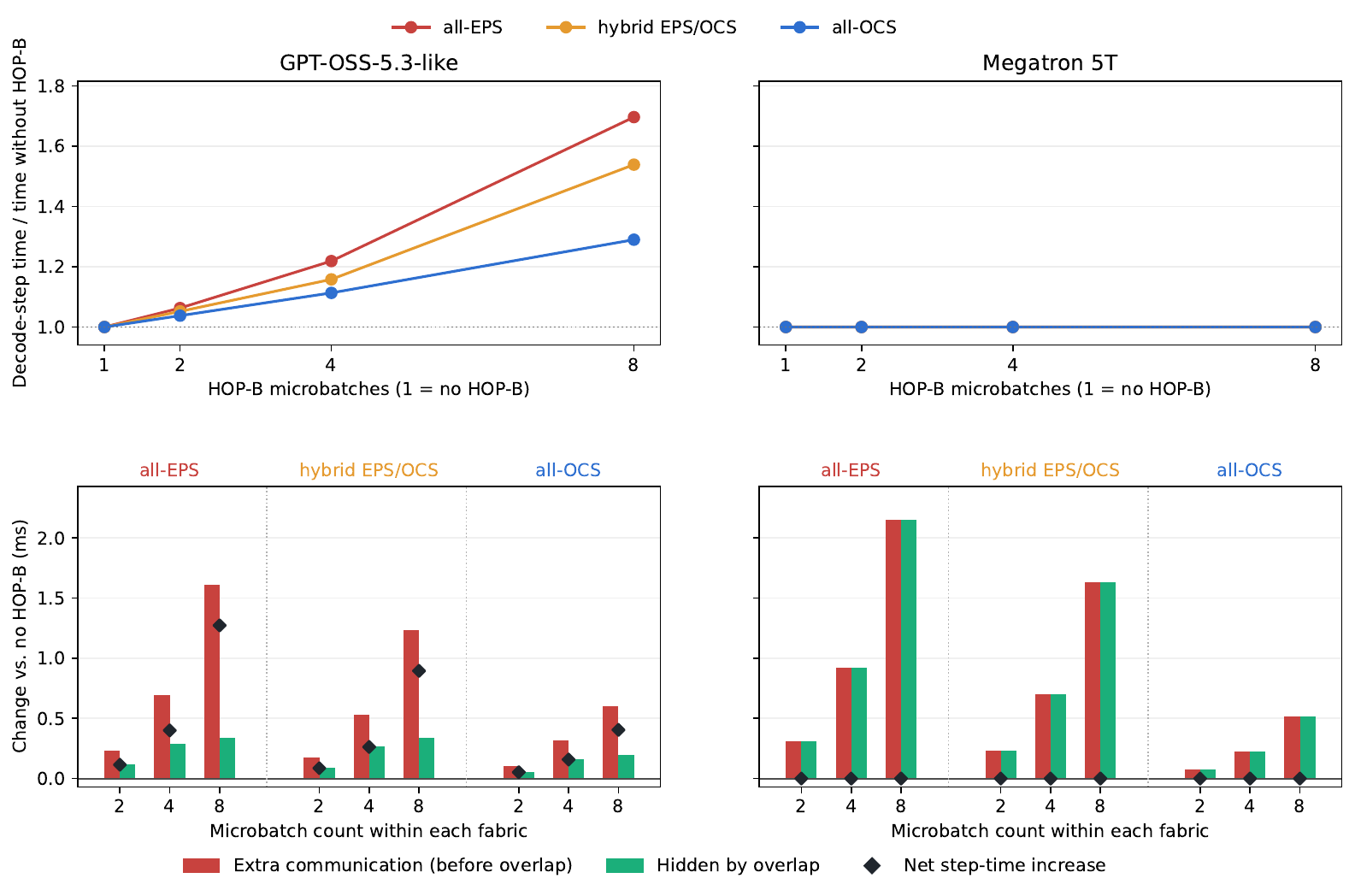}
  \caption{HOP-B sensitivity at batch 16 and 5M context, simulated with \emph{Simlette}, each fabric at its parallelization of Figure~\ref{fig:per-batch-both}. One microbatch is the execution without HOP-B used in Section~\ref{sec:eval}. Top: decode-step time with HOP-B over 2, 4, and 8 microbatches, relative to the step without HOP-B. Note that results for Megatron 5T coincide at approximately 1.0, so red and orange curves are not visible. Bottom: change against the step without HOP-B in communication time before overlap (red), in communication hidden behind compute (green), and in decode-step time (diamonds).}
  \label{fig:hopb}
\end{figure*}

HOP-B~\cite{bhatia2025helix} splits the attention of a decode step into microbatches and overlaps the KV-parallel all-to-all of one microbatch with the attention computation of the next. The QKV projection, the FFN, and their collectives, i.e., the all-reduce after the output projection, the FFN all-reduce and expert all-gather, and the logit all-gather, still run once on the whole batch and are not overlapped. Figure~\ref{fig:hopb} evaluates it for both models at batch 16 and 5M context, the largest batch and context of Section~\ref{sec:eval}, which give HOP-B the most computation to overlap with.

Microbatching does not reduce the bytes of the all-to-all, but it issues the all-to-all once per microbatch in every layer. Its payload is divided among the calls, but each additional call adds the fixed latency of a collective, which dominates for the small payloads of decode (Table~\ref{tab:decode-demand}), so the communication time before overlap grows with the number of microbatches. HOP-B can hide this communication only behind attention, the all-to-all of the last microbatch stays exposed, and the decode step becomes shorter only if the hidden communication exceeds the added communication.

For GPT-OSS-5.3-like, it does not. A decode step computes for only 0.77\,ms, half of its layers attend to a 128-token window, and HOP-B hides at most 0.34\,ms of communication, while eight microbatches add 1.6\,ms of communication on all-EPS. Against the step without HOP-B, the decode step grows by 6, 22, and 70\,\% on all-EPS with 2, 4, and 8 microbatches, and by 4, 11, and 29\,\% on all-OCS, whose collectives have the lowest latency per call.
For Megatron 5T, attention over the 5M-token context dominates the 8.8\,ms of computation per step and hides all added all-to-alls, but the exposed communication does not shrink: the all-to-all of the last microbatch takes as long as the single all-to-all without HOP-B, and the other collectives are not overlapped. The decode-step time is therefore unchanged on every fabric.

For the evaluated model-parallel configurations, microbatching with HOP-B does not improve decode-step time relative to a single microbatch. 
For GPT-OSS-5.3-like, the additional collective startup costs exceed the
communication hidden through overlap. 
For Megatron 5T, the additional communication is almost entirely hidden,
but the remaining exposed communication is unchanged.
The single-microbatch configuration is therefore optimal or equivalent among the evaluated schedules, supporting its use in Section~6.

\subsection{Searching the Parallelization Space}
\label{app:gxnet-search}

On $N$ accelerators, a Helix split (KVP, TPA, EP, TPE) satisfies $\mathrm{KVP} \cdot \mathrm{TPA} = \mathrm{EP} \cdot \mathrm{TPE} = N$, with power-of-two degrees, TPA dividing the number of KV heads, EP dividing the number of experts, and TPE dividing the expert width. At $N = 512$, this gives 72 splits for Megatron 5T and 45 for GPT-OSS-5.3-like.

For Figure~\ref{fig:per-batch-both}, we choose a split per cell, i.e., per model, context (32K and 5M tokens), batch size (1, 2, 4, 8, and 16), and fabric. \textsc{gxnet} with the analytical network model of Section~\ref{sec:analytical-model} times every split in every cell: 2160 points for Megatron 5T and 1350 for GPT-OSS-5.3-like, each simulated in under a second. The analytical model ignores contention and credit stalls (Section~\ref{sec:analytical-model}), so we re-simulate with \emph{Simlette} every split whose analytical interactivity is within 7\,\% of the best split of its fabric in that cell, and report the best \emph{Simlette} result per cell. This amounts to 758 points for Megatron 5T and 338 for GPT-OSS-5.3-like, with a median simulation time of about 10\,minutes per point.
\section{Collective algorithms and time consumption in the Network Simulator}
\label{app:collectives}

This appendix details how Simlette models collective communication on the evaluated network fabrics. It describes the simulator’s mapping stages, derives the analytical timing model, and explains the collective schedules and their consistency checks. It then presents simulation optimizations, hardware measurements used to parameterize latency, and the physical fabric configurations used in the evaluation.

\subsection{\emph{Simlette} Layers}
\label{app:simlette-layers}

\emph{Simlette} separates the OCS-specific problem of mapping an ML workload to the physical network fabric into four layers, applied in the following order. The first is the topology: the physical host graph $H=(V,E)$ of devices, switches, and the ports that connect them, under the per-accelerator port budget $L$ introduced in Section~\ref{sec:port-allocation}. The second is worker allocation: the mapping of logical ranks onto this physical graph, i.e., the embedding of a logical overlay (Section~\ref{sec:logical-topology}) into the accelerators and links that will actually carry its traffic. The third is the collective, which supplies the schedule defined in Section~\ref{sec:topology-aware-collectives}: the ordered sequence of phases, the rank pairs and edges used at each phase, and the payload carried on each. This layer jointly searches over the collective and its logical topology, keeps the lowest-cost pair, and thereby determines the number of phases $h_i$. The fourth is routing, which takes the chosen logical topology together with the physical topology and the allocation, and maps it onto concrete ports, paths, and circuit assignments. These four layers stay independent in the implementation.

\subsection{Derivation of the Analytical Model}
\label{app:analytical-derivation}

We compute the terms of Eq.~\ref{eq:hockney-axis} from the schedule of Section~\ref{sec:topology-aware-collectives}. Let $b_{k,e}$ be the amount of data, in bytes, sent on logical edge $e$ during phase $k$. All transfers of a phase start at the same time, and the phase ends when the transfer on the most loaded edge arrives. This transfer takes the latency $\lambda$ of one hop, plus the time to put its data on the edge. The edge is made of $p$ parallel circuits of bandwidth $R$ each, expressed in bytes per second (400\,Gb/s corresponds to $R = 50$\,GB/s), so sending $b_{k,e}$ takes $b_{k,e}/(pR)$ and phase $k$ lasts $\lambda + \max_e b_{k,e}/(pR)$. Adding the $h$ phases gives Eq.~\ref{eq:hockney-axis}, with the bottleneck volume
\begin{equation}
  v = \sum_{k=1}^{h} \max_e b_{k,e},
  \label{eq:volume}
\end{equation}
i.e., the sum, over all phases, of the data carried by the most loaded edge of each phase. This assumes that phases do not overlap, and that no other collective uses the same links at the same time.

\subsection{Collective Algorithms for Workload-static Switching}
\label{app:collective-algorithms}

Each collective algorithm in \emph{Simlette} computes $h$, the schedule phase count, and $v$, the bottleneck data volume, from its own phases, for a buffer size $\sigma$ on a logical graph of $N$ ranks.

\subsubsection{All-reduce schedules}
\label{app:allreduce-schedules}

\emph{Simlette} implements the all-reduce schedules of Appendix~\ref{app:allreduce}: the one-shot schedule, which reaches every rank within the diameter of the graph, and its data-volume-optimized variant (one-shot DVO), as well as the coset-factored and RS+AG schedules and their parallel variants, which exchange over one coset layer per phase (Tables~\ref{tab:allreduce} and~\ref{tab:opoints}).

\paragraph{Example: Analytical completion time of the all-reduce algorithms}
\label{app:cost-vs-msg}

\begin{figure}[t]
\centering
\includegraphics[width=\columnwidth]{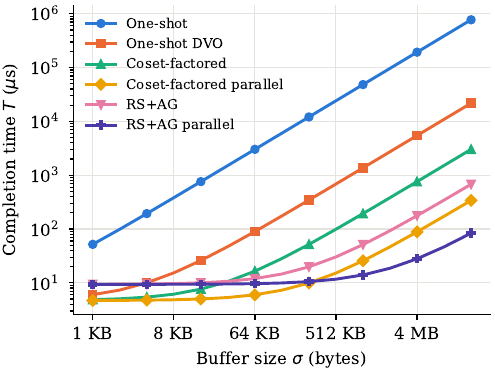}
\caption{Completion time from Eqs.~\eqref{eq:hockney-axis} and~\eqref{eq:volume} for an all-reduce on 512 accelerators
connected as a hypercube without extra generators, with $L=9$ ports, $p=1$,
$\lambda = 0.52\,\mu$s, and $R = 400$\,Gb/s.}
\label{fig:cost-vs-msg}
\end{figure}

Figure~\ref{fig:cost-vs-msg} shows the completion time of the all-reduce algorithms on $N = 512$ accelerators, from 1\,KiB to 16\,MiB. For small buffers the latency term $h\lambda$ is the largest, and the time depends mostly on the number of phases. For example, the two RS+AG algorithms take 18 phases, and their latency term is $18 \times 0.52 = 9.4\,\mu$s. The other algorithms take only 9 phases, and their latency term is $9 \times 0.52 = 4.7\,\mu$s.

For large buffers the bandwidth term $v/(pR)$ is the largest, and the algorithm that sends the least data is the fastest. Coset-factored sends the full buffer in each of its 9 phases. RS+AG sends only a part of the buffer in each phase, and about twice the buffer in total. RS+AG takes twice as many phases, but sends less data. The two curves cross at the buffer size where the extra phases of RS+AG cost as much time as the extra data of coset-factored. Below this size coset-factored is faster, and above it RS+AG is faster. 
Their parallel variants keep the same number of phases, but send 9 times less data on the most loaded link.

One-shot is the slowest at every size. At each phase, a rank sends to each neighbor every contribution it has received so far. This number grows from one phase to the next, and the same contributions cross the same links many times.

\paragraph{Topology-schedule interaction}
\label{app:fold-vs-multiplicity}

\begin{figure}[t]
  \centering
  \includegraphics[width=\columnwidth]{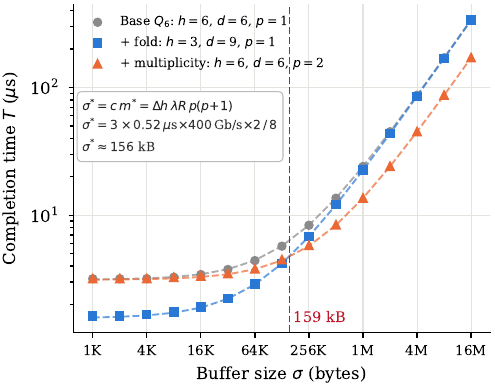}
  \caption{Topology-schedule interaction at network level. Completion time of one all-reduce among 64 accelerators on a single-level OCS fabric (400\,Gb/s per link), as a function of the buffer size $\sigma$ of the all-reduce. Base topology $Q_6$, $+$fold (fewer phases $h$, larger degree $d$), and $+$multiplicity ($p=2$ parallel links per logical edge). The dashed line marks the simulated crossover between $+$fold and $+$multiplicity (159\,kB), close to the buffer size $\sigma^\star = c\,m^\star \approx 156$\,kB predicted by Eq.~\eqref{eq:fold-width-crossover}.}
  \label{fig:fold_vs_multiplicity}
\end{figure}

Figure~\ref{fig:fold_vs_multiplicity} shows the topology-schedule interaction at network level. The collective is a single all-reduce over $N=64$ accelerators, with buffer sizes from 1\,KiB to 16\,MiB in powers of two. The schedule is coset-factored parallel, and the physical fabric is a
single-level all-OCS with a radix of 13 ports per accelerator, of which $L=12$ are used for circuits, and a bandwidth of $R=400$\,Gb/s per link.

\subsubsection{All-to-all schedules}
\label{app:a2a-schedules}

\emph{$r$-Bruck} (radix-$r$ Bruck all-to-all on a block fold, \emph{$r$-Bruck} takes $h = \ell$ phases on a fold of $\ell$ layers. The all-to-all is uniform: every rank starts and ends with the same payload $\sigma$, so $v$ is the same for all ranks. $r$-Bruck implements the Helix all-to-all of Table~\ref{tab:decode-demand} in our evaluation.

\subsection{Analytical Model against \emph{Simlette}}
\label{app:analytic_vs_simlette}

\begin{figure}[t]
  \centering
  \includegraphics[width=\linewidth]{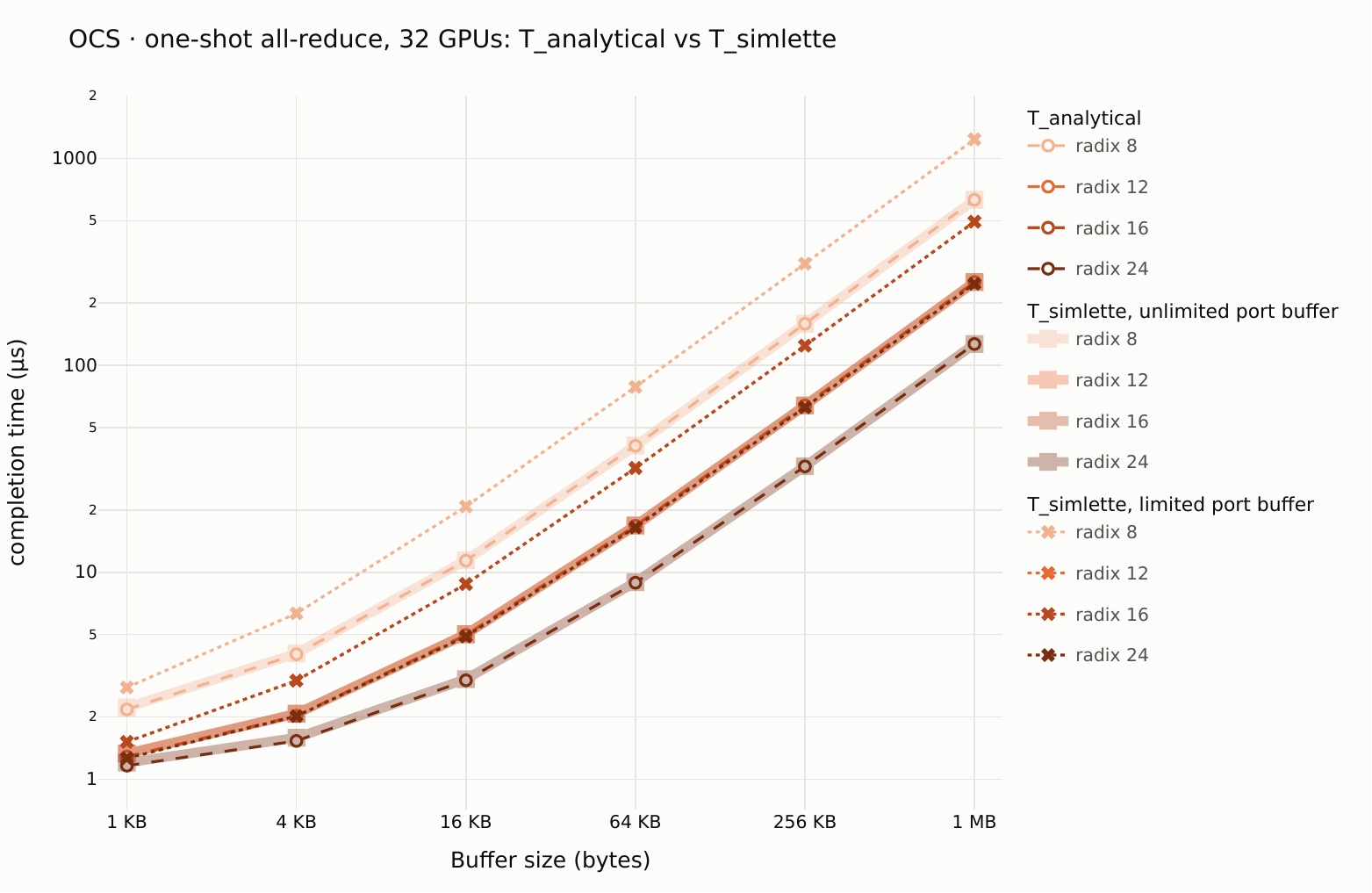}
  \caption{One-shot all-reduce on 32 accelerators with OCS: analytical model against
  \emph{Simlette} with an unlimited and a limited buffer per port. Colors show the number of
  ports per accelerator.}
  \label{fig:credits_ocs}
\end{figure}

\begin{figure}[t]
  \centering
  \includegraphics[width=\linewidth]{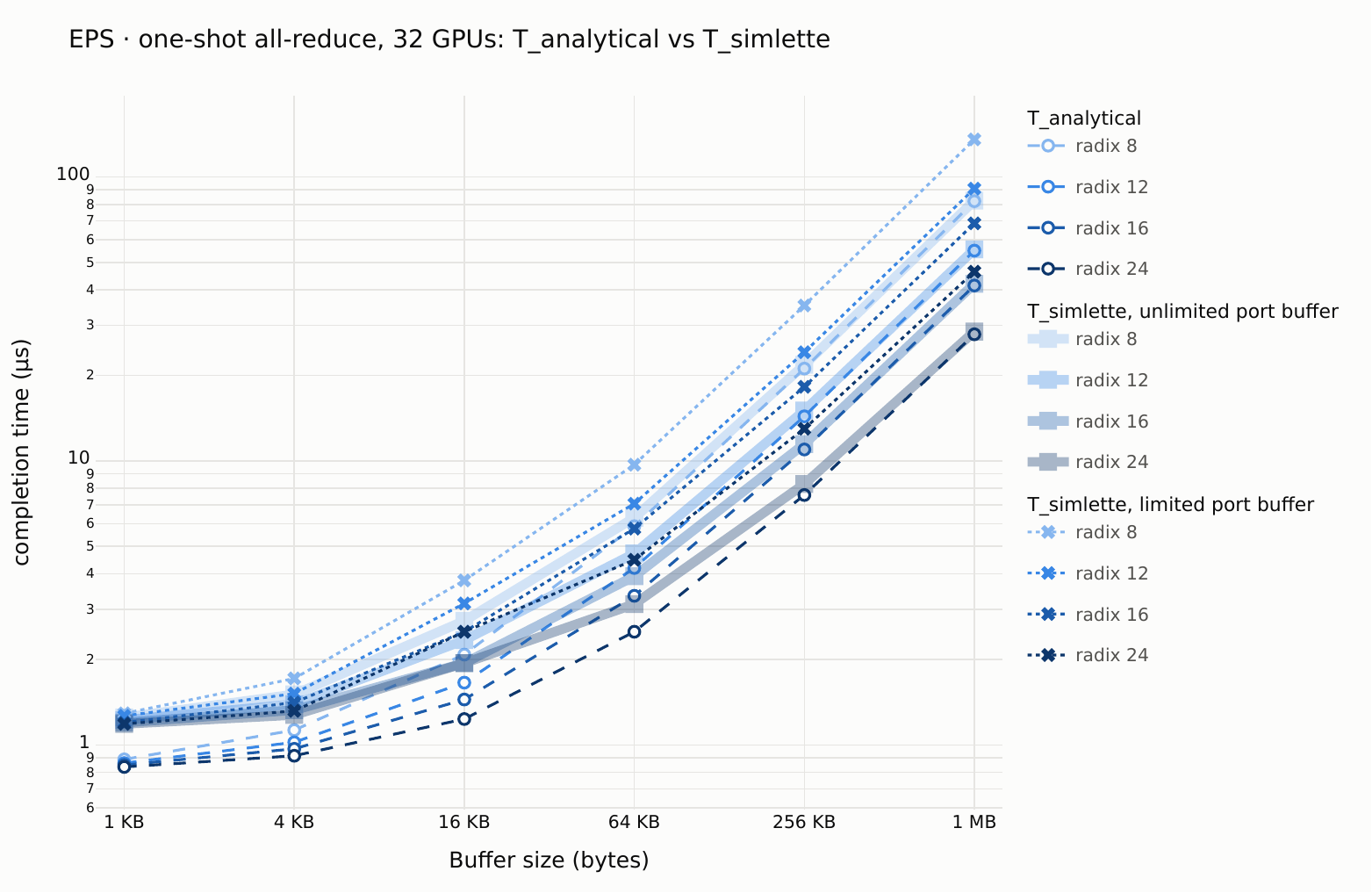}
  \caption{One-shot all-reduce on 32 accelerators with EPS: analytical model against \emph{Simlette} with an unlimited and a limited buffer per port.}
  \label{fig:credits_eps}
\end{figure}

\begin{figure}[t]
  \centering
  \includegraphics[width=\linewidth]{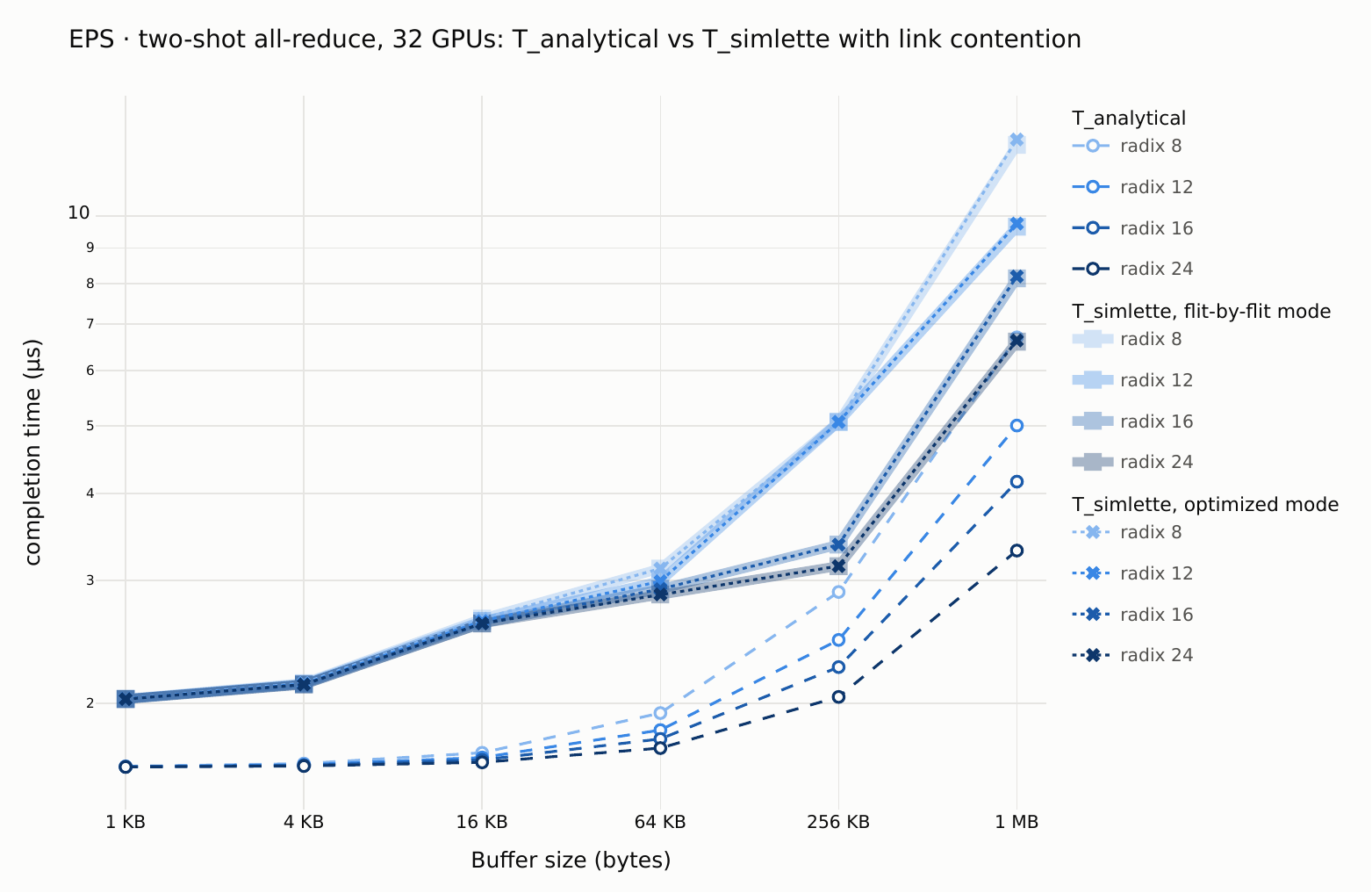}
  \caption{Two-shot all-reduce on 32 accelerators with EPS and an unlimited buffer:
  analytical model against \emph{Simlette} in flit-by-flit and optimized modes.}
  \label{fig:contention}
\end{figure}

The analytical model describes what we expect when nothing gets in the way, so without congestion \emph{Simlette} should give the same results. To compare the two, we run single all-reduces on 32 accelerators connected by a single-level fabric with 400\,Gb/s links, with buffer sizes from 1\,KiB to 1\,MiB and 8, 12, 16 or 24 ports per accelerator. All runs use persistent kernels, so a collective is complete when its last data arrives, as in the analytical model.

\paragraph{OCS.} Figure~\ref{fig:credits_ocs} uses the one-shot all-reduce, whose schedule never sends two flows on the same link at the same time, so there is no link contention. With an unlimited buffer, \emph{Simlette} and the analytical model give the same results for all buffer sizes and radices, which shows that the simulator's behavior is consistent with the model. If the buffer on each port is too small to keep a link busy for one hop, senders have to wait for credits, and \emph{Simlette} is up to $1.9\times$ slower than the analytical model.

\paragraph{EPS: credit stalls.} Figure~\ref{fig:credits_eps} uses the same one-shot all-reduce on EPS, again without link contention. With an unlimited buffer, \emph{Simlette} and the analytical model agree within 5\,\% for small buffers and for buffers of 256\,KiB and larger. For 16-64\,KiB, however, \emph{Simlette} is up to 29\,\% slower: a buffer this small yields per-port messages of only a few flits, which cannot be divided evenly across the links, so the busiest link carries more data than the analytical model assumes. With the buffer of every port reduced below what one hop needs to keep its link busy, senders wait for credits, which makes \emph{Simlette} 1.6-1.7$\times$ slower than the analytical model. For example, with a 1\,MiB buffer and 8 ports, the analytical model gives 82.1\,$\mu$s and \emph{Simlette} 135.0\,$\mu$s: \emph{Simlette} is 52.9\,$\mu$s slower, and its most delayed port spends 53.4\,$\mu$s waiting for credits.

\paragraph{EPS: link contention.} Figure~\ref{fig:contention} uses the two-shot all-reduce on EPS with an unlimited buffer, so there are no credit stalls. Two-shot consists of a reduce-scatter followed by an all-gather. In each phase, all accelerators send to the others in the same order, so they all target the same destination at the same time, and the switch port leading to that destination becomes congested. From 16\,KiB onwards, \emph{Simlette} is 1.5-2$\times$ slower than the analytical model, which counts the right amount of data on each link but cannot capture the time flows spend waiting for each other. The optimized mode of \emph{Simlette} (Appendix~\ref{app:sim-modes}) gives almost the same results as the flit-by-flit mode, with less than 2\,\% difference.

\subsection{\emph{Simlette} Simulation Procedure Optimizations}
\label{app:sim-modes}

Most of the latency along a hop is not paid at the point where it is decided in the simulation. When a switch or a scale-up port adds its share of delay, that delay is added to a timestamp carried on the message, while the operation handing the message to the next stage happens immediately, at the same simulated instant. Simulated time only moves forward at a few points: right before a port puts a message on the link, during serialization, and when a flit reaches its destination.

This deferred accounting matters because several flits are usually in flight at once: by the time a port is ready to send one, part of the delay it owes may already have elapsed while other flits were being processed elsewhere, so only the remaining gap, if any, still needs to be paid as a real wait. \emph{Simlette} computes latency this way so that it always knows exactly how much of that gap is left to catch up, and never pays for the latency twice. The simulation stays exact even though every delay is resolved only when needed.

\begin{figure}[t]
  \centering
  \includegraphics[width=\linewidth]{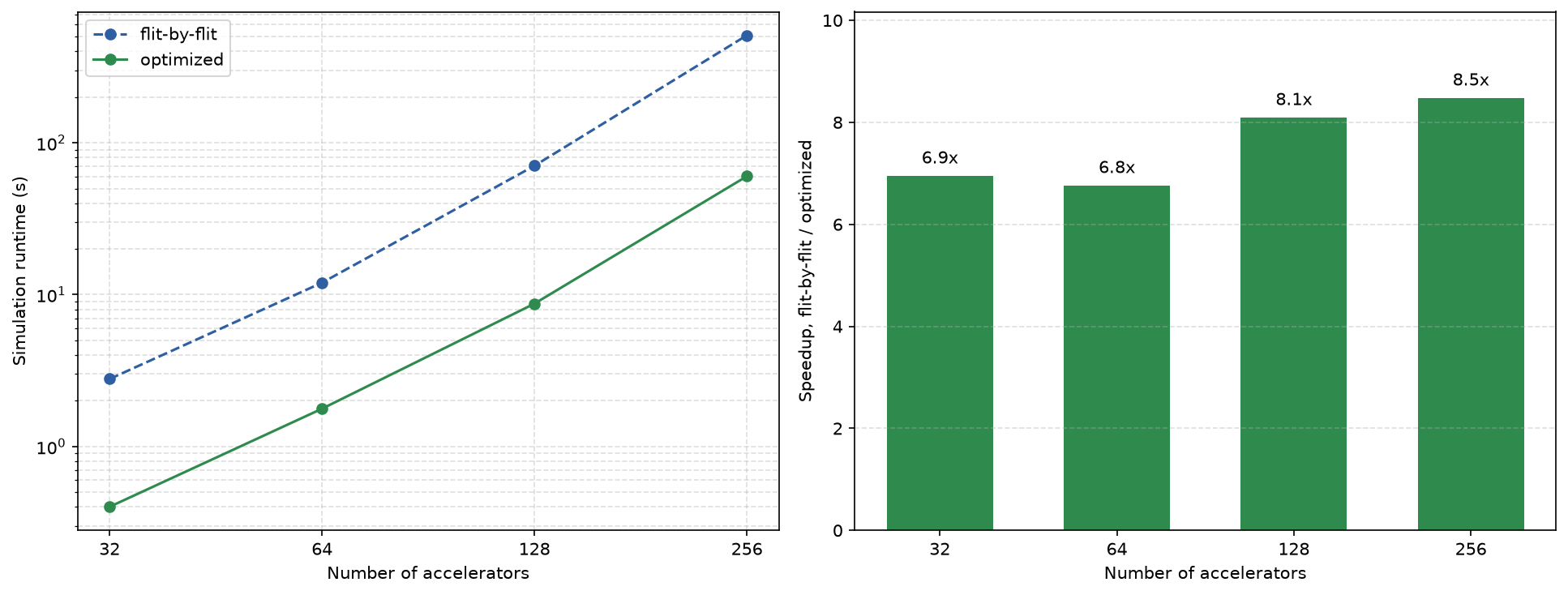}
  \caption{Left: simulation runtime for flit-by-flit vs.\ optimized
  mode, as a function of accelerator count, on a fixed-radix EPS single-level topology (8 ports per accelerator) with a 40\,KiB all-reduce buffer. Right: the resulting speedup of optimized mode over flit-by-flit. Runtimes measured on a single core of an AMD Ryzen 9 9950X desktop processor.}
  \label{fig:runtime_speedup_vs_gpu}
\end{figure}

Simulating a message flit by flit means creating one event for every flit a device sends, rather than a single event for the whole message. At scale, this overhead itself costs real time to execute. A single all-reduce on a simple single-level topology with 256 accelerators and a 40\,KiB buffer already takes about 14 minutes to simulate flit by flit. Larger buffers and more devices in the topology push this further, up to several hours. This is why \emph{Simlette} also supports an optimized mode: the entire message is handed to the link in a single event, and the delay it would have accumulated flit by flit is charged at once on arrival. This way, only one event is created per message per transfer, instead of one event per flit per transfer. A flit is 640 bytes and a message can be tens of kilobytes, so flit-by-flit mode schedules dozens of events where the optimized mode schedules one; with the 40\,KiB buffer of this experiment, a message spans up to 64 flits.
As shown in Figure~\ref{fig:contention}, both modes give almost the same completion times. Figure~\ref{fig:runtime_speedup_vs_gpu} compares the runtime of both modes as a function of accelerator count, on a fixed-radix topology (8 ports per accelerator) with a 40\,KiB buffer, and shows the resulting speedup directly. From 32 to 256 accelerators, the optimized mode is consistently $7$-$9\times$ faster. For the 256-accelerator topology, this cuts runtime from 827 seconds down to 90 seconds.

\paragraph{Determinism}
\label{app:determinism}

The simulation is currently fully deterministic: given the same topology, routing, and workload, every run produces the exact same completion time, with no randomness in link timing, arrivals, or failures. This makes results easy to reproduce and attribute to a single cause, but it does not capture the variability real hardware shows, so the completion times we report should be read as a best case rather than a distribution. This is particularly true for EPS based fabrics, where network-internal dynamics such as arbitration are well known to cause collective completion tail latencies.

\subsection{Scale-up Latency Profiles}
\label{app:latency-profile}

\begin{figure}[t]
  \centering
  \includegraphics{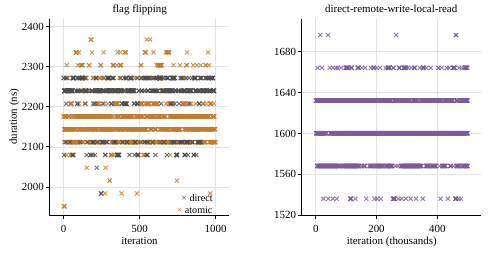}
  \caption{Measured durations of the ping-pong kernel over many iterations; left: flag-flipping implementation (direct and atomic variants), right: direct-remote-write-local-read implementation.}
  \label{fig:latency-measurements}
\end{figure}

\paragraph{End-to-end measurements.}
We characterize accelerator-to-accelerator communication latency using several ping-pong kernel implementations.
Figure~\ref{fig:latency-measurements} illustrates the substantial implementation dependence of measured round-trip times.
For model parameterization, we use the minimum observed latencies as an optimistic estimate of the hardware communication path.

\begin{figure*}[t]
  \centering
  \includegraphics[width=0.9\linewidth]{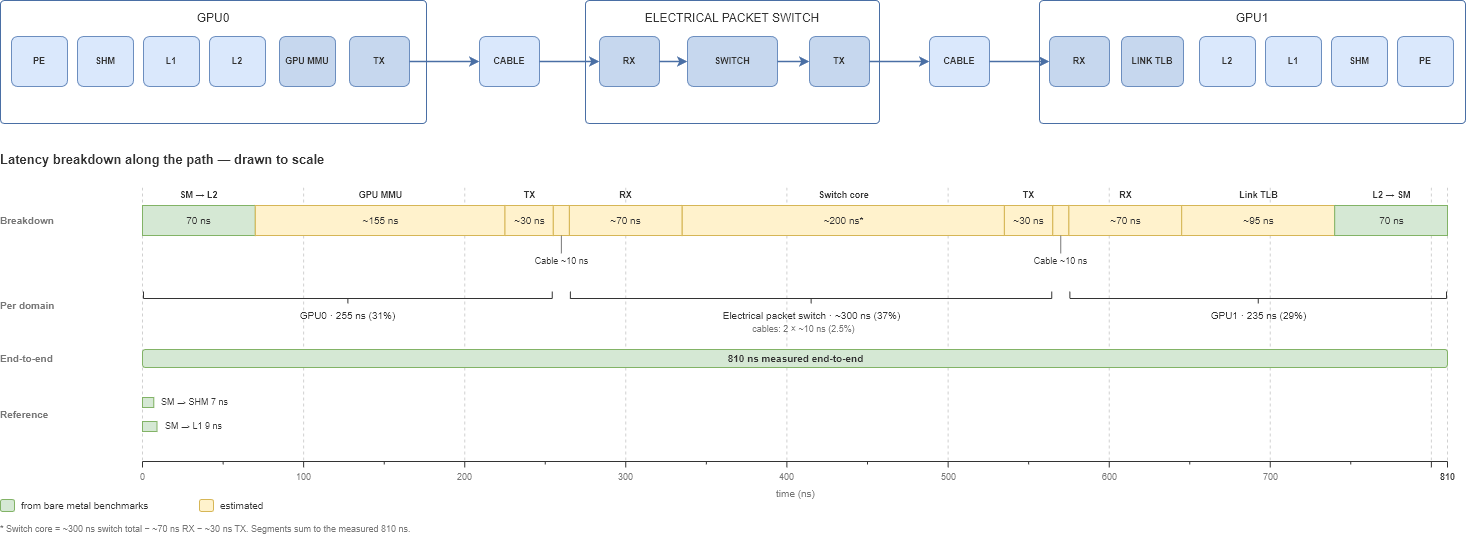}
  \caption{Cumulative latency from the SM of GPU0 to each memory scope, all
  arrows sharing one origin (B300, GPU global timer, 30{,}000-sample average).
  (1)~shared memory, 14~ns; (2)~L1, 17~ns; (3)~L2, 140~ns; (4)~remote access to
  the L2 of GPU1 through two NVLink ports, two cables and one NVSwitch stage,
  1610~ns end-to-end (ping-pong, direct-remote-write-local-read). The on-chip stages
  account for under 10\,\% of the round trip; the remainder is fabric traversal
  and remote-access handling.}
  \label{fig:latency-profile}
\end{figure*}

\paragraph{Latency decomposition.}
Figure~\ref{fig:latency-profile} shows the latency component decomposition of network transactions on scale-up networks as modeled by the analytical network backend of \textsc{gxnet} and as used for the parametrization of our flit-based network simulator interfacing with \textsc{gxnet}. Note that this is not official NVIDIA data. The bare metal measurements carried out in Section~\ref{sec:grounding} can only provide a limited number of known values to a set of equations that will then be underspecified. Specifically, allocating latency to the subcomponents ``GPU MMU'', ``TX'', ``RX'', ``SWITCH'' (core switch latency) and ``LINK TLB'' is a matter of educated guesses. We believe that budgeting more latency into RX rather than TX sections is well justified by the FEC decode overhead alone. 

Results for collective completion times in inference decode, especially for low-batch/high-interactivity cases, will be very susceptible to rebalancing spent latency between TX/RX stages and switch core. A lower switch pin-to-pin latency than the unloaded 300~ns assumed in this paper~\citepalias[Sec.~10.9.2]{ualinkspec} would translate into a smaller performance advantage for all-OCS based network fabrics - and this for two separate reasons: first, the latency difference between EPS and OCS switches as such would reduce. More importantly however, OCS fabrics rely on relaying information through the accelerator itself. A lower EPS pin-to-pin latency directly translates into higher accelerator ingress/egress latency because the absolute PE-to-PE latency must stay the same.

However, the opposite is of course also true: accelerators with very low ingress/egress latencies will emphasize the advantage of all-OCS fabrics. In light of recent industry trends to offload processing elements from collective communication tasks and introduce hardware based ``collective acceleration engines'' into the architectures (e.g., TPU~8i~\cite{google2026tpu8}), the importance to lower switch pin-to-pin latencies in the network will grow in future systems.

\paragraph{Memory-hierarchy latency.}
The accelerator-specific terms of the scale-up latency decomposition cover the path from the
PE through the cache hierarchy to the scale-up port. We anchor them by replicating
the pointer-chasing results of Table~3 in~\cite{luo2025hopper} with the
open-source microbenchmarks accompanying that paper, run on H100 and
B300 (Table~\ref{tab:latency}). On H100 we measure $\approx$41~cycles (L1),
$\approx$29~cycles (shared memory) and $\approx$276~cycles (L2), matching the
published H800 values for shared memory and L2 and slightly exceeding the
published L1 latency. On B300 the same binaries yield $\approx$34,
$\approx$28.6 and $\approx$277~cycles. Reading the GPU global timer from PTX
instead of the cycle counter, and averaging over 30{,}000 samples, gives
$\approx$14.0~ns, $\approx$16.7~ns and $\approx$139.7~ns for shared memory, L1
and L2 respectively, without assuming an SM clock frequency. Two observations
matter for the model: on-SM and L2-hit latency is effectively unchanged across
two GPU generations, so it can be carried into future-accelerator scenarios
with little risk; and L2, the deepest on-chip stage a remote access traverses,
sits below 150~ns and is therefore a small fraction of the
accelerator-to-accelerator path (Figure~\ref{fig:latency-profile}).

\begin{table*}[t]
  \centering
  \caption{Pointer-chasing hit latency in clock cycles. Published values from
  Table~3 of~\cite{luo2025hopper}; H100 and B300 measured with the same benchmark
  binaries.}
  \label{tab:latency}
  \footnotesize
  \setlength{\tabcolsep}{2.5pt}
  \begin{tabular}{@{}lrrrrr@{}}
    \toprule
     & RTX4090 & A100 & H800 & H100 & B300 \\
     & \cite{luo2025hopper} & \cite{luo2025hopper} & \cite{luo2025hopper} & (ours) & (ours) \\
    \midrule
    L1 cache      & 32.0  & 33.0      & 32.0      & 41.1  & 34.1  \\
    Shared memory & 30.1  & 29.0      & 29.0      & 29.0  & 28.6  \\
    L2 cache      & 273.0 & 202.8--408 & 264.5--502 & 276.2 & 276.8 \\
    \bottomrule
  \end{tabular}
\end{table*}

\paragraph{Scale-up latency parametrization variants}

The latency profile break-down presented above has to work with a set of assumptions for the per-component latencies on the data path. In order to study sensitivity of our results to latency attribution variations, we choose three different attributions that either emphasize or de-emphasize the contribution of the EPS switch core (Table~\ref{tab:latency-budget}). 

\begin{table}[t]
  \centering
  \caption{Remote-access latency budget (ns) under low, nominal and high
  switch-latency assumptions.}
  \label{tab:latency-budget}
  \footnotesize
  \setlength{\tabcolsep}{3pt}
  \begin{tabular}{@{}lrrrrrr@{}}
    \toprule
    Case & \multicolumn{1}{c}{Switch} & \multicolumn{1}{c}{Switch}
         & \multicolumn{1}{c}{MMU}    & \multicolumn{1}{c}{GPU}
         & \multicolumn{1}{c}{Link}   & \multicolumn{1}{c}{Total} \\
         & \multicolumn{1}{c}{total}  & \multicolumn{1}{c}{core}
         & \multicolumn{1}{c}{+\,TLB} & \multicolumn{1}{c}{MMU}
         & \multicolumn{1}{c}{TLB}    & \\
    \midrule
    Low     & 200 & 100 & 350 & 217 & 133 & 810 \\
    Nominal & 300 & 200 & 250 & 155 &  95 & 810 \\
    High    & 500 & 400 &  50 &  31 &  19 & 810 \\
    \bottomrule
  \end{tabular}

  \vspace{2pt}
  \parbox{\columnwidth}{\scriptsize
  Fixed in all cases (260\,ns): SM$\to$L2 70, TX 30, cable 10, cable 10,
  RX 70, L2$\to$SM 70.
  Switch total $=$ RX 70 $+$ core $+$ TX 30.
  MMU\,$+$\,TLB $= 550 - $ switch total, split at the nominal ratio 155:95
  (62\,\%\,/\,38\,\%).}
\end{table}

\subsection{Physical Fabrics}
\label{app:physical-fabrics}
\label{app:network-topology}

\begin{figure*}[t]
  \centering
  \subcaptionbox{Hybrid EPS/OCS: level-1 EPS (32 down / 8 up, 4:1), level-2 OCS.\label{fig:fabric-hybrid}}{%
    \includegraphics[width=\textwidth]{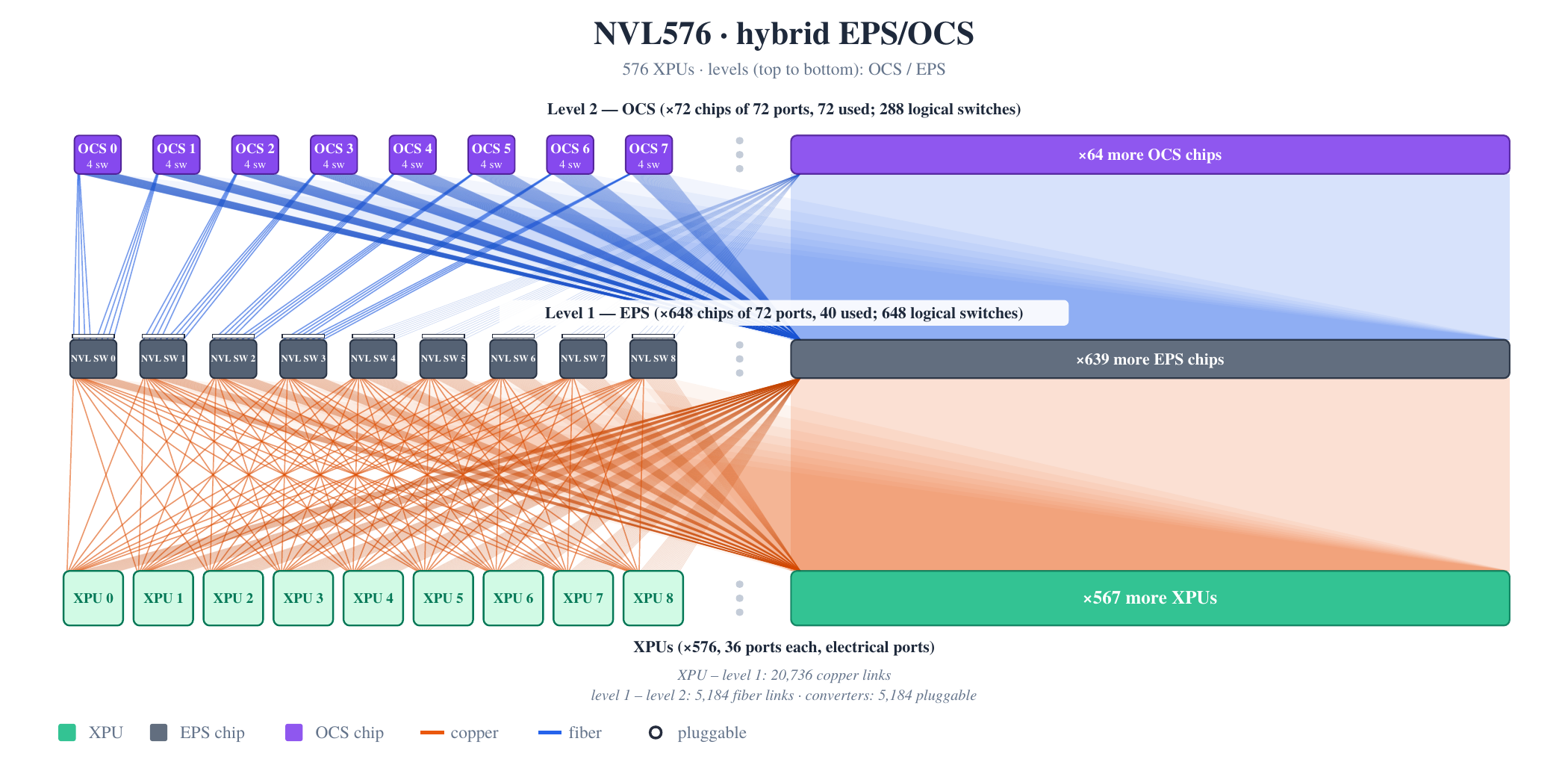}}\\[4pt]
  \subcaptionbox{All-OCS: non-blocking, OCS on both levels, CPO at the accelerators.\label{fig:fabric-allocs}}{%
    \includegraphics[width=\textwidth]{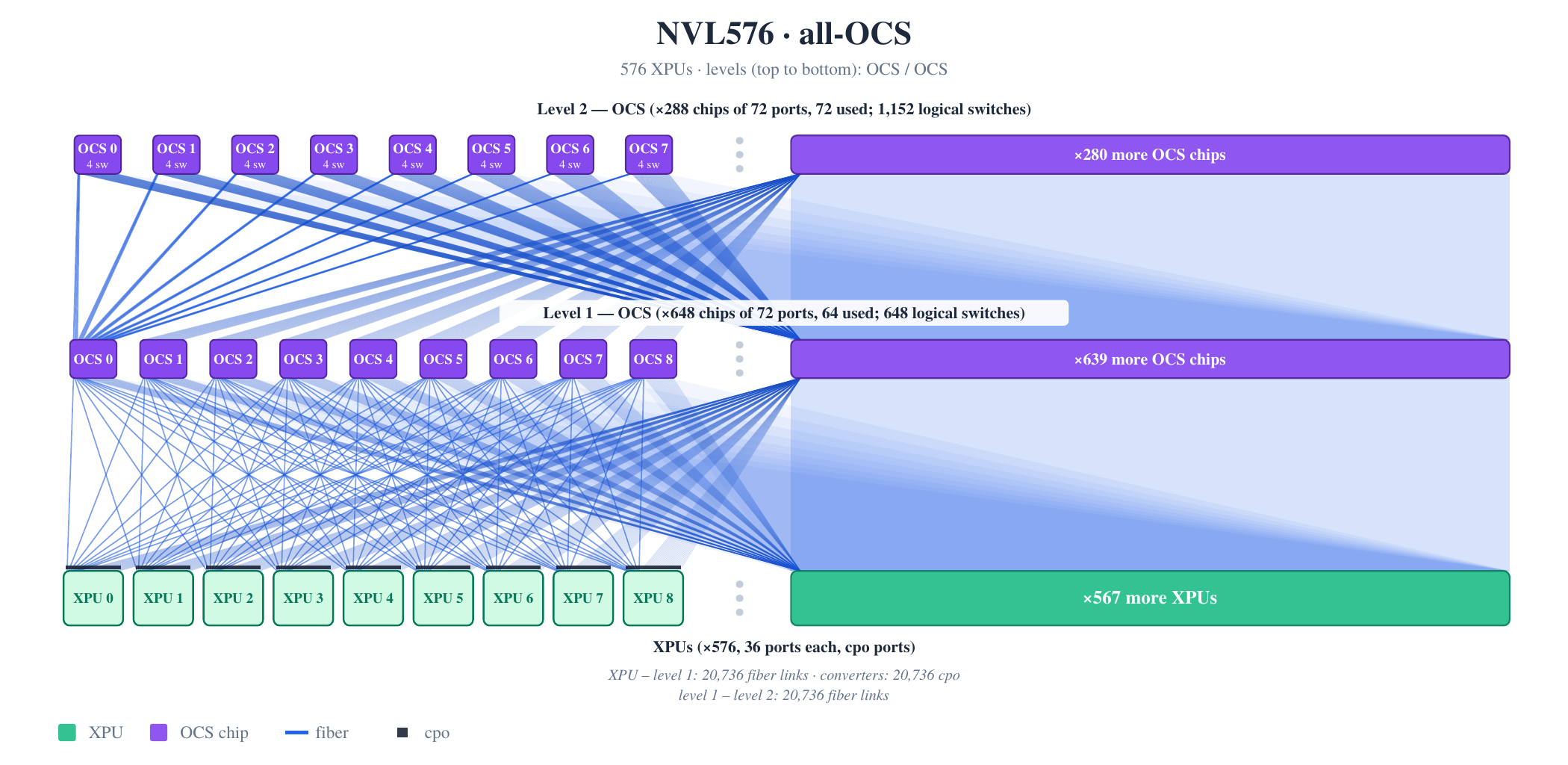}}
  \caption{Physical scale-up fabrics of the 576-accelerator system: 36 ports per accelerator at 400\,Gb/s, switching devices of radix 72. The all-EPS fabric has the topology of (a) with EPS on both levels.}
  \label{fig:physical-fabrics}
\end{figure*}

We evaluate three scale-up fabric families as reference points for where optical circuits enter the accelerator-to-accelerator path. 
In all cases, each accelerator exposes \(L\) links of bandwidth \(R\). 
We describe both packet- and circuit-switched fabrics by the paths available to collective schedules: hop count, per-hop latency, available bandwidth, and port consumption.

The \emph{all-EPS} baseline represents a conventional electronic-packet-switched scale-up fabric. 
It provides flexible any-to-any connectivity, but traffic pays the latency of the electronic switching stages it traverses. 
The \emph{hybrid EPS/OCS} fabric keeps local connectivity electronic while replacing the level-2 switches with deployment-time optical circuits, making it an incremental path from current EPS fabrics toward optical scale-up. 
The \emph{all-OCS} fabric connects accelerator links directly to an optical circuit-switched fabric whose circuits are planned before serving starts and kept fixed throughout deployment.

The fabrics therefore differ in runtime routing flexibility and in how much electronic switching latency remains on accelerator-to-accelerator paths. 
In the all-OCS case, the full port budget \(L\) can be assigned to the logical overlays induced by decode, either to reduce serialized communication phases, widen selected logical edges, or support multiple phase-specific overlays over the same accelerator set.

Concretely, each accelerator exposes $L=36$ links of $R=400$\,Gb/s and all switching devices have a radix of 72, following NVL72-class systems. We instantiate a two-level electronic packet switch (EPS) based network (all-EPS), the same network with the devices on level 2 replaced by OCS (hybrid EPS/OCS), and a third topology in which all switching devices are OCS (all-OCS). In the all-EPS and hybrid fabrics, each level-1 switch uses 32 ports towards the accelerators and 8 towards level 2, i.e., a 4:1 oversubscription towards the second level, since a fully subscribed EPS spine with pluggable conversion exceeds a reasonable power envelope. The all-OCS fabric instead uses 32 ports in each direction and is thus non-blocking: its OCS devices require no E/O conversion and consume so little power per port (Appendix~\ref{app:power-model}) that the full uplink bandwidth still fits into a lower system power than either EPS-based fabric. Consequently, part of the all-OCS advantage for collective groups that span level 2 stems from this higher level-2 bandwidth rather than from the switching technology alone; the energy-efficiency results of Section~\ref{sec:eval} account for the full port count of each fabric. 

Note that while the EPS based fabric uses all traits well-known to EPS based scale-up networks such as multi-path routing and packet spraying in accelerators and switches, the all-OCS fabric runs on the collective schedules presented in Section~\ref{sec:topology-aware-collectives}, which in turn make use of the topologies described in Section~\ref{sec:logical-topology}. The hybrid EPS/OCS fabric configures the OCS on the topmost level once at deployment, using the same planner to place workers such that bandwidth demand is balanced across the level-2 OCS devices, and otherwise relies on the same collective operations as implemented by current inference frameworks such as TensorRT-LLM~\cite{trtllm}, vLLM~\cite{kwon2023vllm} or SGLang~\cite{zheng2024sglang}.
\section{Models and Accelerators}
\label{app:models}

This appendix documents the model architectures, accelerator parameters, and power assumptions underlying the evaluation in Section~\ref{sec:eval}, with model and accelerator specifications summarized in Tables~\ref{tab:model-props} and~\ref{tab:acc-props}. It also presents energy-efficiency results across the GPT-OSS family and examines the compute cost model against measured kernel durations, clarifying the assumptions and limitations of the projected accelerator results.

\subsection{Models}
\label{app:model-details}

\begin{table*}[htbp]
\centering
\caption{Model properties as used in the analyses of this paper. Active parameters include the embedding and LM head. Attention: full, or interleaved sliding-window (SW, 128-token window) and full layers. KV cache per token counts only layers whose cache grows with context.}
\label{tab:model-props}
\small
\setlength{\tabcolsep}{4pt}
\begin{tabular}{@{}lrrrrllrrrr@{}}
\toprule
& \multicolumn{2}{c}{Parameters} & & & \multicolumn{2}{c}{Attention} & \multicolumn{2}{c}{Experts} & & \\
\cmidrule(lr){2-3}\cmidrule(lr){6-7}\cmidrule(lr){8-9}
Model & Total & Active & Layers & Hidden & Type & \makecell[l]{Q / KV heads\\$\times$ dim}
  & \makecell[r]{Total /\\top-$k$} & Width & Vocab & \makecell[r]{KV cache\\per token} \\
\midrule
Megatron 235B & 0.23\,T & 17\,B & 96 & 1536 & GQA, full
  & $64 / 16 \times 96$ & 128 / 8 & 4096 & 51{,}200 & 144\,KiB \\
Megatron 1T & 1.10\,T & 38\,B & 128 & 2048 & GQA, full
  & $128 / 64 \times 32$ & 256 / 8 & 5440 & 51{,}200 & 256\,KiB \\
Megatron 5T & 5.00\,T & 207\,B & 128 & 8192 & GQA, full
  & $256 / 128 \times 64$ & 256 / 8 & 6144 & 51{,}200 & 1{,}024\,KiB \\
\addlinespace
GPT-OSS-120B & 0.12\,T & 5.7\,B & 36 & 2880 & GQA, SW / full
  & $64 / 8 \times 64$ & 128 / 4 & 2880 & 201{,}088 & 9\,KiB \\
GPT-OSS-1T-like & 1.05\,T & 21\,B & 64 & 4608 & GQA, SW / full
  & $72 / 8 \times 64$ & 256 / 4 & 4608 & 201{,}088 & 16\,KiB \\
GPT-OSS-5.3-like & 4.97\,T & 95\,B & 96 & 8192 & GQA, SW / full
  & $128 / 16 \times 64$ & 256 / 4 & 8192 & 201{,}088 & 48\,KiB \\
\bottomrule
\end{tabular}
\end{table*}

\begin{figure*}[htbp]
    \centering
    \subcaptionbox{Megatron 5T MoE\label{fig:model-megatron}}[0.48\linewidth]{%
        \includegraphics[width=\linewidth]{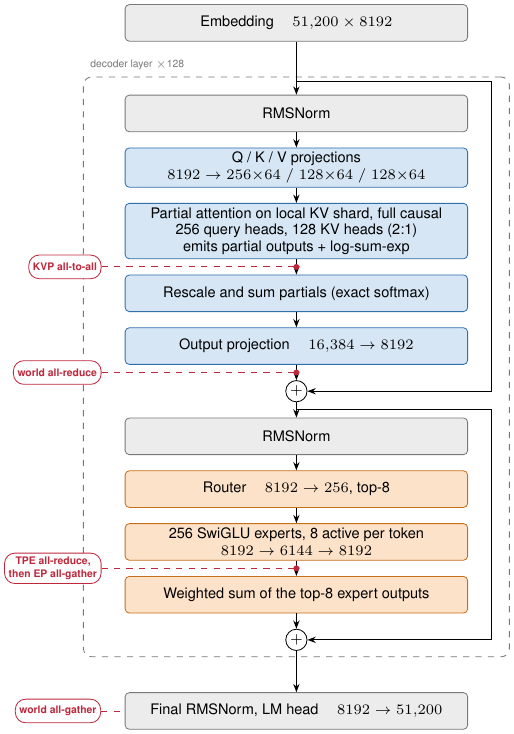}}\hfill
    \subcaptionbox{GPT-OSS-5.3-like (extrapolated from GPT-OSS-120B)\label{fig:model-gpt}}[0.48\linewidth]{%
        \includegraphics[width=\linewidth]{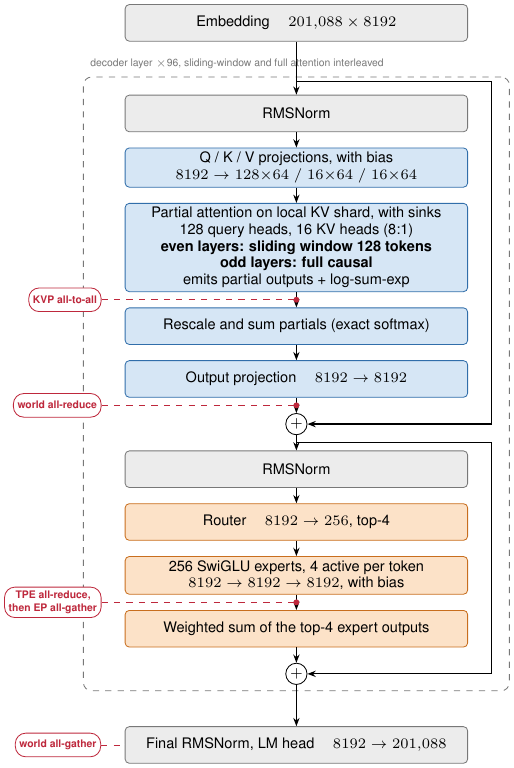}}
    \caption{One decoder layer of each model, with the tensor dimensions of its projections. Red markers show where Helix parallelism issues each collective of Table~\ref{tab:decode-demand}: the KV-parallel all-to-all that exchanges the partial attention outputs and log-sum-exp statistics each rank computes over its KV shard, before they are rescaled into the exact softmax result; the all-reduce over all accelerators after the output projection; the FFN tensor-parallel all-reduce followed by the expert-parallel all-gather, between the experts and their weighted combination; and one all-gather of the vocabulary-sharded logits per decode step.}
    \label{fig:model-graphs}
\end{figure*}

Figures~\ref{fig:model-megatron} and~\ref{fig:model-gpt} show one decoder layer of each model. GPT-OSS-5.3-like is our extrapolation of the open-weight GPT-OSS-120B to the 5T scale. It keeps the GPT-OSS architecture recipe, namely interleaved sliding-window and full attention with a 128-token window and attention sinks, eight query heads per KV head, experts as wide as the hidden size with four active per token, and the 201{,}088-token vocabulary. It scales the hidden size from 2880 to 8192, the depth from 36 to 96 layers, the attention from 64 query and 8 KV heads to 128 and 16, and the expert count from 128 to 256.

The two models share their overall shape. Both are pre-norm decoders of about 5T parameters with a hidden size of 8192, in which grouped-query attention is followed by a mixture-of-experts FFN of 256 SwiGLU experts. Under Helix parallelism both issue the same five collectives at the same points of the layer, so the deployment-static structure of Table~\ref{tab:decode-demand} is identical between them apart from the layer count. Because the hidden size is the same, the all-reduces and the expert all-gather also carry the same 4\,KiB per token in both models.

The models differ above all in their attention, which decides how a decode step responds to context length.
\begin{itemize}[leftmargin=*,itemsep=1pt,topsep=2pt]
\item \emph{Interleaved sliding-window attention.} GPT-OSS-5.3-like alternates between layers that attend only to the most recent 128 tokens and layers with full causal attention. The KV cache and attention time of a sliding-window layer do not depend on context length, so only 48 of its 96 layers read the whole context. Every one of Megatron's 128 layers does. Removing the sliding window from GPT-OSS-5.3-like in \textsc{gxnet} doubles the part of its attention time that grows with context, from 0.39 to 0.77\,ms per decode step between 32K and 5M context at batch 16.
\item \emph{Narrow grouped-query attention.} The full-attention layers of GPT-OSS-5.3-like keep 16 KV heads, eight query heads each. Megatron keeps 128 KV heads under 256 query heads, which also makes its attention 16{,}384 wide, twice the hidden size. Giving GPT-OSS-5.3-like Megatron's 128 KV heads multiplies its context-dependent attention time by eight, to 3.1\,ms in the same setting.
\item \emph{KV cache.} Together, the two points above give Megatron 1\,MiB of KV cache per token of context against 48\,KiB for GPT-OSS-5.3-like, a factor of 21. At 5M tokens one Megatron user occupies 5.2\,TB of HBM and one GPT-OSS-5.3-like user 246\,GB. This is what bounds the evaluated batch to 16 for Megatron on 512 accelerators, and it is why decode attention time grows much faster with batch and context for Megatron than for GPT-OSS-5.3-like (Section~\ref{sec:eval}).
\item \emph{Experts.} Megatron's experts are fine-grained (6144 wide, 0.75 of the hidden size) with eight active per token; those of GPT-OSS-5.3-like are as wide as the hidden size with four active. Megatron therefore activates 207\,B parameters per token against 95\,B.
\item \emph{Vocabulary and details.} GPT-OSS-5.3-like has a four times larger vocabulary, which only enlarges the LM head and the final logit all-gather. Its attention sinks and projection biases do not change the communication structure.
\end{itemize}

\subsection{Energy Efficiency across the GPT-OSS Family}
\label{app:gpt-family}

\begin{figure}[t]
  \centering
  \includegraphics{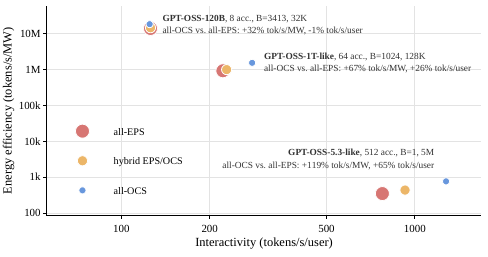}
  \caption{Energy efficiency vs.\ interactivity across models of the GPT-OSS family, analogous to Figure~\ref{fig:model-scaling}.}
  \label{fig:model-scaling-gpt}
\end{figure}

Figure~\ref{fig:model-scaling-gpt} repeats the comparison of Figure~\ref{fig:model-scaling} for three models of the GPT-OSS family: the open-weight GPT-OSS-120B on 8 accelerators at 32K context and batch 3413, GPT-OSS-1T-like on 64 accelerators at 128K context and batch 1024, and GPT-OSS-5.3-like on 512 accelerators at 5M context and batch 1.
GPT-OSS-1T-like is an intermediate extrapolation of GPT-OSS-120B built like GPT-OSS-5.3-like, with a hidden size of 4608, 64 layers, 72 query and 8 KV heads, and 256 experts (Table~\ref{tab:model-props}).
All-OCS fabrics improve throughput per unit energy by 32\,\%, 67\,\% and 119\,\%, respectively.
For GPT-OSS-120B, whose deployment spans only eight accelerators, interactivity is practically unchanged ($-1$\,\%), so the gain stems from system power alone; for GPT-OSS-1T-like, interactivity improves by 26\,\%, and for GPT-OSS-5.3-like, whose decode step is dominated by communication (Figure~\ref{fig:step-breakdown}), by 65\,\%.

\subsection{Accelerators}
\label{app:accelerators}

\begin{table*}[htbp]
\centering
\caption{Accelerator properties as used in the analyses of this paper. FP4 is the tensor-core peak that the roofline model scales by a shape-dependent efficiency. Every execution window between collectives additionally pays the in-graph launch overhead once (Appendix~\ref{app:gxnet-cost}).}
\label{tab:acc-props}
\small
\begin{tabular}{lrrrrrrr}
\toprule
Accelerator & FP4 peak & \makecell[r]{HBM\\capacity} & \makecell[r]{HBM\\bandwidth} & \makecell[r]{Scale-up\\ports} & \makecell[r]{Per-port bandwidth\\(unidirectional)} & \makecell[r]{Kernel\\launch} & \makecell[r]{In-graph\\launch} \\
\midrule
NVIDIA B200 & 20\,PFLOP/s & 186\,GB & 8\,TB/s & 18 & 400\,Gb/s & 2.5\,$\mu$s & 0.8\,$\mu$s \\
NVIDIA RX00 (Rubin-like) & 50\,PFLOP/s & 288\,GB & 22\,TB/s & 36 & 400\,Gb/s & 2.5\,$\mu$s & 0.8\,$\mu$s \\
\bottomrule
\end{tabular}
\end{table*}


RX00 is a projected accelerator modeled on announced Rubin figures (as announced at GTC in March 2026 \cite{nvidia2026verarubinnvl72}), so its kernel durations cannot be measured. Its parameters in Table~\ref{tab:acc-props} are peak throughput and bandwidth; the efficiency functions of Appendix~\ref{app:gxnet-cost} are a step approximation of curves measured on H100, which we use unchanged for B200 and RX00.

The cost model is a roofline with size-dependent efficiencies and no fixed cost per kernel. Operators between two collectives form one execution window that pays a single in-graph launch overhead, which assumes they run back to back as in a persistent kernel~\cite{cheng2025mpk}. The model therefore gives a lower bound on compute time rather than a prediction of a particular kernel implementation.

We choose a roofline deliberately. Predicting kernel durations more precisely requires either profiling operators on the target accelerator, as in Vidur~\cite{agrawal2024vidur}, or models fitted to measurements of existing kernels, such as PipeWeave~\cite{zhang2026pipeweave}. The first is impossible for an unreleased accelerator. The second generalizes to held-out GPUs of architectures seen in training, e.g., from H20 and H800 to H100 and H200, but has not been shown to carry over to a new architecture generation, and reports no separate error for the small batch sizes of decode, where durations depend on kernel setup, tiling, and occupancy rather than on FLOPs and bytes. Such models can be more accurate on the hardware they were fitted to, but their error on a new architecture has neither a known size nor a known sign. The error of a roofline has a known sign: it underestimates.

We compared the model with Nsight Systems traces of GPT-OSS-120B decode on four B200 with tensor parallelism and CUDA-graph replay, at batch sizes 1 to 16. The decode kernels there read a few to a few tens of MB per accelerator and reach 12 to 53\,\% of the peak HBM bandwidth, below the efficiencies of Appendix~\ref{app:gxnet-cost}, so the H100 curves remain optimistic one generation later. Even under CUDA-graph replay, no kernel finishes in less than about 3\,$\mu$s. These decode kernels therefore take 1.4 to 5 times as long as their roofline estimate, and more for the smallest ones. A persistent-kernel implementation avoids part of this per-kernel cost, which is the execution our fused windows assume.

Since the model underestimates compute time, the share of communication in a decode step, and with it the relative OCS gain of Section~\ref{sec:eval}, is an upper estimate, while the absolute reduction of collective completion time per step does not depend on the compute model. For existing accelerators, \textsc{gxnet} can replace the roofline estimate of any operator with a measured kernel duration without changing the graph or the sharding.

\subsection{Power Model}
\label{app:power-model}

Our power models are based on publicly available data. In conjunction with the topologies described in Appendix~\ref{app:physical-fabrics}, we always assume that moving between scale-up network levels 1 and 2 requires optical cabling (due to rack-to-rack distance) and that all-OCS based topologies make use of CPO at the accelerator. 
The power figures we use for the system components are given in Table~\ref{tab:power}, together with their origin wherever applicable.

\begin{table*}[t]
\centering
\caption{Component power assumptions of the power model (TDP basis, not load-aware).
Data rates are unidirectional; every port is bidirectional and carries one transceiver module.}
\label{tab:power}
\small
\begin{tabular}{@{}l l S[table-format=4.2] l p{6.2cm}@{}}
\toprule
Component & Variant & {Value} & Unit & Derivation / source \\
\midrule
\multirow{3}{*}{XPU (incl.\ host share)}
  & B200 (GB200 NVL72) & 1454  & W & $(P_\text{rack}-9P_\text{sw})/72$, $P_\text{rack}=\SI{120}{kW}$~\cite{nv-gb200,nv-mc}, $P_\text{sw}=\SI{1.7}{kW}$ \\
  & B300 (GB300 NVL72) & 1663  & W & as above, $P_\text{rack}=\SI{135}{kW}$~\cite{nv-mc,lenovo-gb300} \\
  & Rubin (VR200 NVL72) & 2631 & W & as above, $P_\text{rack}=\SI{220}{kW}$ (Max-P)~\cite{sa-vr,smc-vr}, $P_\text{sw}=\SI{3.4}{kW}$ (4 ASICs/tray~\cite{nv-rubin}) \\
\midrule
\multirow{3}{*}{EPS (scale-up), per port}
  & \SI{400}{G} & 11.81 & W & switch tray: $P_\text{sw}/(2\times72)$; cross-check~\cite{bcm-tu,ti-th5,ipi-th5} \\
  & \SI{800}{G} & 15.35 & W & extrapolated: $2\times$ rate, $-35\,\%$ energy/bit~\cite{arista-x5,arista-x6} \\
  & \SI{1600}{G} & 19.95 & W & extrapolated: $2\times$ rate, $-35\,\%$ energy/bit \\
\midrule
\multirow{4}{*}{OCS, per port}
  & \SI{200}{G} & 1.36 & W & \multirow{4}{=}{internal projections; unidirectional, two fibers per port} \\
  & \SI{400}{G} & 1.50 & W & \\
  & \SI{800}{G} & 1.65 & W & \\
  & \SI{1600}{G} & 1.82 & W & \\
\midrule
\multirow{3}{*}{Pluggable module}
  & \SI{400}{G} DR4 & 9 & W & vendor datasheet, max~\cite{cisco-400g} \\
  & \SI{800}{G} DR4 & 17 & W & \cite{sa-cpo} \\
  & \SI{1.6}{T} & 30 & W & \cite{sa-gb200} \\
\midrule
\multirow{3}{*}{CPO, per port}
  & \SI{400}{G} (\SI{6.9}{pJ/bit}) & 2.75 & W & Broadcom Bailly, \SI{5.5}{W} per \SI{800}{G}~\cite{bcm-bailly} \\
  & \SI{800}{G} (\SI{5.6}{pJ/bit}) & 4.50 & W & NVIDIA Quantum-X Photonics, 4--\SI{5}{W} per \SI{800}{G}~\cite{sa-cpo} \\
  & \SI{1600}{G} (\SI{4.4}{pJ/bit}) & 7.00 & W & Broadcom TH6 Davisson, \SI{3.5}{W} per \SI{800}{G}~\cite{tnp-cpo} \\
\bottomrule
\end{tabular}
\end{table*}

Table~\ref{tab:power} lists the power figures used by the model. All values are
thermal-design (maximum) figures, so the model gives an upper bound on power
independent of traffic. Per-XPU power is not the accelerator TDP. We take the
designed rack power of an NVL72-class system~\cite{nv-mc}, subtract the scale-up switch trays, and
divide by the 72 accelerators. The resulting figure therefore includes each
XPU's share of host CPU, NICs, DPU, fans and voltage-regulation losses.

No public figure exists for the GB200/GB300 switch tray, so we assume
\SI{1.7}{kW}. We cross-check this value by scaling the published power of a
\SI{51.2}{Tb/s} scale-up Ethernet switch ASIC (\SI{800}{W}~\cite{bcm-tu}) to the
two \SI{28.8}{Tb/s} switch ASICs per tray~\cite{sa-gb200}. We then apply the
system-to-ASIC overhead observed for Tomahawk~5 (\SI{913}{W} system without
optics~\cite{ipi-th5} versus about \SI{500}{W} ASIC~\cite{ti-th5}). This yields
about \SI{1.64}{kW}, within 4\,\% of our assumption.

No rack power has been published for Vera Rubin NVL72. We use the
\SI{220}{kW} Max-P estimate of~\cite{sa-vr}, which agrees with the
\SI{227}{kW} per-rack cooling sizing of an OEM reference design~\cite{smc-vr}.
The rack's power shelves (\(4\times\SI{110}{kW}\), N+1) define a provisioning
ceiling, not a TDP. The Rubin tray holds four switch ASICs instead of two~\cite{nv-rubin}. Each ASIC
keeps the per-ASIC bandwidth of \SI{28.8}{Tb/s}~\cite{sa-vr}, so we double the
tray power. Per \SI{400}{G}-equivalent port, the scale-up switching power is
therefore the same as for GB200.

EPS ports at \SI{800}{G} and \SI{1.6}{T} are extrapolated from the
\SI{400}{G} anchor assuming a 35\,\% energy-per-bit reduction per doubling of
port rate. This is the reduction observed between the 25.6 and
\SI{51.2}{Tb/s} generations of switch systems measured on the same basis
(typical power, excluding optics)~\cite{arista-x5,arista-x6}.

OCS figures are internal projections for a switch whose ports each use two unidirectional fibers. The power per port is \SI{1.5}{W} at \SI{400}{G}.

Every switching device, EPS or OCS, is evaluated with all of its physical ports,
whether a port is used or not. On the level-1 switches of the EPS-based
fabrics, which use 40 of their 72 ports, this is conservative, but the per-port
figure amortizes device-level power, such as that of the switch ASIC core,
which does not scale down with unused ports.

Optical links attach either through one pluggable transceiver per port or
through co-packaged optics (CPO). For CPO we use the energy per bit published
for switch-side optical engines, including the external laser source, and
multiply by the unidirectional rate. We assume the same CPO energy per bit when
an accelerator attaches directly to the level-1 OCS switches.
\section{Sensitivity Analysis}
\label{app:sensitivity}

This appendix examines the sensitivity of the evaluation results to network bandwidth, accelerator radix, switch and endpoint latency, and communication precision. These analyses assess how the relative performance and energy efficiency of the evaluated fabrics depend on the baseline configuration and modeling assumptions.

\begin{figure*}[htpb]
  \centering
  \includegraphics{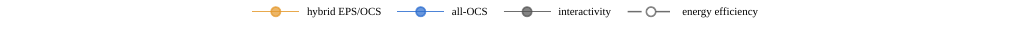}\\[-2pt]
  \includegraphics{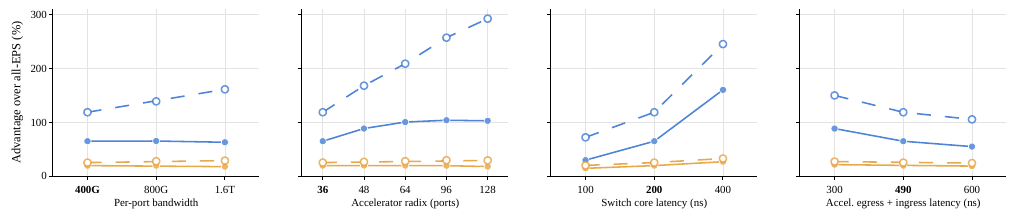}
  \caption{Sensitivity of the advantage of hybrid EPS/OCS and all-OCS fabrics over all-EPS for GPT-OSS-5.3-like at batch 1 and 5M context on 512 of 576 accelerators. Solid: interactivity; dashed: energy efficiency (tokens/s/MW). Bold tick labels mark the baseline configuration of Section~\ref{sec:eval}; the parallelization is re-optimized for every point.}
  \label{fig:sensitivity}
\end{figure*}

Figure~\ref{fig:sensitivity} varies one fabric or accelerator parameter at a time for GPT-OSS-5.3-like at batch 1 and 5M context, the latency-bound regime of Section~\ref{sec:eval}. The parallelization is re-optimized for every point, and all four sweeps share the baseline of Section~\ref{sec:eval}, with a 64.8\,\% interactivity and a 118.7\,\% energy-efficiency advantage of all-OCS over all-EPS. Since energy efficiency is throughput per system power, the energy advantage is the interactivity advantage multiplied by the ratio of all-EPS system power to that of the respective fabric, which is 1.33 for all-OCS at the baseline.

\paragraph{Per-port bandwidth.} At batch 1, the interactivity advantage of all-OCS is nearly insensitive to bandwidth, staying between 62.8\,\% and 65.2\,\% between 400\,Gb/s and 1.6\,Tb/s. Since the all-EPS baseline is oversubscribed towards its second level, it carries a bandwidth-bound component that shrinks with per-port bandwidth, whereas the non-blocking all-OCS fabric has little to gain. The energy advantage of all-OCS nevertheless grows from 119\,\% to 161\,\%, as the power ratio rises from 1.33 to 1.60: per rate doubling, EPS port power grows by 30\,\% and pluggable power almost doubles, while OCS port power grows by only 10\,\% and the energy per bit of CPO decreases (Appendix~\ref{app:power-model}). The hybrid fabric does not benefit, since it retains electronic level-1 switches and pluggables on every link between the two levels.

\paragraph{Accelerator radix.} Co-packaged optics allow the accelerator radix to grow. For inference decode, additional ports pay off substantially at first: the interactivity advantage of all-OCS rises from 65\,\% at 36 ports to 88\,\% at 48 ports and 100\,\% at 64 ports, but then saturates at 104\,\% and 103\,\% for 96 and 128 ports. Along the way, the planner changes the parallelization of all-OCS: it keeps the baseline split up to 48 ports (32/16/2/256 in KVP/TPA/EP/TPE), spreads the FFN over a single tensor-parallel group of all 512 accelerators (32/16/1/512) at 64 and 128 ports, and uses 32/16/8/64 at 96 ports, so additional ports buy freedom of parallelization as much as speed. The saturation follows from the schedulable frontier of Appendix~\ref{app:frontier}: phase counts drop in discrete steps, each further step costs considerably more ports than the previous one, e.g., two phases on an axis of 64 accelerators cost 14 ports and one phase 63, and ports between two steps buy no phase reduction. At a fixed system size, more accelerator radix therefore affords larger scale-up domains, or links for network-attached memory, rather than higher interactivity directly. In energy efficiency, however, the advantage of all-OCS keeps growing to 293\,\%, as the power ratio rises from 1.33 to 1.94 with every port added to the packet-switched fabric.

\paragraph{Switch latency attribution.} Our measurements (Section~\ref{sec:grounding}) cannot separate accelerator from switch latency, and the unloaded 300\,ns pin-to-pin latency of the UALink specification~\cite{ualinkspec} may be exceeded under load. We therefore keep the 810\,ns per switch hop of Table~\ref{tab:latency-budget} constant and shift latency between accelerator and switch core. Since the all-OCS path bypasses the switch core, every nanosecond moved into the switch shortens it: the interactivity advantage grows from 30\,\% to 65\,\% and 160\,\% for switch core latencies of 100, 200, and 400\,ns, at a constant power ratio of 1.33. The results are thus highly sensitive to where latency is spent, which explains the effort of recent architectures to keep latency-critical traffic within a single switch level: Jalape\~no bifurcates its links into separate single-level domains for tensor and expert parallelism~\cite{openai2026jalapeno}, a design-time counterpart of the stacked-link allocation of Appendix~\ref{app:stacked-link}.

\paragraph{Accelerator latency.} Reducing accelerator egress and ingress latency benefits both fabrics, but all-OCS considerably more, since its per-hop latency consists almost entirely of accelerator latency: its interactivity advantage changes by about 9 to 12 percentage points per 100\,ns, from 88\,\% at 300\,ns to 55\,\% at 600\,ns. The hybrid fabric gains almost nothing, as it runs the same packet-switched collective schedules over the same electronic level-1 paths as all-EPS. Collective acceleration engines~\cite{google2026tpu8} that shorten the accelerator-side path therefore directly increase the advantage of workload-static OCS, and so does lighter forward error correction: an EPS switch terminates FEC at every hop, whereas a passive optical path is terminated once at its endpoints. Scale-Up Ethernet names RS-272 as a candidate to reduce receive latency~\citep[Sec.~3.5.4]{broadcom2025suerm104}, with link-level retransmission as the hardware-based safety net for residual errors.

\paragraph{Communication precision.}
Our baseline models activation communication in FP4, consistent with the assumed activation storage precision (Appendix~\ref{app:gxnet-passes}).
This is a forward-looking assumption: many serving collectives still communicate in BF16, including the default MoE combine paths in DeepEP~\cite{deepep} and TensorRT-LLM~\cite{trtllm-blog18}.
However, lower-precision transport is already supported for MoE dispatch (FP8/NVFP4) and optionally for combine (NVFP4) ~\cite{deepep,trtllm-blog14,trtllm-blog18}.

To assess sensitivity to communication precision, we repeat the evaluation of Figure~\ref{fig:per-batch-both} using BF16 activation transfers, quadrupling the bytes per activation element relative to FP4, while Helix softmax statistics are communicated in FP32.
For each precision, we independently repeat the full parallelization search of Appendix~\ref{app:gxnet-search}, selecting the best-performing configuration for each workload and fabric.
We account for the resulting communication volumes but do not model precision-conversion overheads or numerical accuracy effects.

All-OCS continues to outperform all-EPS in every configuration, with an interactivity advantage of 9.6--75.7\,\%, compared with 7.3--81.6\,\% under FP4. 
Up to batch size 8, the advantage generally decreases by up to 16 percentage points. 
At batch size 16, it instead increases by 1--3 points, consistent with additional bandwidth pressure on the oversubscribed EPS fabric. 
The hybrid fabric changes by at most 3 points, except for Megatron 5T at 32K context and batch size 16, where its advantage decreases from 12.5 to 5.7\,\%.

Thus, the relative benefits depend on communication precision, but the conclusions of Section~\ref{sec:eval} remain unchanged. 
The HOP-B findings (Appendix~\ref{app:hopb}) are also robust: repeated
all-to-all startup costs continue to outweigh overlap savings for GPT-OSS-5.3-like, whose decode-step time under BF16 increases by 58\,\% on all-EPS and 24\,\% on all-OCS with eight microbatches.

\end{document}